\documentclass[11pt,reqno]{amsart}

\usepackage[T1]{fontenc}
\usepackage[utf8]{inputenc}
\usepackage[english]{babel}

\usepackage[a4paper,margin=1.15in]{geometry}
\usepackage{microtype}
\usepackage{newpxtext}
\usepackage{mathtools}
\usepackage{amssymb}
\usepackage{newpxmath}
\usepackage{bm}
\usepackage{braket}
\usepackage{extpfeil}
\input{Ustmry.fd}
\DeclareFontShape{U}{stmry}{m}{n}{
  <5><6><7><8><9><10> gen * stmary
  <5.5> stmary5
  <10.95><12><14.4><17.28><20.74><24.88> stmary10
}{}
\allowdisplaybreaks
\numberwithin{equation}{section}

\usepackage{array}
\usepackage{tabularx}

\usepackage[shortlabels]{enumitem}
\usepackage[title]{appendix}
\usepackage[all,poly]{xy}
\usepackage{tikz}
\usetikzlibrary{arrows.meta,positioning}

\usepackage{xcolor}
\definecolor{linkblue}{RGB}{0,72,128}
\usepackage[
colorlinks=true,
linkcolor=linkblue,
citecolor=linkblue,
urlcolor=linkblue
]{hyperref}
\usepackage[nameinlink,noabbrev]{cleveref}

\theoremstyle{plain}
\newtheorem{thm}{Theorem}[section]

\newtheorem{lemma}[thm]{Lemma}

\newtheorem{corollary*}[thm]{Corollary*}
\newtheorem{proposition}[thm]{Proposition}
\newtheorem{proposition*}[thm]{Proposition*}

\theoremstyle{definition}

\newtheorem{defn-thm}[thm]{Definition-Theorem}

\theoremstyle{remark}
\newtheorem{remark}[thm]{Remark}

\newcommand{\Rmnum}[1]{\uppercase\expandafter{\romannumeral #1}}

\tikzset{
	solid black dot/.style={
		circle,
		fill=black,
		inner sep=1.5pt
	}
}

\def\d{\mathrm{d}}

\begin{document}

\raggedbottom
	
	 \title{On Emergent Geometry of the Generalized BGW Theory}
	\author{Dingkang Xiong}
	\address{Department of Mathematical Sciences, Tsinghua University, Beijing 100084, China}
	\email{xdk22@mails.tsinghua.edu.cn}
	\author{Jian Zhou}
	\address{Department of Mathematical Sciences, Tsinghua University, Beijing 100084, China}
	\email{jianzhou@mail.tsinghua.edu.cn}
	
	\begin{abstract}
			We present a systematic study of the mathematical theory of the generalized Brézin-Gross-Witten (BGW) theory using methods from statistical physics. We prove the multidifferentials of the generalized BGW theory are given by the Eynard-Orantin topological recursion in a generic background with general coupling constants. Furthermore, we characterize the transition between the Witten-Kontsevich model and the generalized BGW theory. This approach yields a new proof of the Okuyama-Sakai conjecture, further elucidating the deep structural connections between these two fundamental models.
		\end{abstract}
	\maketitle
	
	\setcounter{tocdepth}{1}
	\tableofcontents
	
	\section{Introduction}
	
	\subsection{Emergent geometry of the Witten-Kontsevich model}
	Gromov-Witten theory serves as the mathematical foundation for physical theories of topological sigma models coupled to two-dimensional topological gravity. 
	For a given sigma model, the theory characterizes the partition function, or equivalently the free energy, expressed as a formal power series in an infinite set of variables, known as coupling constants in the physical literature.
	
	Discussing functions of infinitely many variables is often mathematically cumbersome. In statistical physics, physicists have developed some systematic approaches to handle systems with very large or infinite degrees of freedom. In a series of papers \cite{zhouTopological1DGravity2014,zhouEmergentGeometryMirror2015,zhouPhaseTransitionTwoDimensional2024,zhouEmergentGeometryUnifications2025}, the second-named author has proposed to borrow the notions in statistical physics to interpret and develop various results in Gromov-Witten theory. 
	This paper is the outcome of applications to the generalized BGW theory of this proposal.
	
	The following formalisms from statistical physics have been applied to Gromov-Witten.
	The first is  the Landau-Ginzburg formalism of mean field theory which include the introduction of order parameters, Landau-Ginzburg potential and Landau-Ginzburg equation \cite{dijkgraafMeanFieldTheory1990}.
	The second  is Wilson's renormalization theory \cite{wilsonRenormalizationGroupCritical1975} in which the notion of theory space with all possible coupling constants and renormalized coupling constants are introduced.
	The third is the notion of  emergence phenomena in complex systems introduced by Anderson \cite{andersonMoreDifferentBroken1972}.
	
	The prototypical case is the Gromov--Witten theory of a point, also referred to as the Witten-Kontsevich model. The genus-$g$ free energy is defined as follows:
	\begin{equation}
		F_g^{WK}(\mathbf{t}):=\sum_{n_0,n_1,n_2,\cdots} \langle \tau_0^{n_0}\tau_1^{n_1}\tau_{2}^{n_2}\cdots\rangle_g \frac{t_0^{n_0}}{n_0!}\frac{t_1^{n_1}}{n_1!}\frac{t_2^{n_2}}{n_2!}\cdots.
	\end{equation} 
	The correlators $\langle \tau_{a_1}\tau_{a_2}\cdots\tau_{a_n}\rangle_g$ are interpreted by Witten \cite{wittenTopologicalSigmaModels1988,wittenTwodimensionalGravityIntersection1990} as the intersection numbers of $\psi$-classes on the Deligne-Mumford compactification of the moduli space $\overline{M}_{g,n}$:
	\begin{equation}
		\langle \tau_{a_1}\tau_{a_2}\cdots\tau_{a_n}\rangle_g:=\int_{\overline{M}_{g,n}} \psi_1^{a_1}\psi_2^{a_2}\cdots\psi_{n}^{a_n}.
	\end{equation} 
	As conjectured by Witten in \cite{wittenTwodimensionalGravityIntersection1990} and proved by Kontsevich in \cite{kontsevichIntersectionTheoryModuli1992}, the partition function $Z^{WK}(\mathbf{t};\lambda):=\exp\big(\sum_{g=0}^{\infty} \lambda^{2g-2}F^{WK}_g(\mathbf{t})\big) $ is a tau-function of the KdV hierarchy.
	
	In the literature,  Dijkgraaf and Witten \cite{dijkgraafMeanFieldTheory1990} first introduced mean field theory into topological sigma models. They identified specific second derivatives of the genus-zero free energy as the theory's order parameters, subsequently proving that all second derivatives of genus-zero free energy are functions solely of these parameters.
	Furthermore, they derived the integrable systems satisfied by the order parameters. 
	In the Witten-Kontsevich model, the order parameter is defined as $u_0:=\partial^2 F^{WK}_0/\partial t_0^2$,
	and one then has the following constitutive relations \cite{dijkgraafMeanFieldTheory1990}: 
	\begin{equation}\label{cons of WK}
		\frac{\partial^2 F^{WK}_0}{\partial t_n\partial t_m}=\frac{u_0^{n+m+1}}{(n+m+1)n!m!}
	\end{equation}
	for any $n$ and $m$.  
	The order parameters are determined by the string equations, which can be viewed as the equations for the Landau-Ginzburg theory. 
	In the Witten-Kontsevich model, there is only one order parameter, and the Landau-Ginzburg potential is 
	\begin{equation}
		W^{WK}(x):=-\frac{1}{2}x^2+\sum_{n=0}^{\infty} t_n\frac{x^{n+1}}{(n+1)!}.
	\end{equation}
	The order parameter $u_0$ is the critical point of the function $W^{WK}(x)$; that is
	\begin{equation}
		\frac{\d W^{WK}}{dx}(u_0)=-u_0+\sum_{n=0}^{\infty}t_n\frac{u_0^n}{n!}=0.
	\end{equation}
	It is called the Landau-Ginzburg equation for the order parameter $u_0$. 
	
	Using the constitutive relation, Dijkgraaf and Witten dealt with the genus-zero free energy of the Witten-Kontsevich model. They also provided a formula for genus-one free energy \cite{dijkgraafMeanFieldTheory1990}. 
	For generalizations to higher genera which involve derivatives of the order parameters, 
	see e.g. \cite{eguchiGenusExpansionTopological1995} and \cite{dubrovinNormalFormsHierarchies2001}.
	
	On the other hand, Itzykson and Zuber \cite{itzyksonCombinatoricsModularGroup1992} introduced a set of variables $\mathbf{I}=\{I_0,I_1,I_2,\cdots\}$ defined by
	\begin{equation}
		I_n:=\sum_{m=0}^{\infty}t_{n+m}\frac{u_0^m}{m!}.
	\end{equation}
	In particular, by the Landau-Ginzburg equation, one obtains $I_0=u_0$.
	With these variables, they made a conjecture on the shape of $F^{WK}_{g}$ with $g\geq 2$:
	\begin{equation}
		F_g^{WK}=\sum_{l_2+2l_3+\cdots+(3g-3)l_{3g-2}=3g-3} \langle \tau_2^{l_2}\tau_{3}^{l_3}\cdots \tau_{3g-2}^{l_{3g-2}}\rangle_g \prod_{j=2}^{3g-2}\frac{1}{l_j!}\tilde{I}_j^{l_j},
	\end{equation}
	where $\tilde{I}_j=(1-I_1)^{-\frac{2j+1}{3}}I_j$.
	
	In this formula, the free energy is simply a polynomial in finitely many variables.  
	The second-named author \cite{zhouTopological1DGravity2014} interpreted $I_n$
	as new coupling constants emerging naturally from the renormalization process of the Landau-Ginzburg potential $W^{WK}(x)$.     Specifically, the variables $I_1,I_2,\cdots$ are the coefficients in the Taylor expansion of the Landau-Ginzburg potential at the critical point $x=I_0$; i.e.,
	\begin{equation}
		W^{WK}(x)=W^{WK}(I_0)-\frac{1}{2}(x-I_0)^2+\sum_{n=1}^{\infty}I_{n}\frac{(x-I_0)^{n+1}}{(n+1)!!}.
	\end{equation}
	Thus, we call $\mathbf{I}=\{I_0,I_1,\cdots\}$ the \textbf{renormalized coordinates} of the Witten-Kontsevich model. See \cite{zhouTopological1DGravity2014} for a justification that it can be viewed as a coordinate transformation.
	
	The Itzykson-Zuber Ansatz can be proved based on this perspective \cite{zhouTopological1DGravity2014, zhangItzyksonZuberAnsatz2019}. 
	The core of the proof relies on rewriting the Virasoro constraints \cite{dijkgraafLoopEquationsVirasoro1991} for the Witten-Kontsevich model in terms of the $I$-coordinates. 
	
	The next step is the emergence phenomena. It can be presented by the Eynard-Orantin topological recursion theory.
	The theory was proposed by Eynard and Orantin \cite{eynardInvariantsAlgebraicCurves2007}, and they were inspired by results in the theory of matrix models (with significant contributions by Chekhov \cite{chekhovFreeEnergyTopological2006,chekhovHermiteanMatrixModel2006}). 
	The recursion begins with the data of a Riemann surface equipped with two meromorphic functions and a Bergman kernel; this object is called the spectral curve of the model. The Eynard-Orantin topological recursion then produces a set of multidifferentials that encode the correlators of the model. 
	
	For the Witten-Kontsevich model, it has been shown that the spectral curve is the Airy curve \cite{eynardInvariantsAlgebraicCurves2007,zhouTopologicalRecursionsEynard2013}; i.e., the Riemann sphere with the following data:
	\begin{equation}
		x(z)=\frac{1}{2}z^2,\ y(z)=z,\ B_0(z_1,z_2)=\frac{\d z_1\d z_2}{(z_1-z_2)^2}.
	\end{equation} 
	Through the Eynard-Orantin topological recursion, one obtains the multidifferentials on the Riemann sphere
	\begin{equation}
		\begin{aligned}
			\omega_{g,n}(z_1,z_2,\cdots,z_n)=\sum_{a_1,a_2,\cdots,a_n\geq 0}\langle\tau_{a_1}\tau_{a_2}\cdots\tau_{a_n}\rangle_g\prod_{i=1}^{n}\frac{(2a_i+1)!!\d z_i}{z_i^{2a_i+2}}.
		\end{aligned}
	\end{equation}
	This encodes the intersection numbers that serve as the correlators of the Witten-Kontsevich model. 
	
	It is a good place to explain how the notion of emergence appears in the setting of Gromov-Witten theory. 
	Traditionally, a reconstruction approach has been taken in the computations of Gromov-Witten theory.
	For the Witten-Kontsevich case, all intersection numbers $\langle\tau_{a_1}\tau_{a_2}\cdots\tau_{a_n}\rangle_g$ can be recursively determined starting with $\langle\tau_0^3\rangle_0= 1$ and $\langle\tau_1\rangle_1$.
	In general, one starts with genus zero correlators of primary operators which defines a Frobenius manifold in the sense of Dubrovin \cite{dubrovinGeometry2DTopological1996},
	and reconstruct the whole theory by Dubrovin-Zhang formalism \cite{dubrovinNormalFormsHierarchies2001} or Givental quantization and Teleman's reconstruction theorem \cite{giventalGromovWittenInvariantsQuantization2001,telemanStructure2DSemisimple2012} when the Frobenius manifold is semisimple.
	The second-named author proposed to understand the Eynard-Orantin topological recursion from an emergence point of view.
	To understand the appearance of the spectral curve, he introduced a \textbf{special deformation} of the original spectral curve using the genus-zero one-point functions (derivatives of the genus-zero free energy) in a series of works \cite{zhouEmergentGeometryMirror2015,zhouEmergentGeometryMatrix2019,yangGrothendiecksDessinsDenfants2023,zhouFatThinEmergent2018}. For the Witten-Kontsevich model, the deformed spectral curve is
	\begin{equation} \label{eqn:Spec-def} y^{WK}(w)=w^{\frac{1}{2}}-\sum_{n=0}^{\infty}\frac{t_n}{(2n-1)!!}w^{n-\frac{1}{2}}-\sum_{n=0}^{\infty}\frac{(2n+1)!!}{w^{n+\frac{3}{2}}}\frac{\partial F^{WK}_0}{\partial t_n}.
	\end{equation} 
	Note in \eqref{eqn:Spec-def} all coupling constants $t_n$'s are involved. 
	The deformed spectral curve cannot be determined by only finitely many coupling constants.
	This reflects the title of Anderson's paper: More is different.
	As a reformulation of \cite{zhangItzyksonZuberAnsatz2019} Theorem 5.4.1, it has been shown that  in \cite{zhouEmergentGeometryUnifications2025} this spectral curve can be written as 
	\begin{equation}
		\begin{aligned}
			y^{WK}(w)&=(w-2I_0)^{\frac{1}{2}}-\sum_{n=1}^{\infty} \frac{I_{n}}{(2n-1)!!}(w-2I_0)^{n-\frac{1}{2}},\\
			&=z(1-\sum_{n=0}^{\infty}I_{n+1}\frac{z^{2n}}{(2n+1)!!}),
		\end{aligned}
	\end{equation}
	with $w=z^2+2I_0$. 
	Moreover, the multidifferentials generated by this spectral curve via the Eynard-Orantin topological recursion (with $x=\frac{1}{2}w=\frac{1}{2}z^2+I_0$ and the same Bergman kernel $B_0(z_1,z_2)$) are given by \cite{zhouEmergentGeometryUnifications2025}
	\begin{equation}
		\begin{aligned}
			W_{g,n}^{WK}(z_1,z_2,\cdots,z_n;\mathbf{t})&=\sum_{a_1,a_2,\cdots,a_n\geq 0}\frac{\partial^n F_g^{WK}}{\partial t_{a_1}\cdots\partial t_{a_n}}\prod_{i=1}^n\frac{(2a_i+1)!!\d x_i}{(2x_i)^{a_i+\frac{3}{2}}}.
		\end{aligned}
	\end{equation}
	When the parameters $\mathbf{t}=\{t_0,t_1,\cdots\}$ are set to zero,  the multidifferential $W_{g,n}$ reduces to the original multidifferential $\omega_{g,n}$. 
	
	One may specialize the parameters $\mathbf{I}$ to specific values to obtain a new spectral curve that encodes the information of another theory. This process can be interpreted as a phase transition of theories. It has been shown in \cite{zhouPhaseTransitionTwoDimensional2024,zhouEmergentGeometryUnifications2025} that certain special values of $\mathbf{I}$ correspond to specific $(p,q)$- minimal models in conformal field theory.
	
	Similar to emergent phenomena in statistical physics, a rich geometric structure emerges from a theory with infinitely many degrees of freedom. 
	The second-named author has termed this phenomenon \textbf{Emergent Geometry}. 
	In this paper we will focus on the emergence of spectral curves.
	For emergence of other geometric structures, see \cite{zhouEmergentGeometryKP2024}.
	
	\subsection{Emergent geometry of the generalized BGW theory and main results}
	
	The Brézin-Gross-Witten (BGW) model was introduced in lattice gauge theory as a unitary matrix integral by Brézin and Gross \cite{brezinExternalFieldProblem1980} and by Gross and Witten \cite{grossPossibleThirdorderPhase1980}. It later became one of the basic examples of non-eigenvalue unitary matrix models. Mironov, Morozov and Semenoff showed that the BGW partition function can be represented as a generalized Kontsevich matrix integral \cite{mironovUnitaryMatrixIntegrals1996}. From this point of view the model admits a natural one-parameter deformation (an analogue of the 't Hooft coupling constant), usually called the generalized BGW partition function. This deformation has been studied from several complementary perspectives: cut-and-join operators and Virasoro constraints \cite{alexandrovCutandjoinDescriptionGeneralized2019}, isomonodromic deformations and the Painlevé XXXIV hierarchy \cite{bertolaBrezinGrossWittenTauFunction2019}, Schur $Q$-polynomial and BKP descriptions \cite{alexandrovGeneralizedBrezinGross2023,liuQpolynomialExpansionBrezinGrossWitten2022,wangBKPaffineCoordinatesEmergent2025}, the Eynard-Orantin topological recursion (the ordinary case is in \cite{doTopologicalRecursionBessel2018} and the generalized case is in \cite{wangBKPaffineCoordinatesEmergent2025,zhouEmergentGeometryUnifications2025}), the relation with the Hodge integrals \cite{yangHodgeBGWCorrespondence2024}, and the relation with the Witten-Kontsevich model \cite{wangConnectionKontsevichWittenBrezinGrossWitten2022,okuyamaJTSupergravityBrezinGrossWitten2020,yangHodgeBGWCorrespondence2024,yangNewProofOkuyama2023,zhouEmergentGeometryUnifications2025,zhouRelating2026}. The ordinary BGW partition function also has an enumerative interpretation through Norbury's $\Theta$-classes on $\overline{M}_{g,n}$ (conjectured by Norbury in \cite{norburyNewCohomologyClass2023}, proved by Chidambaram, Garcia-Failde and Giacchetto in \cite{chidambaramRelations$overlinemathcalM_gn$Negative2025}); in physics it appears naturally in JT supergravity and super Weil-Petersson volumes \cite{stanfordJTGravityEnsembles2020a,okuyamaJTSupergravityBrezinGrossWitten2020, alexandrovSuperVolumesKdV2024}. These developments make the generalized BGW theory a close analogue of the Witten-Kontsevich model, even though it does not arise directly from a topological sigma model.
	
	In this article we use the following normalization of the generalized BGW partition function:
	\begin{equation}
		Z^{gB}(\mathbf{t};S;\lambda)=\exp\sum_{g=0}^{\infty}\lambda^{2g-2}F_g^{gB}(\mathbf{t};S),
	\end{equation}
	where $\mathbf{t}=\{t_0,t_1,\cdots\}$ are formal variables, $S$ is the analogue of the 't Hooft parameter, and $\lambda$ is the string coupling constant. 
	The partition function is uniquely characterized by the Virasoro constraints \cite{alexandrovCutandjoinDescriptionGeneralized2019}
	\begin{equation}
		L_m^{gB}Z^{gB}(\mathbf{t};S;\lambda)=0,\qquad m\geq0,
	\end{equation}
	where
	\begin{equation}
		\begin{aligned}
			L_m^{gB}:=&\sum_{k=0}^{\infty}\frac{(2k+2m+1)!!}{(2k-1)!!}(t_k-\delta_{k,0})\frac{\partial}{\partial t_{k+m}}\\
			&+\frac{\lambda^2}{2}\sum_{a+b=m-1}(2a+1)!!(2b+1)!!\frac{\partial^2}{\partial t_a\partial t_b}
			+\left(\frac{1}{8}+\frac{S^2}{2\lambda^2}\right)\delta_{m,0}.
		\end{aligned}
	\end{equation}
	These operators satisfy
	\begin{equation}
		[\frac{1}{2}L_n^{gB},\frac{1}{2}L_m^{gB}]=(n-m)\frac{1}{2}L_{n+m}^{gB},\qquad n,m\geq0.
	\end{equation}
	
	The analogy with the Witten-Kontsevich model becomes sharper at genus zero. In Zhang's PhD thesis \cite{zhangApplication2020}, the generalized BGW theory was shown to possess an order parameter $Q:=\partial^2 F_0^{gB}/\partial S\partial t_0$.	Zhang proved that the generalized BGW theory, although not a standard Gromov-Witten theory, has the same mean-field and Landau-Ginzburg structure that Dijkgraaf and Witten found for topological sigma models \cite{dijkgraafMeanFieldTheory1990}.

	In this article, for the purposes of emergent geometry, however, the more natural order parameter is (see Section \ref{section LG theory of gBGW})
	\begin{equation}
		J_0:=\frac{\partial^2F_0^{gB}}{\partial t_0^2}=\frac{Q^2}{2}.
	\end{equation}
	With this choice, the Landau-Ginzburg equation takes the form
	\begin{equation}
		J_0=\sum_{n=1}^{\infty}\frac{(2n-1)t_{n-1}}{n!}J_0^n+\frac{\partial F_0^{gB}}{\partial t_0}.
	\end{equation}
	The corresponding Landau-Ginzburg potential is
	\begin{equation}
		W^{gB}(x):=-\frac{1}{2}x^2+\frac{\partial F_0^{gB}}{\partial t_0}x+\sum_{n=1}^{\infty}\frac{(2n-1)t_{n-1}}{(n+1)!}x^{n+1}.
	\end{equation}
	Expanding this potential around the critical point $x=J_0$ produces the renormalized coordinates 
	\begin{equation}
		J_n:=\sum_{m=0}^{\infty}(2n+2m-1)\frac{t_{n+m-1}}{m!}J_0^m,\qquad n\geq1.
	\end{equation}
	This is the renormalized coordinate system that plays the role of the Itzykson-Zuber coordinates in the generalized BGW theory.
	
	Let us now recall the known topological-recursion results for the BGW and generalized BGW theories.
	For the ordinary BGW theory, Do and Norbury \cite{doTopologicalRecursionBessel2018} considered the Bessel curve
	\begin{equation}
		x(z)=\frac{1}{2}z^2,\qquad y(z)=\frac{1}{z},\qquad 
		B_0(z_1,z_2)=\frac{\d z_1\d z_2}{(z_1-z_2)^2}.
	\end{equation}
	They proved the Eynard-Orantin topological recursion of this spectral curve produces the correlators $\langle\tau_{i_1}\cdots\tau_{i_n}\rangle^{BGW}_{g}$ of the BGW theory.

	The enumerative meaning of these BGW correlators is supplied by Norbury's cohomology classes $\Theta_{g,n}\in H^*(\overline{M}_{g,n},\mathbb{Q})$ \cite{norburyNewCohomologyClass2023}. Norbury conjectured and Chidambaram, Garcia-Failde and Giacchetto proved in \cite{chidambaramRelations$overlinemathcalM_gn$Negative2025} that
	\begin{equation}
		\langle\tau_{i_1}\cdots\tau_{i_n}\rangle^{BGW}_{g}:= \frac{\partial^n F^{BGW}_g}{\partial t_{i_1}\cdots\partial t_{i_n}}|_{\mathbf{t}=0}
		=	\int_{\overline{M}_{g,n}}\Theta_{g,n}\prod_{j=1}^n\psi_j^{i_j}.
	\end{equation}
	 
	 In \cite{zhouEmergentGeometryUnifications2025}, the second-named author considered the special deformation for the Bessel curve
	 \begin{equation}
	 	y^{BGW}(w)= w^{-\frac{1}{2}}-\sum_{n=0}^{\infty} \frac{t_n}{(2n-1)!!}w^{n-\frac{1}{2}}.
	 \end{equation} 
	 It has no negative power part other than $w^{-\frac{1}{2}}$ since the genus-zero free energy of the BGW theory vanishes. Then, he proved the Eynard-Orantin topological recursion for this spectral curve
	 \begin{equation}
	 	y^{BGW}(z)=\frac{1}{z}-\sum_{n=0}^{\infty} \frac{t_n}{(2n-1)!!}z^{2n-1}, x(z)=\frac{1}{2}z^2, B_0(z_1,z_2)=\frac{\d z_1\d z_2}{(z_1-z_2)^2}
	 \end{equation}
	 produces the multidifferentials 
	 \begin{equation}
	 	W_{g,n}^{BGW}=\sum_{i_1,\cdots,i_n}\frac{\partial^n F^{BGW}_g}{\partial t_{i_1}\cdots\partial t_{i_n}} \prod_{k=1}^{n} \frac{(2i_k+1)!!\d z_k}{z_k^{2i_k+2}}.
	 \end{equation}
	This result is related to the Eynard-Orantin topological recursion for super Weil--Petersson volumes on moduli spaces of super Riemann surfaces \cite{xiongEYNARDORANTINTOPOLOGICALRECURSIONS,ahmedDiscreteAnalogue$mathscrN2$2026}.

	For the generalized BGW theory, 
	Alexandrov conjectured \cite{alexandrovCutandjoinDescriptionGeneralized2019} that the spectral curve for it is defined by 
	\begin{equation}
		x^2y^2-x+\frac{S^2}{4}=0,
	\end{equation} 
	with the Bergman kernel given by the genus-zero two-point functions. 
	
	On the other hand, Wang--Yang--Zhang \cite{wangBKPaffineCoordinatesEmergent2025} proved that the Eynard-Orantin topological recursion for the spectral curve
	\begin{equation}	
		x^2y^2=x^2+S^2,
	\end{equation}
	with the Bergman kernel given by the genus-zero two-point functions produces the multidifferentials encoding the correlators of the generalized BGW theory.

	In \cite{zhouEmergentGeometryUnifications2025}, the second-named author proved that these two spectral curves are equivalent. He considered the special deformation for the generalized BGW theory  by the genus-zero one-point functions:
	\begin{equation}
		y^{gB}(w;\mathbf{t};S):=w^{-\frac{1}{2}}-\sum_{n=0}^{\infty}\frac{t_n}{(2n-1)!!}w^{n-\frac{1}{2}}
		-\sum_{n=0}^{\infty}\frac{(2n+1)!!}{w^{n+\frac{3}{2}}}\frac{\partial F_0^{gB}}{\partial t_n}.
	\end{equation}
	Setting the variables $\mathbf{t}$ to zero gives the spectral curve
	\begin{equation}
		y^{gB}(w;\mathbf{t}=0;S)=\frac{\sqrt{w-S^2}}{w}=\frac{z}{z^2+S^2},
	\end{equation}
	where $w=z^2+S^2$. With $x=\frac{1}{2}w=\frac{1}{2}z^2+\frac{1}{2}S^2$ and the standard Bergman kernel $B_0(z_1,z_2)$, he proved that the Eynard-Orantin topological recursion for this spectral curve produces
	\begin{equation}
		W^{gB}_{g,n}(z_1,\cdots,z_n;\mathbf{t}=0;S)=\sum_{i_1,\cdots,i_n\geq 0} \frac{\partial^n F^{gB}_g(\mathbf{t};S)}{\partial t_{i_1}\cdots\partial t_{i_n}}\big|_{\mathbf{t}=0}\prod_{k=1}^{n}\frac{(2i_k+1)!!\d x_k}{(2x_k)^{i_k+\frac{3}{2}}}.
	\end{equation}
	Alexandrov and Norbury \cite{alexandrovSuperVolumesKdV2024} provided a geometric description for the correlators given by this spectral curve.
	
	In this article, we continue to consider the special deformation $y^{gB}(w;\mathbf{t};S)$ with general coupling constants $\mathbf{t}$. By the result of \cite{zhangApplication2020}, we derive the following form for the deformed spectral curve: (see \textbf{Proposition \ref{spec curve in J}})
	\begin{equation}\label{special def in J}
		\begin{aligned}
			y^{gB}=&\frac{\sqrt{w-2J_0}}{w}(1-\sum_{n=0}^{\infty}\frac{(w-2J_0)^{n}}{(2n+1)!!}J_{n+1})\\
			=&\frac{z}{z^2+2J_0}(1-\sum_{n=0}^{\infty}\frac{z^{2n}}{(2n+1)!!}J_{n+1}),
		\end{aligned}
	\end{equation}
	with the coordinate transformation $w=z^2+2J_0$. Generalizing the method of \cite{zhouEmergentGeometryUnifications2025}, we prove the Eynard-Orantin topological recursion for this special deformation with
	\begin{equation}\label{x and B}
	 x(z)=\frac{1}{2}z^2+J_0, \quad B_0(z_1,z_2)=\frac{\d z_1\d z_2}{(z_1-z_2)^2}
	\end{equation}
	 produces the multidifferentials of the generalized BGW theory in a generic background: 
	 \begin{thm}[\textbf{Theorem \ref{EO TR foe gB}}]
	 	 The Eynard-Orantin topological recursion for the spectral curve (\ref{special def in J}) and (\ref{x and B}) produces the multidifferentials:
	 	\begin{equation}
	 		\begin{aligned}
	 			W^{gB}_{g,n}(z_1,\cdots,z_n;\mathbf{t};S)=\sum_{i_1,\cdots,i_n\geq 0} \frac{\partial^n F^{gB}_g(\mathbf{t};S)}{\partial t_{i_1}\cdots\partial t_{i_n}}\prod_{k=1}^{n}\frac{(2i_k+1)!!\d x_k}{(2x_k)^{i_k+\frac{3}{2}}}.
	 		\end{aligned}
	 	\end{equation}
	 \end{thm}
	
	Based on this result, we further consider the transition between the Witten-Kontsevich model and the generalized BGW theory. After the coordinate transformation between two renormalized coordinates 
		\begin{equation}\label{phase transiftion}
		\begin{aligned}
			I_0&=J_0;\\
			\frac{I_n}{(2n-1)!!}&=(-2J_0)^{-n}\big(1-\sum_{k=1}^{n}\frac{J_k(-2J_0)^{k-1}}{(2k-1)!!} \big)+\delta_{n,1},
		\end{aligned}
	\end{equation}
	the multidifferentials of two theories are equal for all stable $(g,n)$, i.e., for $2g-2+n>0$,
	\begin{equation}
		W^{WK}_{g,n}(z_{[n]};\mathbf{I}(\mathbf{J}))=W^{gB}_{g,n}(z_{[n]};\mathbf{J}).
	\end{equation}
	Furthermore, for genus $g\geq1$, one obtains:
	\begin{equation}
		F^{gB}_g(\mathbf{J})=F^{WK}_g(\mathbf{I}(\mathbf{J}))-F^{WK}_g(\mathbf{I}(S^2)),
	\end{equation}
	with $\mathbf{I}(S^2)$ as in (\ref{I(S)}). It provides an explicit formula for genus $g$ free energy of the generalized BGW theory in intersection numbers for $g\geq 2$: 
	\begin{thm}[\textbf{Theorem \ref{expression of F}}]
		For genus $g\geq 2$, the free energy of the generalized BGW theory has the following expression in the renormalized coordinates $\mathbf{J}=\{J_0,J_1,\cdots\}$:
		\begin{equation*}
			\begin{aligned}
				F^{gB}_g(\mathbf{J})=&F^{WK}_g(\mathbf{I}(\mathbf{J}))-F^{WK}_g(\mathbf{I}(S^2))\\
				=&\sum_{\sum_{j=2}^{3g-2} (j-1)l_j=3g-3} \langle \tau_2^{l_2}\tau_{3}^{l_3}\cdots \tau_{3g-2}^{l_{3g-2}}\rangle_g \big(\prod_{j=2}^{3g-2}\frac{1}{l_j!}\tilde{I}_j^{l_j}(\mathbf{J})-\prod_{j=2}^{3g-2}\frac{1}{l_j!}\tilde{I}_j^{l_j}(S^2)\big),
			\end{aligned}
		\end{equation*}
		where $\tilde{I}(\mathbf{J})$ and $\tilde{I}(S^2)$ are defined in (\ref{tilde I in J}) and (\ref{tilde I in S}).
	\end{thm}
	
	This result provides a new proof of the Okuyama-Sakai Conjecture \cite{okuyamaJTSupergravityBrezinGrossWitten2020} (see Subsection \ref{subsection Okuyama-Sakai}), which has been proved by Yang and Zhang \cite{yangHodgeBGWCorrespondence2024,yangNewProofOkuyama2023}. We find that after the transition (\ref{phase transiftion}), the jet variables of the two theories are equal, i.e.,
	\begin{equation}
		(\frac{\partial}{\partial t_0^{WK}})^{n+2}F^{WK}_0(\mathbf{I}(\mathbf{J}))=(\frac{\partial}{\partial t^{gB}_0})^{n+2} F^{gB}_0(\mathbf{J}), \quad n\geq0.
	\end{equation}
	This is just the jet-variable version for the Okuyama-Sakai conjecture.

For a different approach based on $I$-coordinates and relation to the Witten-Kontsevich model,
see \cite{zhouRelating2026}.

	The remainder of this article is organized as follows. In Section \ref{section gBGW}, we review the basic properties of the generalized BGW theory, including the Virasoro constraints, the constitutive relations, and the special deformation of the spectral curve. In Section \ref{section LG theory of gBGW}, we introduce the Landau-Ginzburg theory of the generalized BGW model, define the renormalized coordinates, and establish the properties of the coordinate transformation between the original and renormalized variables. Section \ref{section Multidifferential} is devoted to the definition of the multidifferentials for the generalized BGW theory, which are related to the quantization of the spectral curve, and a discussion of their recursion relations. In Section \ref{section EO}, we briefly review the Eynard-Orantin topological recursion and prove that the multidifferentials defined in the previous section are indeed generated by the spectral curve of the generalized BGW theory. Section \ref{section finiteness} contains the proof that the free energy of the generalized BGW theory is a finite expression in the renormalized coordinates, using the Eynard-Orantin topological recursion. In Section \ref{section relation}, we investigate the relationship between the Witten-Kontsevich model and the generalized BGW theory through the transition. Section \ref{section application} provides two applications of the emergent geometry of the generalized BGW theory--the Alexandrov conjecture and the Okuyama-Sakai conjecture. Section \ref{section last} summarizes our results and outlines future directions. We also give a new proof of the constitutive relation and the special deformation for the generalized BGW theory in Appendix \ref{appendix identification}.
	
	\vspace{.1in}
	{\bf Declaration of AI Assistance}
	No AI tools were used in developing the mathematical ideas, results, or
	proofs presented in this paper. AI tools were used solely to polish the
	language of the manuscript. 
	
\vspace{.2in}
{\bf Acknowledgements}. The second-named author  is partly supported by NSFC (No.~12371254, No.~11890662, No.~12061131014).  
	
	\section{Generalized BGW Theory}\label{section gBGW}
		
		\subsection{Generalized BGW theory and its Virasoro constraints}
		
			As mentioned in the Introduction, we consider the generalized BGW partition function $Z^{gB}(\mathbf{t};S;\lambda)$. It is uniquely determined by the following Virasoro constraints \cite{alexandrovCutandjoinDescriptionGeneralized2019}  
			\begin{equation}
				L^{gB}_{m} Z^{gB}(\mathbf{t};S;\lambda)=0 \qquad m\geq 0,
			\end{equation}
			where 
			\begin{equation}
				\begin{aligned}
						L^{gB}_{m} &:=\sum_{k=0}^{\infty}\frac{(2k+2m+1)!!}{(2k-1)!!}(t_k-\delta_{k,0})\frac{\partial}{\partial t_{k+m}}\\
						&+\frac{\lambda^2}{2}\sum_{a+b=m-1} (2a+1)!!(2b+1)!!\frac{\partial^2}{\partial t_a \partial t_b}+(\frac{1}{8}+\frac{S^2}{2\lambda^2})\delta_{m,0}.
				\end{aligned}
			\end{equation}
			These operators satisfy the Virasoro commutation relations
			\begin{equation}
				[\frac{1}{2}L^{gB}_n,\frac{1}{2}L_m^{gB}]=(n-m)\frac{1}{2}L^{gB}_{n+m} \quad \forall n,m\geq0.
			\end{equation}
			From these constraints, one can derive the following cut-and-join representation of $Z^{gB}(\mathbf{t};S;\lambda)$ (see \cite{alexandrovCutandjoinDescriptionGeneralized2019}):
			\begin{equation}
				Z^{gB}(\mathbf{t};S;\lambda)=\exp(\frac{1}{\lambda^2}\hat{W}^{(-1)}+\hat{W}^{(0)}+\lambda^2 \hat{W}^{(1)})\cdot 1,
			\end{equation}
			where
			\begin{equation}
				\begin{aligned}
					&\hat{W}^{(-1)}:=\frac{S^2t_0}{2};\\
					&\hat{W}^{(0)}:=\sum_{k,m=0}^{\infty} \frac{(2k+2m+1)!!}{(2k-1)!!(2m-1)!!}t_k t_m\frac{\partial}{\partial t_{k+m}}+\frac{1}{8}t_0;\\
					&\hat{W}^{(1)}:=\frac{1}{2}\sum_{k,m=0}^{\infty}\frac{(2k+1)!!(2m+1)!!}{(2k+2m+1)!!}t_{k+m+1}\frac{\partial^2}{\partial t_k \partial t_m}.
				\end{aligned}
			\end{equation}
			From this representation, one sees that if one assigns
			\begin{equation}
				 \deg t_k=2k \qquad \deg \lambda=-1 \qquad \deg S=-1,
			\end{equation}
			then $Z^{gB}(\mathbf{t};S;\lambda)$ is weighted homogeneous of degree $0$. In particular, if one expands the free energy $F^{gB}(\mathbf{t};S;\lambda)=\log Z^{gB}(\mathbf{t};S;\lambda)$ as
			\begin{equation}
				F^{gB}(\mathbf{t};S;\lambda)=\sum_{g=0}^{\infty} F^{gB}_g(\mathbf{t};S)\lambda^{2g-2},
			\end{equation}
			then $F^{gB}_g(\mathbf{t};S)$ is weighted homogeneous of degree $2g-2$ in $S$ and $\mathbf{t}=\{t_0,t_1,\cdots\}$.
		
			In terms of the free energy, the Virasoro constraint for the generalized BGW theory becomes
			\begin{equation}
				\begin{aligned}
					&(2m+1)!!\frac{\partial F^{gB}}{\partial t_m} =\sum_{k=0}^{\infty} \frac{(2k+2m+1)!!}{(2k-1)!!} t_{k}\frac{\partial F^{gB}}{\partial t_{m+k}}\\
					&+\frac{\lambda^2}{2}\sum_{a+b=m-1}(2a+1)!!(2b+1)!!\big(\frac{\partial F^{gB}}{\partial t_a}\frac{\partial F^{gB}}{\partial t_b}+\frac{\partial^2 F^{gB}}{\partial t_a \partial t_b}\big)\\
					&+(\frac{1}{8}+\frac{S^2}{2\lambda^2})\delta_{m,0}.
				\end{aligned}
			\end{equation}
			For the genus-zero sector, it becomes
			\begin{equation}\label{genus zero part of gB}
				\begin{aligned}
					&(2m+1)!!\frac{\partial F^{gB}_0}{\partial t_m} =\sum_{k=0}^{\infty} \frac{(2k+2m+1)!!}{(2k-1)!!} t_{k}\frac{\partial F^{gB}_0}{\partial t_{m+k}}\\
					&+\frac{1}{2}\sum_{a+b=m-1}(2a+1)!!(2b+1)!!\frac{\partial F^{gB}_0}{\partial t_a}\frac{\partial F^{gB}_0}{\partial t_b}+\frac{S^2}{2}\delta_{m,0}.
				\end{aligned}
			\end{equation}
			When $S\to 0$, the generalized BGW partition function degenerates to the ordinary BGW theory.
			
		\subsection{Constitutive relations and order parameter of generalized BGW theory}	
		
			In Zhang's PhD thesis \cite{zhangApplication2020}, he generalized the constitutive relations of the Witten-Kontsevich model to several other models including the generalized BGW theory.  We summarize the relevant results below.
			
			When restricted to the small phase space--that is, when the parameters $t_n=0$ for all $n=0,1,2,\cdots$-- the Virasoro constraint (\ref{genus zero part of gB}) becomes
			\begin{equation}
				(2m+1)!!\frac{\partial F^{gB}_0}{\partial t_m}\Big|_s =
				\frac{1}{2}\sum_{a+b=m-1}(2a+1)!!(2b+1)!!\frac{\partial F^{gB}_0}{\partial t_a}\Big|_s\frac{\partial F^{gB}_0}{\partial t_b}\Big|_s+\frac{S^2}{2}\delta_{m,0}.
			\end{equation}
			where $|_s$ represents evaluation on the small phase space.
			One then obtains
			\begin{equation}
				\begin{aligned}
					(2m+1)!!\frac{\partial F^{gB}_0}{\partial t_m}\Big|_s&=2^{-(2m+1)}\frac{1}{m+1}\binom{2m}{m}S^{2m+2};\\
					\frac{\partial F^{gB}_0}{\partial t_m}\Big|_s&=\frac{1}{(2m+1)(m+1)!}(\frac{S^2}{2})^{m+1}.
				\end{aligned}
			\end{equation}
			Differentiating the equation with respect to $S$ yields
			\begin{equation}
			\frac{\partial^2 F^{gB}_0}{\partial S\partial t_m}\Big|_{s}=\frac{S}{(2m+1)m!}(\frac{S^2}{2})^{m}.
			\end{equation} 
			Similarly, differentiating the equation (\ref{genus zero part of gB}) with respect to $t_n$ and restricting to the small phase space yields
			\begin{equation}
				\begin{aligned}
					&(2n+1)!!(2m+1)!!\frac{\partial^2 F^{gB}_0}{\partial t_n\partial t_m}\Big|_s= \\
					&\frac{(m+1)(n+1)}{2^{2m+2n+2}}\binom{2m+2}{m+1}\binom{2n+2}{n+1}\frac{S^{2m+2n+2}}{2m+2n+2};\\
					&\frac{\partial^2 F^{gB}_0}{\partial t_n\partial t_m}\Big|_s=\frac{1}{(n+m+1)n!m!}(\frac{S^2}{2})^{n+m+1}.
				\end{aligned}
			\end{equation}
			The constitutive relations refer to the fact that the second derivatives of the genus-zero free energy, viewed as functions of the order parameters, take the same form in the small phase space and in the full parameter space. In the present case, the order parameter is chosen as
			\begin{equation}
				Q:=\frac{\partial^2 F^{gB}_0}{\partial S\partial t_0}.
			\end{equation}
			In the small phase space, it becomes
				$Q\big|_s=S$.
			 
			\begin{thm}[Zhang\cite{zhangApplication2020}]\label{Zhang's main thm}
				The second derivatives of the genus-zero free energy of the generalized BGW theory have the following expressions:
				\begin{equation}
						\frac{\partial^2 F^{gB}_0}{\partial S\partial t_m}=\frac{Q}{(2m+1)m!}(\frac{Q^2}{2})^m;
				\end{equation}
				and 
				\begin{equation}
					\begin{aligned}
						\frac{\partial^2 F^{gB}_0}{\partial t_n\partial t_m}= 
						\frac{1}{(m+n+1)n!m!}(\frac{Q^2}{2})^{m+n+1},
					\end{aligned}
				\end{equation}
				where
				\begin{equation}
					Q=\frac{\partial^2 F^{gB}_0}{\partial S\partial t_0}.
				\end{equation}
			\end{thm}
			We will give a new proof of the constitutive relation of the generalized BGW theory in Appendix \ref{appendix identification}.
			
			\begin{remark}
				We use the different notations from those in \cite{zhangApplication2020}; the relations are
				\begin{equation}
					\frac{2}{(2n+1)!!}t_n=T_{n+1} \text{ for } n\geq 1; \qquad S=T_0; \qquad 4F^{gB}(\mathbf{t};S;\lambda)=F^{B}(\mathbf{T};\lambda).
				\end{equation}
			\end{remark}
			
			The first Virasoro constraint for the genus-zero sector in the equation (\ref{genus zero part of gB}) is
			\begin{equation}
				\frac{\partial F^{gB}_0}{\partial t_0} =\sum_{k=0}^{\infty} (2k+1) t_k\frac{\partial F^{gB}_0}{\partial t_k} +\frac{S^2}{2}.
			\end{equation}
			Differentiating this with respect to $S$ and using the constitutive relation yields
			\begin{equation}\label{Q equ}
				Q=\sum_{k=0}^{\infty} \frac{t_k}{k!}(\frac{Q^2}{2})^kQ+S.
			\end{equation}
			This is the Landau-Ginzburg equation of the order parameter $Q$. The corresponding Landau-Ginzburg potential is
			\begin{equation}
				W(x)=-\frac{1}{2}x^2 + Sx+\sum_{k=0}^{\infty} \frac{t_k}{(2k+2)!!}x^{2k+2}.
			\end{equation}
			Here $Q$ is the critical point of this potential; i.e. $Q$ satisfies
			\begin{equation}
				W'(Q)=0.
			\end{equation}
			The quantity $Q$ is viewed as a formal power series in $S$ and $\mathbf{t}$. Using the Lagrange inversion formula (see Subsection \ref{subsection alexandrov conjecture}), one obtains the expansion of $Q$:
			\begin{equation}\label{Q exp}
				\begin{aligned}
					Q&=\sum_{k=1}^{\infty}\frac{1}{k}\sum_{\substack{l=1\\ l\text{ odd}}}^{k}\binom{k}{l}S^l \sum_{2a_1+2a_2+\cdots+2a_{k-l}=l-1}\frac{t_{a_1}}{(2a_1)!!}\frac{t_{a_2}}{(2a_2)!!}\cdots\frac{t_{a_{k-l}}}{(2a_{k-l})!!} \\
					&=\sum_{n=0}^{\infty} \frac{S^{2n+1}}{2^n}\sum_{m=0}^{\infty} \sum_{\sum_{a} aj_a=n,\sum_a j_a=m}\frac{(2n+m)!}{(2n+1)!}\prod_{a=0}^{\infty} \frac{1}{j_a!}(\frac{t_a}{a!})^{j_a}\\
					&= S\frac{1}{1-t_0}+S^3\frac{t_1}{2(1-t_0)^4}+S^5\frac{t_2}{8(1-t_0)^6}+S^5\frac{3t_1}{4(1-t_0)^7}+o(S^5).
				\end{aligned}
			\end{equation}
			
			With the weighted homogeneity, one can recover $\frac{\partial F^{gB}}{\partial t_n}$ from the second derivatives. Indeed, $F^{gB}_0$ has degree $-2$ and the derivatives $\frac{\partial F^{gB}_0}{\partial t_n}$ have degree $-2-2n$. Thus,
			\begin{equation}
				\begin{aligned}\label{first derivate of genus zero}
					(-2-2n)\frac{\partial F^{gB}_{0}}{\partial t_n}
					=\sum_{m=0}^{\infty}2m t_m\frac{\partial^2 F^{gB}_0}{\partial t_n\partial t_m}-S\frac{\partial^2 F^{gB}_0}{\partial S \partial t_n}
				\end{aligned}
			\end{equation} 
			 \begin{equation}\label{F0 d S}
			 	\frac{\partial F^{gB}_0}{\partial S} =-\sum_{n=0}^{\infty} 2nt_n \frac{\partial^2 F^{gB}_0}{\partial t_n \partial S}+S\frac{\partial^2 F^{gB}_0}{\partial S^2}.
			 \end{equation}
			 Furthermore:
			 \begin{equation}
			 	F^{gB}_0=-\frac{1}{2}\big(\sum_{m=0}^{\infty}2mt_m\frac{\partial F^{gB}_{0}}{\partial t_m}-S\frac{\partial F^{gB}_0}{\partial S}\big).
			 \end{equation}
			 Here the quantity $\frac{\partial^2 F^{gB}_{0}}{\partial S^2}$ can be computed as follows. Substituting the constitutive relations into the equation (\ref{F0 d S}) yields
			 \begin{equation}
			 	\frac{\partial F^{gB}_0}{\partial S} = S\frac{\partial^2 F^{gB}_0}{\partial S^2}-\sum_{n=1}^{\infty} \frac{t_n}{(2n-2)!!(2n+1)}Q^{2n+1}.
			 \end{equation}
			 Differentiating both sides with respect to $S$ yields
			 \begin{equation}\label{d2 F with S}
			 	\frac{\partial^2 F^{gB}_0}{\partial S^2}=\frac{\partial^2 F^{gB}_0}{\partial S^2}+S\frac{\partial^3 F^{gB}_0}{\partial S^3}-\sum_{n=1}^{\infty} \frac{t_n}{(2n-2)!!}Q^{2n}\frac{\partial Q}{\partial S}.
			 \end{equation}
			 Differentiating the Landau-Ginzburg equation for $Q$ (\ref{Q equ}) with respect to $S$ yields
			 \begin{equation}\label{Q der S}
			 	\begin{aligned}
			 		\frac{\partial Q}{\partial S}&=\sum_{k=0}^{\infty}\frac{(2k+1)t_k}{(2k)!!}Q^{2k}\frac{\partial Q}{\partial S}+1 \\
			 		&=\sum_{k=1}^{\infty} \frac{t_k}{(2k-2)!!}Q^{2k}\frac{\partial Q}{\partial S}+\sum_{k=0}^{\infty} \frac{t_k}{(2k)!!}Q^{2k+1}\frac{1}{Q}\frac{\partial Q}{\partial S}+1\\
			 		&=\sum_{k=1}^{\infty} \frac{t_k}{(2k-2)!!}Q^{2k}\frac{\partial Q}{\partial S}+\frac{Q-S}{Q}\frac{\partial Q}{\partial S}+1,
			 	\end{aligned}
			 \end{equation}
			 where the third line uses the Landau-Ginzburg equation (\ref{Q equ}). Substituting this into (\ref{d2 F with S}) yields
			 \begin{equation}
			 	\begin{aligned}
			 		\frac{\partial^3 F^{gB}_0}{\partial S^3}=\frac{1}{Q}\frac{\partial Q}{\partial S}-\frac{1}{S}.
			 	\end{aligned}
			 \end{equation}
			 Integrating with respect to $S$, and using the weighted homogeneity to fix the constant of integration yields
			 \begin{equation}
			 	\frac{\partial^2 F^{gB}_0}{\partial S^2}=\log \frac{Q}{S}.
			 \end{equation}
			 All of these first derivatives and the genus-zero free energy were computed in Zhang's thesis; we summarize the results as follows:
			 \begin{proposition}[Zhang \cite{zhangApplication2020}] 
			 		The genus-zero free energy of the generalized BGW theory and its first derivatives with respect to $S$ and $t_0,t_1,\cdots$ have the following expressions in terms of the order parameter $Q$:
			 		\begin{equation}\label{F0}
			 			\begin{aligned}
			 				&F^{gB}_0(\mathbf{t};S)=\frac{1}{2}S^2\log\frac{Q}{S} +\frac{3}{4}S^2-\\
			 				 &\sum_{n,m=0}^{\infty} \tilde{t}_{n}\tilde{t}_{m}\frac{4m+2n+3}{2(2n+1)(n+m+1)n!m!}(\frac{Q^2}{2})^{n+m+1};
			 			\end{aligned}
			 		\end{equation}
			 		\begin{equation}\label{one point func of genus zero}
			 			\begin{aligned}
			 				\frac{\partial F^{gB}_0}{\partial t_n}=-\sum_{m=0}^{\infty}\tilde{t}_m \frac{2m+1}{(2n+1)n!m!(n+m+1)}(\frac{Q^2}{2})^{m+n+1};
			 			\end{aligned}
			 		\end{equation}
			 		\begin{equation}
			 			\begin{aligned}
			 				\frac{\partial F^{gB}_0}{\partial S}=S\log\frac{Q}{S}+S+\sum_{n=0}^{\infty}\tilde{t}_{n}\frac{Q}{n!(2n+1)}(\frac{Q^2}{2})^{n},
			 			\end{aligned}
			 		\end{equation}
			 		where $\tilde{t}_{n}=t_n-\delta_{n,0}$ for $n=0,1,\cdots$.
			 \end{proposition}
			 \begin{proof}
			 	Although the direct computation is somewhat involved, the result can be verified by differentiation. For example, differentiating the equation (\ref{one point func of genus zero}) with respect to $S$ yields
			 	\begin{equation}
			 		\begin{aligned}
			 			\frac{\partial }{\partial S}RHS=&-\sum_{m=0}^{\infty} \tilde{t}_m \frac{2m+1}{(2n+1)n!m!}(\frac{Q^2}{2})^{m+n}Q\frac{\partial Q}{\partial S}\\
			 			=&-\frac{1}{(2n+1)n!}(\frac{Q^2}{2})^n\big( \sum_{m=0}^{\infty} t_m(\frac{2}{(m-1)!}
			 			+\frac{1}{m!})(\frac{Q^2}{2})^m\frac{\partial Q}{\partial S}\big)\\
			 			&+\frac{1}{(2n+1)n!}(\frac{Q^2}{2})^nQ\frac{\partial Q}{\partial S}.
			 		\end{aligned}
			 	\end{equation}
			 	With the equations (\ref{Q equ}) and (\ref{Q der S}), one obtains
			 	\begin{equation}
			 		\begin{aligned}
			 			\frac{\partial}{\partial S}RHS=&-\frac{1}{(2n+1)n!}(\frac{Q^2}{2})^n\big((Q-S)\frac{\partial Q}{\partial S}+Q(\frac{S}{Q}\frac{\partial Q}{\partial S}-1)\big)\\
			 			&+\frac{1}{(2n+1)n!}(\frac{Q^2}{2})^nQ\frac{\partial Q}{\partial S}\\
			 			=&\frac{1}{(2n+1)n!}(\frac{Q^2}{2})^nQ=\frac{\partial^2F^{gB}_0}{\partial S\partial t_n}.
			 		\end{aligned}
			 	\end{equation}
			 	The other derivatives are similar, using the relation
			 		\begin{equation}\label{Q der for t}
			 		\begin{aligned}
			 			\frac{\partial Q}{\partial t_n}=(1-\sum_{k=0}^{\infty}\frac{(2k+1)t_k}{2^kk!}Q^{2k})^{-1}\frac{Q^{2n+1}}{2^nn!}=\frac{\partial Q}{\partial S} \frac{Q}{n!}(\frac{Q^2}{2})^n.
			 		\end{aligned}
			 	\end{equation}
			 \end{proof}

		\subsection{The spectral curve of the generalized BGW theory}
			From the perspective of emergent geometry, the spectral curve of the generalized BGW theory is defined as
			\begin{equation}\label{def of spectral curve}
				\begin{aligned}
					y^{gB}(w;\mathbf{t};S):=&-K(w;\mathbf{t})-\nabla (w)F^{gB}_0(\mathbf{t};S)\\
					=& w^{-\frac{1}{2}}-\sum_{n=0}^{\infty} \frac{t_n}{(2n-1)!!}w^{n-\frac{1}{2}}-\sum_{n=0}^{\infty}\frac{(2n+1)!!}{w^{n+\frac{3}{2}}} \frac{\partial F^{gB}_0}{\partial t_n},
				\end{aligned}
			\end{equation} 
			where 
			\begin{equation}
				K(w;\mathbf{t}):=-w^{-\frac{1}{2}}+\sum_{n=0}^{\infty} \frac{t_n}{(2n-1)!!}w^{n-\frac{1}{2}}
			\end{equation}
			represents a bosonic field and 
			\begin{equation}
				\nabla(w):=\sum_{n=0}^{\infty}\frac{(2n+1)!!}{w^{n+\frac{3}{2}}}\frac{\partial}{\partial t_n}
			\end{equation}
			is the loop insertion operator.
			
			In Zhang's PhD thesis, he computed this spectral curve and obtained:
			\begin{proposition}[Zhang \cite{zhangApplication2020}]\label{Zhang's spectral curve}
				The spectral curve of the generalized BGW theory has the following form:
				\begin{equation}
					\begin{aligned}
						y^{gB}(w;\mathbf{t};S)=\frac{\sqrt{w-Q^2}}{w}\Big(1-\frac{2}{Q}\sum_{m,k=0}^{\infty}  \frac{t_{m+k}}{(2m+2k-1)!!}\binom{2k}{k}\frac{Q^{2k+1}}{2^{2k+1}}w^{m}\Big),
					\end{aligned}
				\end{equation}
				where $\sqrt{w-Q^2}$ is viewed as the formal power series
				\begin{equation}
					\sqrt{w-Q^2}=w^{\frac{1}{2}}\sqrt{1-\frac{Q^2}{w}}=\sum_{n=0}^{\infty} \binom{\frac{1}{2}}{n}(-Q^2)^nw^{-n+\frac{1}{2}}.
				\end{equation}
			\end{proposition}
			When all parameters $\mathbf{t}$ are set to zero, the spectral curve becomes
			\begin{equation}
				\begin{aligned}
					y^{gB}(w;\mathbf{t}=0;S)=\frac{\sqrt{w-S^2}}{w}=\frac{z}{z^2+S^2},
				\end{aligned}
			\end{equation}
			under the coordinate transformation $w=z^2+S^2$.
			
			Following the work \cite{zhouEmergentGeometryUnifications2025}, it is natural to expand the spectral curve in powers of $w-Q^2$. Then one obtains
			\begin{proposition}\label{spe cur gBGW}
					The spectral curve of the generalized BGW theory can be expressed as
				\begin{equation}\label{spectral curve of gBGW}
					\begin{aligned}
						y^{gB}(w)=\frac{\sqrt{w-Q^2}}{w}\Big(1-\sum_{n=0}^{\infty} \frac{(w-Q^2)^n}{(2n+1)!!} 
						\sum_{m=0}^{\infty} (2n+2m+1) \frac{t_{n+m}}{m!}\frac{Q^{2m}}{2^{m}}
						\Big).
					\end{aligned}
				\end{equation}
			\end{proposition}
			\begin{proof}
				Starting from Proposition \ref{Zhang's spectral curve}, 
				\begin{equation}
					\begin{aligned}
						&y^{gB}(w)=\frac{\sqrt{w-Q^2}}{w}\Big[1-\frac{2}{Q}\sum_{m,k=0}^{\infty}  \frac{t_{m+k}Q_{k+1}}{(2m+2k-1)!!} \sum_{l=0}^{m}\binom{m}{l}(w-Q^2)^l Q^{2(m-l)}\Big]\\
						=&\frac{\sqrt{w-Q^2}}{w}\Big[1\Big.\\
						&\Big.-\frac{2}{Q}\sum_{n=0}^{\infty}(w-Q^2)^n\sum_{m=0}^{\infty} \frac{t_{n+m}}{(2(n+m)-1)!!}\big(\sum_{k+l=m}\frac{1}{2^{2k}} \binom{2k}{k}\binom{n+l}{l}\big)\frac{Q^{2m+1}}{2}\Big]\\
						=&\frac{\sqrt{w-Q^2}}{w}\Big(1-\sum_{n=0}^{\infty}(w-Q^2)^n\sum_{m=0}^{\infty}\frac{t_{n+m}}{(2(n+m)-1)!!}\binom{n+m+\frac{1}{2}}{m}Q^{2m}\Big).
					\end{aligned}
				\end{equation}
			\end{proof}
			In the proof, we used the combinatorial identity 
				\begin{equation}
					\sum_{k=0}^{m}\frac{1}{2^{2k}}\binom{2k}{k}\binom{n+m-k}{m-k}=\binom{n+m+\frac{1}{2}}{m},
				\end{equation}
			which follows from comparing coefficients of $x^m$	in the expansion of
			\begin{equation*}
				(1-x)^{-\frac{1}{2}}(1-x)^{-n-1}=(1-x)^{-n-\frac{3}{2}}.
			\end{equation*}
	
			With the coordinate transformation $w=z^2+Q^2$, the spectral curve takes the form:
			\begin{equation}\label{spec cur of gBGW z}
				y^{gB}(z)=\frac{z}{z^2+Q^2}\Big(1-\sum_{n=0}^{\infty} \frac{z^{2n}}{(2n+1)!!} 
				\sum_{m=0}^{\infty} (2n+2m+1) \frac{t_{n+m}}{m!}\frac{Q^{2m}}{2^{m}}
				\Big).
			\end{equation}
			When $S\to 0$, then $Q\to 0$, and the spectral curve reduces to the BGW theory \cite{zhouEmergentGeometryUnifications2025}:
			\begin{equation}\label{BGW spe cur}
				y^{BGW}(z;\mathbf{t})=\frac{1}{z}-\sum_{n=0}^{\infty} \frac{z^{2n-1}t_n}{(2n-1)!!}.
			\end{equation}
			When $\mathbf{t}=0$, it becomes
			\begin{equation}
				y^{BGW}(z;\mathbf{t}=0)=\frac{1}{z}.
			\end{equation}
			This is the spectral curve for the Eynard-Orantin topological recursion of the ordinary BGW theory \cite{doTopologicalRecursionBessel2018}.
		
			\section{Landau-Ginzburg Theory and renormalization for generalized BGW theory}\label{section LG theory of gBGW}
			\subsection{Renormalized coordinate of Witten-Kontsevich model}
			
			As mentioned in the introduction, the spectral curve of the generalized BGW theory (see (\ref{spectral curve of gBGW}) and (\ref{spec cur of gBGW z})) bears a close resemblance to that of the Witten-Kontsevich model. In \cite{zhouEmergentGeometryUnifications2025}, the spectral curve of the Witten-Kontsevich model is written as
			\begin{equation}\label{spec of WK}
				\begin{aligned}
					y^{WK}:=&w^{1/2}-\sum_{n=0}^{\infty}\frac{t_n}{(2n-1)!!}w^{n-\frac{1}{2}}-\sum_{n=0}^{\infty}\frac{(2n+1)!!}{w^{n+\frac{3}{2}}} \frac{\partial F^{WK}_0}{\partial t_n} \\
					=&(w-2I_0)^{\frac{1}{2}}(1-\sum_{n=0}^{\infty}\frac{(w-2I_0)^{n}}{(2n+1)!!}I_{n+1})\\
					=& z(1-\sum_{n=0}^{\infty}\frac{z^{2n}}{(2n+1)!!}I_{n+1}),
				\end{aligned}
			\end{equation}
			where $w=z^2+2I_0$.
			
			The variables $\mathbf{I}=\{I_0,I_1,\cdots\}$ are the renormalized coordinates of the Witten-Kontsevich model (also called the Itzykson-Zuber coordinates). They were first introduced by Itzykson and Zuber \cite{itzyksonCombinatoricsModularGroup1992}, and have been further elaborated in \cite{zhangItzyksonZuberAnsatz2019}. One significant feature of the $\mathbf{I}$-coordinates is that, in these coordinates, the free energy $F^{WK}_g$ of the Witten-Kontsevich model admits a polynomial expression for $g>1$.
			
			The order parameter $I_0:=\frac{\partial^2 F_0^{WK}}{\partial t^2_0}$ is determined by the Landau-Ginzburg equation of the Witten-Kontsevich model:
			\begin{equation}\label{LG equ of WK}
				I_0=\sum_{n=1}^{\infty}\frac{t^{WK}_n}{n!}I^n_0+t^{WK}_0.
			\end{equation}
			And the higher coordinates $I_n$ are given by the Galilean transformation of the KdV hierarchy \cite{xuGalileanSymmetryKdV2025}:
			\begin{equation}
				I_n=\sum_{m=0}^{\infty}\frac{t_{n+m}^{WK}}{m!} I_{0}^{m}.
			\end{equation}
			In Landau-Ginzburg theory, $I_0$ is the critical point of the Landau-Ginzburg potential:
			\begin{equation}
				W^{WK}(x)=-\frac{1}{2}x^2+t^{WK}_{0}x+\sum_{n=1}^{\infty} \frac{t^{WK}_{n}}{(n+1)!}x^{n+1},
			\end{equation}
			i.e. $I_0$ satisfies 
			\begin{equation}
				(W^{WK})'(I_0)=0.
			\end{equation}
			And $I_n$ are the coefficients of $W^{WK}$ expanded around this critical point $x=I_0$:
			\begin{equation}
				W^{WK}(x)=I_{-1}-\frac{1}{2}(x-I_0)^2+\sum_{n=1}^{\infty}\frac{I_{n}}{(n+1)!}(x-I_0)^{n+1},
			\end{equation}
			where we denote
			\begin{equation}
				I_{-1}:=W^{WK}(I_0)=\sum_{n=0}^{\infty}\frac{t^{WK}_n}{(n+1)!}I_0^{n+1}-\frac{1}{2}I_0^2.
			\end{equation}
		
			From the perspective of the statistical physics, the change of coordinates from $\mathbf{t}=\{t_0,t_1,\cdots\}$ to $\mathbf{I}=\{I_0,I_1,\cdots\}$ constitutes a renormalization procedure. After renormalization, the bare coupling constants $\mathbf{t}$ are replaced by the renormalized coupling constants $\mathbf{I}$. As formal power series in $\mathbf{t}$, the partition function and free energy are not expected to converge; however, they may admit finite expressions in the $\mathbf{I}$-coordinates.
			
			Indeed, one has the following result:
			\begin{thm}[Itzykson-Zuber \cite{itzyksonCombinatoricsModularGroup1992}, Zhou \cite{zhouTopological1DGravity2014}, Zhang-Zhou \cite{zhangItzyksonZuberAnsatz2019}]
				The free energy of the Witten-Kontsevich model takes the following form in $\mathbf{I}=\{I_0,I_1,\cdots\}$ coordinates:
				\begin{equation}
					\begin{aligned}
						F^{WK}_0=\frac{1}{6} I_0^3-\sum_{n=0}^{\infty} \frac{(-1)^nI_0^{n+2}}{(n+2)!}I_n+\frac{1}{2}\sum_{n,m=0}^{\infty} \frac{(-1)^{n+m}I_0^{n+m+1}}{(n+m+1)n!m!}I_nI_m;
					\end{aligned}
				\end{equation}
				\begin{equation}\label{WK g1}
					F_1^{WK}=\frac{1}{24}\log\frac{1}{1-I_1};
				\end{equation}
				\begin{equation}\label{WK Fg >1}
					F_g^{WK}=\sum_{l_2+2l_3+\cdots+(3g-3)l_{3g-2}=3g-3} \langle \tau_2^{l_2}\tau_{3}^{l_3}\cdots \tau_{3g-2}^{l_{3g-2}}\rangle_g \prod_{j=2}^{3g-2}\frac{1}{l_j!}\tilde{I}_j^{l_j},
				\end{equation}
				where $g\geq 2$,
				\begin{equation}\label{tilde I}
					\tilde{I}_j=(1-I_1)^{-\frac{2j+1}{3}}I_j,
				\end{equation}
				and $\langle\tau_2^{l_2}\tau_{3}^{l_3}\cdots \tau_{3g-2}^{l_{3g-2}}\rangle_g$ are the intersection numbers of $\psi$-classes on the moduli space:
				\begin{equation}
					\langle\tau_{a_1}\tau_{a_2}\cdots\tau_{a_n}\rangle_g:=\int_{\bar{M}_{g,n}}\psi_1^{a_1}\psi_2^{a_2}\dots\psi_{n}^{a_n}=\frac{\partial^n F_g^{WK}}{\partial t_{a_1}\partial t_{a_2}\dots\partial t_{a_n}}\big|_{\mathbf{t}=0}.
				\end{equation}
			\end{thm}
			
			This yields an effective method to compute the free energies of the Witten-Kontsevich model recursively via the Virasoro constraints expressed in the $\mathbf{I}$-coordinates \cite{zhangItzyksonZuberAnsatz2019}. The first few genus free energies are
			\begin{equation}\label{WK g2}
				F^{WK}_2=\frac{7}{1440}\tilde{I}^3_2+\frac{29}{5760}\tilde{I}_2\tilde{I}_3+\frac{1}{1152}\tilde{I}_4;
			\end{equation}
			\begin{equation}
				\begin{aligned}
					F^{WK}_{3}=&\frac{245}{20736}\tilde{I}_2^6+\frac{193}{6912}\tilde{I}_2^4\tilde{I}_3 +\frac{205}{13824}\tilde{I}_2^2\tilde{I}_3^2 +\frac{583}{580608}\tilde{I}_3^3 +\frac{53}{6912}\tilde{I}_2^3\tilde{I}_4+\frac{1121}{241920}\tilde{I}_2\tilde{I}_3\tilde{I}_4 \\
					 &+\frac{607}{2903040}\tilde{I}_4^2+\frac{17}{11520}\tilde{I}_2^2\tilde{I}_5 +\frac{503}{1451520}\tilde{I}_3\tilde{I}_5 +\frac{77}{414720}\tilde{I}_2\tilde{I}_6+\frac{1}{82944}\tilde{I}_7.
				\end{aligned}
			\end{equation}
			
			\subsection{Renormalized coordinate of generalized BGW theory}
			In the generalized BGW theory, a similar structure emerges. However, instead of using the Landau-Ginzburg equation for $Q$ (\ref{Q equ}), we must choose a new order parameter and an associated Landau-Ginzburg potential to generate the renormalized coordinates.
			
			Denote the new order parameter
			\begin{equation}\label{J0 exp}
				\begin{aligned}
					J_0:=&\frac{\partial^2 F^{gB}_0}{\partial t^2_0}=\frac{Q^2}{2}\\
					=&S^2\frac{1}{2(1-t_0)^{2}}+S^4\frac{t_1}{2(1-t_0)^5}+S^6(\frac{7t_1^2}{8(1-t_0)^8}+\frac{t_2}{8(1-t_0)^7})+o(S^6).
				\end{aligned}
			\end{equation}
			By Theorem \ref{Zhang's main thm} and the weighted homogeneity relation (\ref{first derivate of genus zero}), one obtains the Landau-Ginzburg equation for the generalized BGW theory:
			\begin{equation}\label{LG equ for J0}
				J_0=\sum_{n=1}^{\infty} \frac{(2n-1)t^{gB}_{n-1}}{n!}J_0^n+\frac{\partial F^{gB}_0}{\partial t_0}.
			\end{equation}  
			The corresponding potential is
			\begin{equation}\label{LG potential of J_0}
				W^{gB}(x)=-\frac{1}{2}x^2+\frac{\partial F^{gB}_0}{\partial t_0} x+\sum_{n=1}^{\infty}\frac{(2n-1)t^{gB}_{n-1}}{(n+1)!}x^{n+1}.
			\end{equation}
			Expanding $W^{gB}$ around its critical point $x=J_0$ yields
			\begin{equation}
				W^{gB}(x)=J_{-1}-\frac{1}{2}(x-J_0)^2+\sum_{n=1}^{\infty}\frac{J_{n}}{(n+1)!}(x-J_0)^{n+1}.
			\end{equation}
			We obtain, for $n=1,2,\cdots$,
			\begin{equation}\label{def of Jn}
				J_n:=\frac{d^{n+1} W^{gB}}{dx^{n+1}}(J_0)+\delta_{n,1}=\sum_{m=0}^{\infty} (2n+2m-1)\frac{t^{gB}_{n+m-1}}{m!}J^{m}_{0},
			\end{equation} 
			and
			\begin{equation}
				J_{-1}:=W^{gB}(J_0)=\sum_{n=1}^{\infty}\frac{(2n-1)t_{n-1}^{gB}}{(n+1)!}J_0^{n+1}+\frac{\partial F^{gB}_0}{\partial t_0}J_0-\frac{1}{2}J_0^2.
			\end{equation}
			\begin{remark}\label{tran from LG}
				This construction can be viewed as the following substitution in equation (\ref{LG equ of WK}):
				\begin{equation}
					t_0^{WK}=\frac{\partial F^{gB}_0}{\partial t_0};\quad t_{n}^{WK}=(2n-1)t_{n-1}^{gB} \quad n=1,2,\cdots.
				\end{equation}
				This offers a new perspective: the genus-zero free energies of the generalized BGW theory and the Witten-Kontsevich model are simply different parametrizations of the same underlying theory; however, after different quantization procedures, the higher-genus free energies diverge. We will return to this viewpoint in Appendix \ref{appendix identification}.
			\end{remark}
		
			With the renormalized coordinates $\mathbf{J}=\{J_0,J_1,\cdots\}$, the constitutive relation becomes:
			\begin{proposition}\label{cons rel of F in J}
				The second derivatives of the genus-zero free energy of the generalized BGW theory have the following expressions:
				\begin{equation}
					\frac{\partial^2 F^{gB}_0}{\partial S^2}=\frac{1}{2}\log\frac{2J_0}{S^2};
				\end{equation}
				\begin{equation}
					\frac{\partial^2 F_0^{gB}}{\partial S\partial t_n}=\frac{\sqrt{2J_0}J_0^n}{(2n+1)n!};
				\end{equation}
				and 
				\begin{equation}\label{cons for J}
					\frac{\partial^2 F^{gB}_{0}}{\partial t_m\partial t_n}=\frac{J_0^{m+n+1}}{(m+n+1)m!n!}.
				\end{equation}
			\end{proposition}
			A new proof of this proposition will be given in Appendix \ref{appendix identification}.
			
			Consequently, the spectral curve (\ref{spec cur of gBGW z}) of the generalized BGW theory takes a form analogous to that of the Witten-Kontsevich model:
			\begin{proposition}\label{spec curve in J}
				The spectral curve of the generalized BGW theory (\ref{spec cur of gBGW z}) can be written as
				\begin{equation}
					\begin{aligned}
						y^{gB}=&\frac{\sqrt{w-2J_0}}{w}(1-\sum_{n=0}^{\infty}\frac{(w-2J_0)^{n}}{(2n+1)!!}J_{n+1})\\
						=&\frac{z}{z^2+2J_0}(1-\sum_{n=0}^{\infty}\frac{z^{2n}}{(2n+1)!!}J_{n+1}),
					\end{aligned}
				\end{equation}
				with $w=z^2+2J_0$.
			\end{proposition}
			
			\subsection{Coordinate transformation between two coordinates}\label{subsection Coordinate transformation between two coordinates}
			We have not yet verified that $\mathbf{J}=\{J_0,J_1,\cdots\}$ can be regarded as a coordinate transformation. Unlike the Witten-Kontsevich model, the Landau-Ginzburg equation for $J_0$ (\ref{LG equ for J0}) contains the term $\frac{\partial F^{gB}_0}{\partial t_0}$. From the expression (\ref{J0 exp}) and the Landau-Ginzburg equation (\ref{LG equ for J0}), one finds
			\begin{equation}
				\begin{aligned}
					\frac{\partial F^{gB}_0}{\partial t_0}&=J_0-\sum_{n=0}^{\infty}\frac{t_n(2n+1)}{(n+1)!}J_0^{n+1}\\
					&=\frac{S^2}{2(1-t_0)}+O(t_n) \qquad \text{for } t_n \to 0.
				\end{aligned}
			\end{equation}
			Therefore, $\frac{\partial F^{gB}_0}{\partial t_0}$ can serve as a replacement parameter for the parameter $S$.
			The variables $\{J_0,J_1,\cdots\}$ also constitute a valid coordinate system, since their relationship with $\{S^2,\mathbf{t}\}$ is upper triangular:
			\begin{equation}
				\begin{aligned}
					J_0&=\frac{S^2}{2}+S^2t_0+\cdots\\
					J_n&=(2n-1)t_{n-1}+(2n+1)t_{n}J_0+\cdots \qquad \text{for } n\geq 1,
				\end{aligned}
			\end{equation}
			where $\cdots$ denotes the terms involving only $t_m$ for $m\geq n+1$. As formal power series, the relationship between $\mathbf{J}$ and $\{S,\mathbf{t}\}$ is invertible.  
			The inverse transformation is given by
			\begin{equation}\label{J to t}
				\begin{aligned}
					t_n=\frac{1}{2n+1}\sum_{m=0}^{\infty} J_{n+m+1}\frac{(-J_0)^m}{m!}.
				\end{aligned}
			\end{equation}
			This identity follows from the inverse of the Galilean transformation (\ref{def of Jn}). 
			\begin{remark}
				For two infinite sequences $\{a_n\}$, $\{b_n\}$ and an indeterminate $q$, if $\{b_n\}$ is the Galilean transformation of $\{a_n\}$ as a formal power series:
				\begin{equation}
					b_n=\sum_{m=0}^{\infty} a_{n+m}\frac{q^m}{m!},
				\end{equation}
				then the inverse Galilean transformation is:
				\begin{equation}
					a_n=\sum_{m=0}^{\infty} b_{n+m}\frac{(-q)^m}{m!}.
				\end{equation}
			\end{remark}
			
			One also has
			\begin{equation}\label{S expre}
				\begin{aligned}
					S&=-\sum_{n=0}^{\infty} \frac{t_n}{(2n)!!}Q^{2n+1}+Q=-\sqrt{2J_0}\sum_{n=0}^{\infty} \frac{t_n}{n!}J_0^{n}+\sqrt{2J_0}\\
					&=\sqrt{2J_0}\big(1-\sum_{n=0}^{\infty}\frac{1}{(2n+1)n!}\sum_{m=0}^{\infty}\frac{J_{n+m+1}(-1)^m J_0^{n+m}}{m!} \big)\\
					&=\sqrt{2J_0}(1-\sum_{n=0}^{\infty}\sum_{k+l=n} \frac{(-1)^k}{2l+1}\binom{n}{l}J_{n+1}\frac{J_0^{n}}{n!})\\
					&=\sqrt{2J_0}\big(1-\sum_{n=0}^{\infty} J_{n+1}\frac{(-2J_0)^{n}}{(2n+1)!!}\big),
				\end{aligned}
			\end{equation}
			where in the third line we have used the combinatorial identity
			\begin{equation}
				\sum_{l=0}^{n} \frac{(-1)^l}{2l+1}\binom{n}{l}=\int_{0}^{1} (1-x^2)^n \d x=\int_{0}^{\frac{\pi}{2}} \cos^{2n+1}(\theta)\d \theta=\frac{(2n)!!}{(2n+1)!!}.
			\end{equation}
			
			The $\mathbf{J}$-coordinates are also weighted homogeneous, with degrees
			\begin{equation}
				\begin{aligned}
					\deg J_n= 2n-2 \qquad \text{for } n=0,1,2,\cdots.
				\end{aligned}
			\end{equation}
			
			By directly differentiating the Landau-Ginzburg equation (\ref{LG equ for J0}) and the defining relation (\ref{def of Jn}), we obtain the derivatives of $\mathbf{J}$ with respect to $S$ and $\mathbf{t}$:
			\begin{equation}\label{jacobian for J0}
				\begin{aligned}
					&\frac{\partial J_0}{\partial S}=\frac{\sqrt{2J_0}}{1-J_1} ; \\
				 &  \frac{\partial J_0}{\partial t_n}=(1-J_1)^{-1}\frac{2}{n!}J_0^{n+1},
				\end{aligned}
			\end{equation}
			and for $n\geq 1$, $m=0,1,\cdots$ 
			\begin{equation}\label{Jac for Jn}
				\begin{aligned}
					\frac{\partial J_n}{\partial S}&=\frac{\sqrt{2J_0}}{1-J_1}J_{n+1};\\
					\frac{\partial J_n}{\partial t_m}&=(1-J_1)^{-1}\frac{2}{m!}J_0^{m+1}J_{n+1}+\delta_{m+1\geq n}\frac{2m+1}{(m+1-n)!}J_0^{m+1-n},
				\end{aligned}
			\end{equation}
			where $\delta_{m+1\geq n}$ equals $1$ when $m+1\geq n$ and equals $0$ when $m\leq n$. 
			
			The derivatives of the transformation from $\mathbf{J}$-coordinates to $\mathbf{t}$-coordinates are:
			\begin{equation}\label{S der with Jn}
				\begin{aligned}
					&\frac{\partial S}{\partial J_0}=(2J_0)^{-\frac{1}{2}}\big(1-\sum_{n=0}^{\infty} \frac{(-1)^{n}J_{n+1}}{(2n-1)!!}(2J_0)^n\big);\\
					&\frac{\partial S}{\partial J_n}=\frac{(-1)^n (2J_0)^{n-\frac{1}{2}}}{(2n-1)!!} \qquad \text{for } n=1,2,\cdots
				\end{aligned}
			\end{equation}
			where $(-1)!!=1$.
			And for $n=0,1,2,\cdots$
			\begin{equation}
				\begin{aligned}
					&\frac{\partial t_n}{\partial J_0}=-\frac{1}{2n+1}\sum_{m=0}^{\infty} J_{n+m}\frac{(-J_0)^{m}}{m!};\\
					&\frac{\partial t_n}{\partial J_m}=\delta_{m\geq n+1}\frac{1}{2n+1}\frac{(-J_0)^{m-n-1}}{(m-n-1)!} \qquad \text{for } m=1,2,\cdots.
				\end{aligned}
			\end{equation}
		
			We will prove the finiteness of $F^{gB}_g$ in $\mathbf{J}$-coordinates for $g\geq 1$ in Section \ref{section finiteness}.
		
			\section{Multidifferentials of generalized BGW theory}\label{section Multidifferential}
			Information about the genus-zero sector is encoded in the spectral curve of the theory, while the higher-genus sectors can be regarded as arising from a quantization of this spectral curve \cite{zhouEmergentGeometryMirror2015,zhouEmergentGeometryUnifications2025}. The full Virasoro constraints correspond precisely to this quantization and provide the recursions for the following definition of the multidifferentials. 
			In this section, we define the multidifferentials for the generalized BGW theory and derive their recursion relations.
			
			\subsection{Quantization of the spectral curve}\label{quantization of the spectral curve}
			For the spectral curve $y^{gB}(w)$  of the generalized BGW theory defined in the equation (\ref{def of spectral curve}),
			we regard it as a classical field and define its quantum field $\beta(w)$ by
			\begin{equation}
				\begin{aligned}
						\beta(w)&=w^{-\frac{1}{2}}-\sum_{n=0}^{\infty}\frac{t_n}{(2n-1)!!}w^{n-\frac{1}{2}}-\lambda^2\sum_{n=0}^{\infty} \frac{(2n+1)!!}{w^{n+\frac{3}{2}}}\frac{\partial}{\partial t_n}\\
						&=-K(w)-\lambda^2\nabla(w).
				\end{aligned}
			\end{equation}
			We regard $t_n$ as a creation operator and $\partial/\partial t_n$ as an annihilation operator, and we define the normal ordering by
			\begin{equation}
				:t_n\frac{\partial}{\partial t_m}:=:\frac{\partial}{\partial t_m} t_n:=t_n\frac{\partial}{\partial t_m}.
			\end{equation}
			It follows that the normal-ordered product of two fields $\beta(z)$ and $\beta(w)$ is
			\begin{equation}
				:\beta(z)\beta(w):=K(z)K(w)+\lambda^2(K(z)\nabla(w)+K(w)\nabla(z))+\lambda^4\nabla(z)\nabla(w).
			\end{equation}
			The difference between the ordinary product of two fields $\beta(z)\beta(w)$ and the normal-ordered product of them is then given by
			\begin{equation}\label{diff for prod fie}
				\begin{aligned}
					\beta(z)\beta(w)-:\beta(z)\beta(w):&=\sum_{n=0}^{\infty} \big[\lambda^2 \frac{(2n+1)!!}{z^{n+\frac{3}{2}}}\frac{\partial}{\partial t_n}, \frac{t_n}{(2n-1)!!}w^{n-\frac{1}{2}}\big]\\
					&=\sum_{n=0}^{\infty}\lambda^2(2n+1)\frac{w^{n-\frac{1}{2}}}{z^{n+\frac{3}{2}}}\\
					&=\frac{2\lambda^2}{(z-w)^2}(\frac{z}{w})^{\frac{1}{2}}-\frac{\lambda^2}{(z-w)}\frac{1}{\sqrt{zw}}.
				\end{aligned}
			\end{equation}
		
			The regularized product $\beta(w)^{\odot 2}$ of the field $\beta(w)$ is a standard notion in conformal field theory; it encodes the operator product expansion (OPE). It is defined as the sum of the normal-ordered product $:\beta(w)^2:$ and the non-singular component of (\ref{diff for prod fie}) in the limit $z\to w$.
			
			 Let $z=w+\epsilon$ and take $\epsilon\to 0$. Then
			\begin{equation}
				\begin{aligned}
					&\frac{2\lambda^2}{(z-w)^2}(\frac{z}{w})^{\frac{1}{2}}-\frac{\lambda^2}{(z-w)}\frac{1}{\sqrt{zw}}=\frac{2\lambda^2}{\epsilon^2}(1+\frac{\epsilon}{w})^{\frac{1}{2}}-\frac{\lambda^2}{\epsilon w}(1+\frac{\epsilon}{w})^{-\frac{1}{2}}\\
					&=\frac{2\lambda^2}{\epsilon^2}(1+\frac{1}{2}\frac{\epsilon}{w}-\frac{1}{8}\frac{\epsilon^2}{w^2}+O(\epsilon^3))-\frac{\lambda^2}{\epsilon w}(1-\frac{1}{2}\frac{\epsilon}{w}+O(\epsilon^2))\\
					&=\frac{2\lambda^2}{\epsilon^2}+\frac{1}{4}\frac{\lambda^2}{w^2}+O(\epsilon).
				\end{aligned}
			\end{equation}
			The non-singular component of the equation (\ref{diff for prod fie}) is $\frac{\lambda^2}{4w^2}$.
			Hence, the regularized product is
			\begin{equation}
				\beta(w)^{\odot 2}=:\beta(w)^2:+\frac{\lambda^2}{4w^2}.
			\end{equation}
			It has been shown that the partition function of the BGW theory is annihilated by the negative part of this regularized product (after an appropriate shift in the degree of $w$) \cite{zhouEmergentGeometryUnifications2025}:
			\begin{equation}
				[\beta(w;\lambda)^{\odot 2}]_{\leq -2}Z^{BGW}(\mathbf{t};\lambda)=0,
			\end{equation}
				where $[\ \cdot\ ]_{\leq -2}$ denotes the truncation of the Laurent series:
			\begin{equation}\label{cutoff notation}
				[\sum_{n=-\infty}^{\infty}a_n w^n]_{\leq-2}=\sum_{n=-\infty}^{-2}a_n w^n.
			\end{equation}
			This is equivalent to the Virasoro constraints of the BGW theory:
			\begin{equation}
				\begin{aligned}
						&\lambda^2\nabla(w) F^{BGW}(\mathbf{t};\lambda)\cdot \lambda^2\nabla(w) F^{BGW}(\mathbf{t};\lambda)+\lambda^4 \nabla(w)\nabla(w)F^{BGW}(\mathbf{t};\lambda)\\
					+&2\lambda^2 [K(w;\mathbf{t})\nabla(w)F^{BGW}(\mathbf{t};\lambda)]_{\leq -2}+\frac{\lambda^2}{4w^2}=0.
				\end{aligned}
			\end{equation}
			By adding the term $\frac{S^2}{w^2}$ to the equation, one obtains the Virasoro constraints of the generalized BGW theory:
			\begin{equation}
				[\beta(w;\lambda)^{\odot 2}+\frac{S^2}{w^2}]_{\leq -2} Z^{gB}(\mathbf{t};S;\lambda)=0,
			\end{equation}
			and
			\begin{equation}\label{Vir1}
				\begin{aligned}
					&\lambda^2\nabla(w) F^{gB}(\mathbf{t};S;\lambda)\cdot \lambda^2\nabla(w) F^{gB}(\mathbf{t};S;\lambda)+\lambda^4 \nabla(w)\nabla(w)F^{gB}(\mathbf{t};S;\lambda)\\
					+&2\lambda^2 [K(w;\mathbf{t})\nabla(w)F^{gB}(\mathbf{t};S;\lambda)]_{\leq -2}+\frac{\lambda^2}{4w^2}+\frac{S^2}{w^2}=0,
				\end{aligned}
			\end{equation}

			\begin{remark}
				 It appears that $S$ may be the first member of an additional sequence of parameters. We leave this possibility for future investigation.
			\end{remark}
			
			\subsection{Multidifferentials of the generalized BGW theory}
			Following the work \cite{zhouEmergentGeometryUnifications2025}, we define the multidifferentials of the generalized BGW theory in $w$ as
			\begin{equation}
				\begin{aligned}
					&W^{gB}_{g,n}(w_1,w_2,\cdots,w_n;\mathbf{t};S)\\
					&:=\nabla(w_1)\nabla(w_2)\cdots\nabla(w_n) F^{gB}_{g}(\mathbf{t};S;\lambda)\d w_1\d w_2\cdots\d w_n \\
					&=\sum_{i_1,i_2,\cdots,i_n\geq 0}\frac{\partial}{\partial t_{i_1}}\frac{\partial}{\partial t_{i_2}}\cdots\frac{\partial}{\partial t_{i_n}}F^{gB}_{g}\prod_{k=1}^{n} \frac{(2i_k+1)!!\d w_k}{w_k^{i_k+\frac{3}{2}}}
				\end{aligned}
			\end{equation}			
			
			From \ref{Vir1}, we obtain the recursion, for $n=1$,
			\begin{equation}\label{n=1 tr}
				\begin{aligned}
					\frac{1}{2\d  w}\sum_{g_1+g_2=g}W^{gB}_{g_1,1}(w)\cdot W^{gB}_{g_2,1}(w)+\frac{1}{2\d w}W^{gB}_{g-1,2}(w,w)\\
					+[K(w)W^{gB}_{g,1}(w)]_{\leq -2}+\frac{\delta_{g,1}}{8w^2}\d w+\frac{\delta_{g,0}S^2}{2w^2}\d w=0.
				\end{aligned}
			\end{equation}
			Acting with several $\nabla(w_i)$ on this equation, we obtain the recursions, for $n\geq 2$,
			\begin{equation}\label{n>1 tr}
				\begin{aligned}
					\frac{1}{2\d w}\sum_{\substack{g_1+g_2=g\\ I_1\sqcup I_2=[n]}}W^{gB}_{g_1,|I_1|+1}(w,w_{I_1})W^{gB}_{g_2,|I_2|+1}(w,w_{I_2})+\frac{1}{2\d w}W^{gB}_{g-1,n+2}(w,w,w_{[n]})\\
					+[K(w)W^{gB}_{g,n+1}(w,w_{[n]})]_{\leq -2}+\sum_{i=1}^{n}[\nabla(w_i)K(w)\d w_i W^{gB}_{g,n}(w,w_{[n]_i})]_{\leq-2}=0,
				\end{aligned}
			\end{equation}
			where $[n]=\{1,2,\cdots,n\}$, $[n]_i=[n]-\{i\}$ and $w_{I}=\{w_i\}_{i\in I}$. 
			
			Two special initial data are $(g,n)=(0,1),(0,2)$, $W^{gB}_{0,1}$ is just a component of the spectral curve
			\begin{equation}
				W^{gB}_{0,1}(w)=(-K(w)-y^{gB}(w))\d w.
			\end{equation}
			Using Proposition \ref{cons rel of F in J}, $W_{0,2}^{gB}$ is given by
				\begin{equation}\label{w02}
				\begin{aligned}
					W^{gB}_{0,2}(w_1,w_2)=
					& \sum_{n,m=0}^{\infty}(2n+1)!!(2m+1)!!\frac{\partial^2 F^{gB}_0}{\partial t_n\partial t_m} \frac{\d w_1}{w_1^{n+\frac{3}{2}}}\frac{\d w_2}{w_2^{m+\frac{3}{2}}} \\
					=&\sum_{n,m=0}^{\infty}\frac{(2n+1)!!(2m+1)!!}{(m+n+1)n!m!}J^{n+m+1}_0\frac{\d w_1}{w_1^{n+\frac{3}{2}}}\frac{\d w_2}{w_2^{m+\frac{3}{2}}}\\
					=&\frac{1}{(w_1-w_2)^2}\Big(\frac{w_1+w_2-4J_0}{\sqrt{w_1-2J_0}\sqrt{w_2-2J_0}}-\frac{w_1+w_2}{w_1^{\frac{1}{2}} w_2^{\frac{1}{2}}}\Big)\d w_1\d w_2.
				\end{aligned}
			\end{equation}
			To pass from the second line to the third, one may differentiate both sides with respect to $J_0$ and verify that they coincide, sharing the same initial value $0$ at $J_0=0$; that is,
			\begin{equation}
				\begin{aligned}
						\frac{\partial}{\partial J_0} W^{gB}_{0,2}(w_1,w_2)=&\sum_{n,m=0}^{\infty} \frac{(2n+1)!!}{n!}\frac{J_0^n}{w_1^{n+\frac{3}{2}}}\frac{(2m+1)!!}{m!}\frac{J_0^{m}}{w_2^{m+\frac{3}{2}}} \d w_1\d w_2\\
						=&\frac{\d w_1 }{(w_1-2J_0)^{\frac{3}{2}}}\frac{\d w_2}{(w_2-2J_0)^{\frac{3}{2}}},
				\end{aligned}
			\end{equation}
			where we have used the combinatorial identity
			\begin{equation}
				\sum_{n=0}^{\infty} \frac{(2n+1)!!}{n!} x^n=\sum_{n=0}^{\infty} \binom{-\frac{3}{2}}{n}(-2)^n x^n=(1-2x)^{-\frac{3}{2}}.
			\end{equation}
			
			However, recursions (\ref{n=1 tr}) and (\ref{n>1 tr}) are not yet in an effective form, because $W^{gB}_{g,n+1}$ itself appears in the first summation. We rewrite \eqref{n=1 tr} as
			\begin{equation}\label{real recursion n=1}
					\begin{aligned}
					[(y^{gB}(w)\d  w)W^{gB}_{g,1}(w)]_{\leq -2}=&
					\frac{1}{2}\sum_{\substack{g_1+g_2=g\\g_1,g_2>0}}W^{gB}_{g_1,1}(w)\cdot W^{gB}_{g_2,1}(w)\\
					&+\frac{1}{2}W^{gB}_{g-1,2}(w,w)
					+\frac{\delta_{g,1}}{8w^2}(\d w)^2+\frac{\delta_{g,0}S^2}{2w^2}(\d w)^2.
				\end{aligned}
			\end{equation}
			This constitutes a genuine recursion, since the right-hand side no longer involves the term $W^{gB}_{g,1}$. Moreover, the meromorphic 1-form $\omega=y\d x$ that appears in the Eynard-Orantin topological recursion (see the next section) emerges on the left-hand side if we set $x=\frac{1}{2}w$. 
			
			To establish a connection with the Eynard-Orantin topological recursion, we must also extract the contribution of $W^{gB}_{0,2}$ in (\ref{n>1 tr}). A direct computation yields
			\begin{equation}
					\begin{aligned}
						\nabla(w_i)K(w)=&\sum_{n=0}^{\infty}\frac{(2n+1)!!}{w_i^{n+\frac{3}{2}}}\frac{\partial}{\partial t_n}(-w^{-\frac{1}{2}}+\sum_{m=0}^{\infty}\frac{t_m}{(2m-1)!!}w^{m-\frac{1}{2}})\\
						=&\sum_{n=0}^{\infty} (2n+1)w_i^{-\frac{3}{2}}w^{-\frac{1}{2}}(\frac{w}{w_i})^{n}=
						w^{-\frac{1}{2}}w_i^{-\frac{1}{2}}\frac{w+w_i}{(w-w_i)^2}.
					\end{aligned}
			\end{equation}
			Thus,
			\begin{equation}\label{B0}
				\begin{aligned}
					W^{gB}_{0,2}(w,w_i)+\nabla(w_i)K(w)\d w\d w_i=&\frac{1}{(w-w_i)^2}\frac{w+w_i-4J_0}{\sqrt{w-2J_0}\sqrt{w_i-2J_0}}\d w\d w_i \\
					=&4\frac{z^2+z^2_i}{(z^2-z^2_i)^2}\d z\d z_i,
				\end{aligned}
			\end{equation}
			with the coordinate transformation $w=z^2+2J_0$. The bidifferential appearing in the last line is a variant of the Bergman kernel used in the Eynard-Orantin topological recursion (see \cite{zhouTopologicalRecursionsEynard2013}). We denote it by 
			\begin{equation}
				B(w,w_i)=\frac{1}{4}W^{gB}_{0,2}(w,w_i)+\frac{1}{4}\nabla(w_i)K(w)\d w\d w_i.
			\end{equation}
			Be careful, $W_{0,2}^{gB}(w,w_i)$ is a Puiseux series in $w$ and $w_i$ whose order is at most $-3/2$, while $\nabla(w_i)K(w)$ is a Puiseux series in $w$ of positive order and in $w_i$ of negative order. The term $(w-w_i)^{-2}$ in the expression (\ref{B0}) should be understood via the expansion
			\begin{equation}\label{B exp}
				(w-w_i)^{-2}=w_i^{-2}(1-\frac{w}{w_i})^{-2}=\sum_{n=0}^{\infty}(n+1)\frac{w^n}{w_i^{n+2}}.
			\end{equation}
			
			With this, the equation (\ref{n>1 tr}) becomes
			\begin{equation}\label{real re for n>1}
				\begin{aligned}
					[(y^{gB}(w)\d w)W^{gB}_{g,n+1}(w,w_{[n]})]_{\leq -2}=
					 \frac{1}{2}\sum^s_{\substack{g_1+g_2=g\\ I_1\sqcup I_2=[n]}}W^{gB}_{g_1,|I_1|+1}(w,w_{I_1})W^{gB}_{g_2,|I_2|+1}(w,w_{I_2})\\
					+\frac{1}{2}W^{gB}_{g-1,n+2}(w,w,w_{[n]})+\sum_{i=1}^{n} 4[B(w,w_i)W_{g,n}^{gB}(w,w_{[n]_i})]_{\leq-2},
				\end{aligned}
			\end{equation}
			where $\sum^s$ indicates that the summation only includes the stable terms $(g,n)\ne(0,1),(0,2)$.
			The only exception to the recursion (\ref{real re for n>1}) is $W_{0,3}^{gB}$, which we compute separately. By definition,
			\begin{equation}
				\begin{aligned}
					W^{gB}_{0,3}(w_0,w_1,w_2)=\nabla(w_0)W^{gB}_{0,2}(w_1,w_2)\d w_0
				\end{aligned}
			\end{equation}
			Observe that the coefficients of $W^{gB}_{0,2}(w_1,w_2)$ depend only on $J_0$ (by the equation (\ref{w02})); thus using the relation (\ref{jacobian for J0}), we have
			\begin{equation}
				\begin{aligned}
					&W^{gB}_{0,3}(w_0,w_1,w_2)=\nabla(w_0)W^{gB}_{0,2}(w_1,w_2)\d w_0\\
					=&\sum_{n=0}^{\infty} \frac{(2n+1)!!}{w_0^{n+\frac{3}{2}}}\frac{\partial J_0}{\partial t_n}\frac{\partial}{\partial J_0}W^{gB}_{0,2}(w_1,w_2)\d w_0\\
					=&\sum_{n=0}^{\infty} \frac{(2n+1)!!}{w_0^{n+\frac{3}{2}}}\frac{1}{1-J_1}\frac{2}{n!}J_0^{n+1}\frac{\d w_1}{(w_1-2J_0)^{\frac{3}{2}}}\frac{\d w_2}{(w_2-2J_0)^{\frac{3}{2}}}\d w_0\\
					=& \frac{2J_0}{1-J_1}\frac{\d w_0}{(w_0-2J_0)^{\frac{3}{2}}}\frac{\d w_1}{(w_1-2J_0)^{\frac{3}{2}}}\frac{\d w_2}{(w_2-2J_0)^{\frac{3}{2}}}.
				\end{aligned}
			\end{equation}

			\section{Eynard-Orantin topological recursion of generalized BGW theory}\label{section EO}
			In this section, we show that the recursion formulas (\ref{real recursion n=1}) and (\ref{real re for n>1}) lead naturally to the Eynard-Orantin topological recursion, and that the spectral curve defined by the special deformation in the context of emergent geometry coincides with the spectral curve in the sense of Eynard-Orantin theory.
			
			\subsection{Eynard-Orantin topological recursion with only one ramification point}
			We will use the version of the Eynard-Orantin topological recursion that involves only one ramification point. For brevity, we give only a brief outline here; for the general theory, we refer the reader to \cite{eynardInvariantsAlgebraicCurves2007}.
			
			The Eynard-Orantin topological recursion with one ramification point generates a sequence of multidifferentials $\{\omega_{g,n}\}$ from the initial data of a (local) spectral curve. In the local Airy coordinate for this curve, the initial data are
			\begin{equation}
				\begin{aligned}
					x(z)=\frac{1}{2}z^2+x(a); \qquad y(z)=a_0+a_1 z+\cdots \text{ with } a_1\ne 0; \\
					B(z_1,z_2)=B_0(z_1,z_2)+\text{regular terms}=\frac{\d z_1\d z_2}{(z_1-z_2)^2}+\text{regular terms}.
				\end{aligned}
			\end{equation} 
			After setting $\omega_{0,1}=y\d x,\ \omega_{0,2}=B(z_1,z_2)$, then the recursion for $2g-2+n>0$ is 
			\begin{equation}
				\begin{aligned}
						\omega_{g,n+1}(z_0,z_1,\cdots,z_n)=&\frac{1}{2}\text{res}_{z=0}\frac{K(z_0,z)}{\omega_{0,1}(z)-\omega_{0,1}(\bar{z})}
						\big(
						\omega_{g-1,n+2}(z,\bar{z},z_{[n]})\\
						&+\sum^\prime_{\substack{g_1+g_2=g\\ I_1\sqcup I_2=[n]}}\omega_{g_1,|I_1|+1}(z,z_{I_1})\omega_{g_2,|I_2|+1}(\bar{z},z_{I_2})
						\big),
				\end{aligned}
			\end{equation}
			where $K$ is the recursion kernel defined via the Bergman kernel: 
			\begin{equation}
				K(z_0,z):=\int_{z_2=\bar{z}}^{z}B(z_0,z_2),
			\end{equation}
			 $\bar{z}=-z$ is the involution of $z$ in the local Airy coordinate, and $\sum^{\prime}$ indicates that the term $\omega_{0,1}$ is omitted in the summation.
			
			When the Bergman kernel is the standard planar Bergman kernel $B_0$, the recursion kernel simplifies to
			\begin{equation}
				K_0(z_0,z)=\frac{2z\d z_0}{z_0^2-z^2}.
			\end{equation}
			 Moreover, one can show that, apart from the case of $\omega_{1,1}$, the standard Bergman kernel used in the recursion can be replaced by the bidifferential (see \cite{zhouTopologicalRecursionsEynard2013})
			 \begin{equation}
			 B(z_1,z_2)=\frac{z_1^2+z_2^2}{(z_1^2-z_2^2)^2}\d z_1 \d z_2.
			 \end{equation}
		
			\subsection{Multidifferentials in shifted coordinates}
			To derive the Eynard-Orantin topological recursion, we must solve the truncation equations (\ref{real recursion n=1}) and (\ref{real re for n>1}). Although the spectral curve and multidifferentials admit a convenient expression in the coordinate $z=\sqrt{w-2J_0}$, a direct substitution would disrupt the truncation condition, which is formulated in terms of the order in $w$. However, the coordinate transformation $\tilde{z}=\sqrt{w}$  preserves the truncation condition with respect to the order in $\tilde{z}$.
			
			Thus, by abuse of notation, we define the multidifferentials in $z$ as
			\begin{equation}\label{def of Wz}
				\begin{aligned}
					W^{gB}_{g,n}(z_1,z_2,\cdots,z_n):=&\frac{1}{2^n}W^{gB}_{g,n}(w_1=z_1^2+2J_0,\cdots,w_n=z_n^2+2J_0)\\
						=&\sum_{i_1,i_2,\cdots,i_n\geq 0} \frac{\partial^n F^{gB}_g(\mathbf{t};S)}{\partial t_{i_1}\partial t_{i_2}\cdots\partial t_{i_n}}\prod_{k=1}^{n}\frac{(2i_k+1)!!z_k\d z_k}{(z_k^2+2J_0)^{i_k+\frac{3}{2}}}.
				\end{aligned}
			\end{equation}
			Under the transformation $z^2=\tilde{z}^2-2J_0$, the multidifferentials 
			\begin{equation}
				W^{gB}_{g,n}(\tilde{z}_1,\cdots,\tilde{z}_n):=W^{gB}_{g,n}(z_1=\sqrt{\tilde{z}_1^2-2J_0},\cdots,z_n=\sqrt{\tilde{z}_n^2-2J_0})
			\end{equation}
			satisfy the following truncation equations:
			\begin{equation}\label{re for n=1 final}
				\begin{aligned}
						[y^{gB}(\tilde{z})(\tilde{z}\d \tilde{z})W^{gB}_{g,1}(\tilde{z})]_{\leq -2}=&
						\frac{1}{2}\sum_{\substack{g_1+g_2=g\\g_1,g_2>0}}W^{gB}_{g_1,1}(\tilde{z})\cdot W^{gB}_{g_2,1}(\tilde{z})\\
						&+\frac{1}{2}W^{gB}_{g-1,2}(\tilde{z},\tilde{z})
						+\frac{\delta_{g,1}}{8\tilde{z}^2}(\d \tilde{z})^2+\frac{\delta_{g,0}S^2}{2\tilde{z}^2}(\d \tilde{z})^2,
				\end{aligned}
			\end{equation}
			and
			\begin{equation}\label{re for n>1 final}
				\begin{aligned}
					[y^{gB}(\tilde{z})(\tilde{z}\d \tilde{z})W^{gB}_{g,n+1}(\tilde{z},\tilde{z}_{[n]})]_{\leq -2}=
					\frac{1}{2}\sum^s_{\substack{g_1+g_2=g\\ I_1\sqcup I_2=[n]}}W^{gB}_{g_1,|I_1|+1}(\tilde{z},\tilde{z}_{I_1})W^{gB}_{g_2,|I_2|+1}(\tilde{z},\tilde{z}_{I_2})\\
					+\frac{1}{2}W^{gB}_{g-1,n+2}(\tilde{z},\tilde{z},\tilde{z}_{[n]})+\sum_{i=1}^{n} [B(\tilde{z},\tilde{z}_i)W_{g,n}^{gB}(\tilde{z},\tilde{z}_{[n]_i})]_{\leq-2},
				\end{aligned}
			\end{equation}
			where $[\ \cdot\  ]_{\leq-2}$ means taking the truncation of the Laurent series in $\tilde{z}$ as in (\ref{cutoff notation}), and 
			\begin{equation}\label{B z}
				B(\tilde{z},\tilde{z}_i)=\frac{\tilde{z}^2+\tilde{z}^2_i-4J_0}{(\tilde{z}^2-\tilde{z}^2_i)^2}\frac{\tilde{z}\d \tilde{z}}{\sqrt{\tilde{z}^2-2J_0}}\frac{\tilde{z}_i\d \tilde{z}_i}{\sqrt{\tilde{z}_i^2-2J_0}}.
			\end{equation}
			
			By Proposition \ref{spec curve in J}, the spectral curve of the generalized BGW theory can be expressed as
			\begin{equation}
				y^{gB}(\tilde{z})=\sum_{n=0}^{\infty} y_n(\tilde{z}^2-2J_0)^{n+\frac{1}{2}}\tilde{z}^{-2},
			\end{equation}
			where
			\begin{equation}
				y_n=\delta_{n,0}-\frac{J_{n+1}}{(2n+1)!!}.
			\end{equation}
			Considering only the variable $\tilde{z}$ and expanding 
			\begin{equation}
				\begin{aligned}
					\frac{1}{\tilde{z}^{2m+3}}=&\frac{1}{(\tilde{z}^2-2J_0+2J_0)^{m+\frac{3}{2}}} \\
						=&\frac{1}{(\tilde{z}^2-2J_0)^{m+\frac{3}{2}}}\big(\sum_{k=0}^{\infty}\binom{-m-\frac{3}{2}}{k}(\frac{2J_0}{\tilde{z}^2-2J_0})^k\big),
				\end{aligned}
			\end{equation}
			 we see that the multidifferentials can be expressed in the form
			\begin{equation}
				W^{gB}_{g,n+1}=\sum_{m=0}^{\infty}\frac{a_m \tilde{z}\d \tilde{z}}{(\tilde{z}^2-2J_0)^{m+\frac{3}{2}}},
			\end{equation}
			for some temporary coefficients $a_m$.
			In general, the right-hand side of the truncation equations can also be written in the form
			\begin{equation}
				RHS=\sum_{m=0}^{\infty} \frac{b_m(\tilde{z}\d \tilde{z})^2}{(\tilde{z}^2-2J_0)^{m+2}},
			\end{equation}
			for some temporary coefficients $b_m$.
			Thus, the truncation equation becomes
			\begin{equation}
				[\sum_{n,m=0}^{\infty}y_na_m(\tilde{z}^2-2J_0)^{n-m-1}]_{\leq-2}=\sum_{k=0}^{\infty} \frac{b_k\tilde{z}^2}{(\tilde{z}^2-2J_0)^{k+2}}.
			\end{equation}
			
			We have the following lemma:
			\begin{lemma}
				The truncation equation
				\begin{equation*}
						[\sum_{n,m=0}^{\infty}y_na_m(\tilde{z}^2-2J_0)^{n-m-1}]_{\leq-2}=\sum_{k=0}^{\infty} \frac{b_k\tilde{z}^2}{(\tilde{z}^2-2J_0)^{k+2}}
				\end{equation*}
				admits the residue solution:
				\begin{equation}
					\begin{aligned}
							\sum_{m=0}^{\infty} \frac{a_m\d z_0}{z_0^{2m+2}}=& \text{res}_{z=0}\frac{
							\frac{z\d z_0}{z_0^2-z^2}(z^2+2J_0)\sum_{k=0}^{\infty} \frac{b_k(z\d z)^2}{z^{2k+4}}
							}{ \sum_{n=0}^{\infty} y_n z^{2n+1} z\d z
							}\\
							=&\text{res}_{z=0}\frac{\frac{z\d z_0}{z_0^2-z^2}\sum_{k=0}^{\infty} \frac{b_k(z\d z)^2}{z^{2k+4}}
							}{y(z)\d x(z)
							},
					\end{aligned}
				\end{equation}
				where $y(z)=\frac{1}{z^2+2J_0}\sum_{n=0}^{\infty} y_n z^{2n+1}$ and $x(z)=\frac{1}{2}(z^2+2J_0)$.
			\end{lemma}
			\begin{proof}
				The left-hand side of the truncation equation is just
				\begin{equation}
					\begin{aligned}
						LHS=& \sum_{n-m\leq 0}^{\infty}y_na_m(\tilde{z}^2-2J_0)^{n-m-1}\\
						=&\sum_{k=0}^{\infty} \sum_{m-n=k}y_na_m z^{-2k-2},
					\end{aligned}
				\end{equation}
				where $z^2=\tilde{z}^2-2J_0$, and the right-hand side is 
				\begin{equation}
					RHS=\sum_{k=0}^{\infty} (\frac{b_k}{z^{2k+2}}+\frac{2J_0b_k}{z^{2k+4}})
				\end{equation}
				under the coordinate transformation $z^2=\tilde{z}^2-2J_0$. This yields an infinite-dimensional linear system for the coefficients $a_m$:
				\begin{equation}
					\begin{aligned}
						y_0 a_0+&y_1 a_1+&y_2 a_2+\cdots=&b_0\\
									   &y_0 a_1 +&y_1 a_2+\cdots=&2J_0b_0+b_1\\
									   &                &y_0 a_2+\cdots=&2J_0b_1+b_2\\
									   &				     &\cdots.
					\end{aligned}
				\end{equation}
				Since for generic $y_0=1-J_1$, $y_0\ne 0$, one can write
				\begin{equation}
					\frac{1}{y_0+y_1 z^2+y_2z^4+\cdots}=\frac{1}{y_0}(1+v_1z^2+v_2z^4+\cdots).
				\end{equation}
				We see that the coefficients $y_i$ and $v_i$ are related by
				\begin{equation}
					\begin{aligned}
						&y_1+y_0v_1=0\\
						&y_2+y_1v_1+y_0v_2=0\\
						&y_3+y_2v_1+y_1v_2+y_0v_3=0\\
						&\cdots.
					\end{aligned}
				\end{equation}
				One then computes
				\begin{equation}
					\begin{aligned}
						a_0=&\frac{1}{y_0}(1\cdot b_0+v_1\cdot (2J_0b_0+b_1)+v_2\cdot (2J_0b_1+b_2)+\cdots)\\
						=&\text{res}_{z=0}\frac{z}{y_0}(1+v_1z^2+v_2z^4+\cdots)\\
						&\cdot (b_0 z^{-2}+(2J_0b_0+b_1)z^{-4}+(2J_0b_1+b_2)z^{-6}+\cdots)\d z\\
						=&\text{res}_{z=0}\frac{z(z^2+2J_0)\sum_{k=0}^{\infty} \frac{b_k(z\d z)^2}{z^{2k+4}}}{
						\sum_{n=0}^\infty y_nz^{2n+1}(z\d z)}\\
						=&\text{res}_{z=0}\frac{z\sum_{k=0}^{\infty} \frac{b_k(z\d z)^2}{z^{2k+4}}}{y(z)\d x(z)}.
					\end{aligned}
				\end{equation}
				Similarly,
				\begin{equation}
					a_m=\text{res}_{z=0}\frac{z^{2m+1}\sum_{k=0}^{\infty} \frac{b_k(z\d z)^2}{z^{2k+4}}}{y(z)\d x(z)}.
				\end{equation}
				And 
				\begin{equation}
					\sum_{m=0}^{\infty} \frac{z^{2m+1}}{z_0^{2m+2}}=\frac{z}{z_0^2-z^2}.
				\end{equation}
				It yields the desired formula.
			\end{proof}
			
			With this lemma, for $g\geq1$, the solution of the equation (\ref{re for n=1 final}) is
			\begin{equation}
				\begin{aligned}
						W_{g,1}^{gB}(z_0)=&\frac{1}{2}\text{res}_{z=0}\frac{z\d z_0}{z_0^2-z^2}\frac{1}{y^{gB}(z)\d x(z)}\\
						\cdot&
						\Big(\sum_{\substack{g_1+g_2=g\\g_1,g_2>0}}W^{gB}_{g_1,1}(z)\cdot W^{gB}_{g_2,1}(z)
					+W^{gB}_{g-1,2}(z,z)
					+\frac{\delta_{g,1}}{4
						(z^2+2J_0)^2}(z\d z)^2\Big).
				\end{aligned}
			\end{equation}
			To verify it satisfies the Eynard-Orantin topological recursion, we also need to check the case $W^{gB}_{1,1}$ coincides with the relation.
			It follows that
			\begin{equation}
				W^{gB}_{0,2}(z,z)=\frac{1}{4}\Big(\frac{1}{z^4}-\frac{1}{(z^2+2J_0)^2}\Big)(z\d z)^2,
			\end{equation}
			and
			\begin{equation}
				\begin{aligned}
					W^{gB}_{0,2}(z,z)+\frac{1}{4
						(z^2+2J_0)^2}(z\d z)^2=-\frac{\d z\d (-z)}{(z-(-z))^2}=-B_0(z,-z).
				\end{aligned}
			\end{equation}
			Thus, $W^{gB}_{1,1}$ coincides with the Eynard-Orantin topological recursion. Our multidifferentials differ by a factor $(-1)^n$ from the standard ones of the Eynard-Orantin topological recursion, because we use $\d z$ instead of $\d \bar{z}=-\d z$.
	
			Similarly, for $n\geq 1$ the solution of the equation (\ref{re for n>1 final}) is
			\begin{equation}\label{need remove}
				\begin{aligned}
						W_{g,n+1}^{gB}(z_0,z_{[n]})=&\frac{1}{2}\text{res}_{z=0}\frac{z\d z_0}{z_0^2-z^2}\frac{1}{y^{gB}(z)\d x(z)}\\
					\cdot&
					\Big[
					\sum^{s}_{\substack{g_1+g_2=g\\ I_1\sqcup I_2=[n]}}W^{gB}_{g_1,|I_1|+1}(z,z_{I_1})W^{gB}_{g_2,|I_2|+1}(z,z_{I_2})
					+W^{gB}_{g-1,n+2}(z,z,z_{[n]})\\
					+&2\sum_{i=1}^{n} [B(\tilde{z},\tilde{z}_i)W_{g,n}^{gB}(\tilde{z},\tilde{z}_{[n]_i})]_{\leq-2}\big|_{\tilde{z}^2=z^2+2J_0}
					\Big].
				\end{aligned}
			\end{equation}
			The last term requires careful treatment, as $B(\tilde{z},\tilde{z}_i)$ admits two kinds of expansions.  As explained in (\ref{B exp}), the correct expansion of $B(\tilde{z},\tilde{z}_i)$ is
			\begin{equation}
				\begin{aligned}
					B(\tilde{z},\tilde{z}_i)=\sum_{n=0}^{\infty}(n+1)\frac{\tilde{z}^{2n}}{\tilde{z}_i^{2n+4}}\frac{(\tilde{z}^2+\tilde{z}_i^2-4J_0)\d (\tilde{z}^2)\d (\tilde{z}_i^2)}{4\sqrt{\tilde{z}^2-2J_0}\sqrt{\tilde{z}^2_i-2J_0}}.
				\end{aligned}
			\end{equation}
			Writing $B(\tilde{z},\tilde{z}_i)$ and $W_{g,n}^{gB}$ as
			\begin{equation}
			\begin{aligned}
					B(\tilde{z},\tilde{z}_i)&=\sum_{n=0}^{\infty} B_n (\tilde{z}^2-2J_0)^{n-\frac{1}{2}}\d (\tilde{z}^2) \\
				W^{gB}_{g,n}&=\sum_{m=0}^{\infty} \frac{b_m}{(\tilde{z}^2-2J_0)^{m+\frac{3}{2}}}\d (\tilde{z}^2).
			\end{aligned}
			\end{equation}
			Then 
			\begin{equation}\label{B cutoff}
				\begin{aligned}
					&[B(\tilde{z},\tilde{z}_i)W_{g,n}^{gB}(\tilde{z},\tilde{z}_{[n]_i})]_{\leq-2}\big|_{\tilde{z}^2=z^2+2J_0}\\
					=&\sum_{n,m=0}^{\infty} [b_mB_n (\tilde{z}^2-2J_0)^{n-m-2}]_{\leq-4} (\d (\tilde{z}^2))^2\Big|_{\tilde{z}^2=z^2+2J_0}\\
					=&\sum_{n-m\leq 0}b_m B_nz^{2n-2m-4}(\d (z^2))^2\\
					&+\sum_{n-m=1}b_mB_n[\frac{1}{\tilde{z}^2-2J_0}]_{\leq-4}|_{\tilde{z}^2=z^2+2J_0}(\d (z^2))^2,
				\end{aligned}
			\end{equation}
			where $(\tilde{z}^2-2J_0)^{-1}$ should take the expansion
			\begin{equation}
				\frac{1}{\tilde{z}^2-2J_0}=\frac{1}{\tilde{z}^2}(\frac{1}{1-\frac{2J_0}{\tilde{z}^2}})=\frac{1}{\tilde{z}^2}+\frac{2J_0}{\tilde{z}^4}+\cdots.
			\end{equation}
			As a result, the truncated terms have the expression
			\begin{equation*}
				\begin{aligned}
					\sum_{n-m\geq2}b_mB_nz^{2n-2m-4}(\d (z)^2)^2+\sum_{n-m=1}b_mB_n(z^2+2J_0)^{-1}(\d (z^2))^2.
				\end{aligned}
			\end{equation*}
			One finds that the truncated terms do not contribute to the residue, because the factor $$\frac{z\d z_0}{z_0^2-z^2}\frac{1}{y^{gB}(z)\d x(z)}(\d (z^2))^2$$ is regular at $z=0$, and the truncated terms are also regular at $z=0$.
			We may therefore drop the truncation $[\ \cdot \  ]_{\leq-2}$ of the last term in the equation (\ref{need remove}).
			
			Finally, one also needs to verify that $W^{gB}_{0,3}$  satisfies the Eynard-Orantin topological recursion. This is a direct computation:
			\begin{equation}
				\begin{aligned}
						W^{gB}_{0,3}(z_0,z_1,z_2)=&
						\text{res}_{z=0}\frac{z\d z_0}{z_0^2-z^2}\frac{1}{y^{gB}(z)\d x(z)}B(z,z_1)B(z,z_2)\\
						=& \frac{2J_0}{1-J_1}\frac{\d z_0\d z_1 \d z_2}{z_0^2z_1^2z_2^2}.
				\end{aligned}
			\end{equation}

			In summary, we have proved the following theorem:
			\begin{thm}\label{EO TR foe gB}
					The multidifferentials $W^{gB}_{g,n}(z_{[n]})$ of the generalized BGW theory, defined by
						\begin{equation}
						\begin{aligned}
							W^{gB}_{g,n}(z_1,z_2,\cdots,z_n;\mathbf{t};S)
							:=\sum_{i_1,i_2,\cdots,i_n\geq 0} \frac{\partial^n F^{gB}_g(\mathbf{t};S)}{\partial t_{i_1}\partial t_{i_2}\cdots\partial t_{i_n}}\prod_{k=1}^{n}\frac{(2i_k+1)!!z_k\d z_k}{(z_k^2+2J_0)^{i_k+\frac{3}{2}}},
						\end{aligned}
					\end{equation}
					 satisfy the Eynard-Orantin topological recursion with the spectral curve $y^{gB}(z)$ obtained from the special deformation. More precisely,
					\begin{equation}
						\begin{aligned}
								W_{g,n+1}^{gB}(z_0,z_{[n]})=&\frac{1}{2}\text{res}_{z=0}\frac{z\d z_0}{z_0^2-z^2}\frac{1}{y^{gB}(z)\d x(z)}\\
							\cdot&
							\Big[
							\sum^\prime_{\substack{g_1+g_2=g\\ I_1\sqcup I_2=[n]}}W^{gB}_{g_1,|I_1|+1}(z,z_{I_1})W^{gB}_{g_2,|I_2|+1}(z,z_{I_2})
							+W^{gB}_{g-1,n+2}(z,z,z_{[n]})
							\Big],
						\end{aligned}
					\end{equation} 
					for $2g-2+n>0$, with
					\begin{equation}
						\begin{aligned}
							&x^{gB}(z)=\frac{1}{2}(z^2+2J_0)\\
							&y^{gB}(z)=\frac{z}{z^2+2J_0}\Big(1-\sum_{n=0}^{\infty} \frac{z^{2n}}{(2n+1)!!} J_{n+1}
							\Big),
						\end{aligned}
					\end{equation}
					and the equivalent Bergman kernel on the sphere
					\begin{equation}\label{equivalent Bergman kernel}
						B(z_1,z_2)=\frac{z_1^2+z_2^2}{(z_1^2-z_2^2)^2}\d z_1\d z_2.
					\end{equation}
			\end{thm}
			We use the equivalent Bergman kernel instead of standard $B_0(z_1,z_2)$ to avoid invalidation of the recursion for $W_{1,1}$.
			Some initial multidifferentials are
			\begin{equation}
				W^{gB}_{0,3}(z_1,z_2,z_3)=\frac{2J_0}{1-J_1}\frac{\d z_1\d z_2 \d z_3}{z_1^2z_2^2z_3^2},
			\end{equation}
			\begin{equation}\label{W11}
				W^{gB}_{1,1}(z)=\frac{J_0}{4(1-J_1)}\frac{\d z}{z^4}+\frac{1}{8(1-J_1)}\frac{\d z}{z^2}+\frac{J_0J_2}{12(1-J_1)^2}\frac{\d z}{z^2},
			\end{equation}
			\begin{equation}
				\begin{aligned}
						W^{gB}_{1,2}(z_1,z_2)=&\Big( \frac{5J_0^2}{2(1-J_1)^2}(\frac{1}{z_1^6z_2^2}+\frac{1}{z_1^2z_2^6})+\frac{3J_0^2}{2(1-J_1)^2}\frac{1}{z_1^4z_2^4}\\
						+&(\frac{3J_0}{2(1-J_1)^2}+\frac{J_0^2J_2}{(1-J_1)^3})(\frac{1}{z_1^4z_2^2}+\frac{1}{z_1^2z_2^4})\\
						+& (\frac{1}{8(1-J_1)^2}+\frac{7J_0J_2}{12(1-J_1)^3}+\frac{J_0^2J_3}{6(1-J_1)^3}+\frac{J_0^2J_2^2}{3(1-J_1)^4})\frac{1}{z_1^2z_2^2}\Big) \d z_1 \d z_2.
				\end{aligned}
			\end{equation}
			With Mathematica, we obtain
			\begin{equation}\label{W2,1}
				\begin{aligned}
					&W^{gB}_{2,1}(z)= \frac{105}{16}J_0^3\tilde{J}^3_1\frac{\d z}{z^{10}}+(\frac{203}{32}J_0^2\tilde{J}_1^3+\frac{203}{48}J_0^3\tilde{J}_1^4J_2)\frac{\d z}{z^8}\\
					&+\big(\frac{107}{64}J_0\tilde{J}_1^3+\frac{359}{96}J_0^2\tilde{J}_1^4J_2+\frac{7}{4}J_0^3\tilde{J}_1^5J_3+\frac{29}{48}J_0^3\tilde{J}_1^4J_3\big)\frac{\d z}{z^6}\\
					&+\big(\frac{9}{128}\tilde{J}_1^3+\frac{55}{64}J_0\tilde{J}_1^4J_2+\frac{17}{12}J_0^2\tilde{J}_1^5J_2^2+\frac{7}{12}J_0^3\tilde{J}_1^6J_2^3\\
					&+\frac{81}{160}J_0^2\tilde{J}_1^4J_3+\frac{29}{60}J_0^3\tilde{J}_1^5J_2J_3+\frac{1}{16}J_0^3\tilde{J}_1^4J_4\big)\frac{\d z}{z^4}\\
					&+\Big( \frac{3}{128}\tilde{J}_1^4J_2+\frac{55}{192}J_0\tilde{J}_1^5J_2^2+\frac{17}{36}J_0^2\tilde{J}_1^6J_2^3+\frac{7}{36}J_0^3\tilde{J}_1^7J_2^4\\
					&+\frac{107}{960}J_0\tilde{J}_1^4J_3+\frac{301}{720}J_0^2\tilde{J}_1^5J_2J_3+\frac{5}{18}J_0^3\tilde{J}_1^6J_2^2J_3+\frac{29}{720}J_0^3\tilde{J}_1^5J_3^2\\
					&+\frac{29}{480}J_0^2\tilde{J}_1^4J_4+\frac{11}{180}J_0^3\tilde{J}_1^5J_2J_4+\frac{1}{144}J_0^3\tilde{J}_1^4J_5
					\Big)\frac{\d z}{z^2},
				\end{aligned}
			\end{equation}
			where $\tilde{J}_1=(1-J_1)^{-1}$.
			
			\begin{remark}
				In particular, when all parameters $\mathbf{t}=\{t_0,t_1,\cdots\}$ are set to zero, the multidifferentials $W^{gB}_{g,n}(z_{[n]};\mathbf{t}=0;S)$ reduce to the standard multidifferentials of the generalized BGW theory obtained from the Eynard-Orantin topological recursion \cite{wangBKPaffineCoordinatesEmergent2025,zhouEmergentGeometryUnifications2025}. The corresponding spectral curve is
				\begin{equation}
					y=\frac{z}{z^2+S^2}.
				\end{equation}
				
				It has also been shown in \cite{zhouEmergentGeometryUnifications2025} that when $S=0$, the multidifferentials $W^{gB}_{g,n}(z_{[n]};\mathbf{t};S=0)$ coincide with those of the ordinary BGW theory $W^{BGW}_{g,n}(z_{[n]};\mathbf{t})$, since they share the same spectral curve (\ref{BGW spe cur}):
				\begin{equation*}
					y^{BGW}(z;\mathbf{t})=\frac{1}{z}-\sum_{n=0}^{\infty} \frac{t_n}{(2n-1)!!}z^{2n-1},
				\end{equation*}
				and the same remaining data of the Eynard-Orantin topological recursion relation.
			\end{remark}

			\section{Finiteness of free energy in the renormalized coordinate}\label{section finiteness}
			In the previous section, we proved that the spectral curve arising from the Eynard-Orantin topological recursion for the generalized BGW theory coincides with the spectral curve defined by the special deformation. According to the general theory of Eynard-Orantin topological recursion, the multidifferential $W_{g,n}$ depends only on finitely many coefficients of the expansion of the spectral curve at its ramification points for fixed $(g,n)$. We will use this property to show that, for genus $g\geq 1$, the free energy of the generalized BGW theory admits a finite expression in the $\mathbf{J}$-coordinates.
			
			\subsection{Proof of finiteness} 
			By finiteness, we mean that for $g\geq 1$, the free energy $F_g^{gB}$ can be expressed as a function of $S$ and only finitely many of the renormalized coordinates $J_0,J_1,J_2,\cdots$.
			
			We begin with the Eynard-Orantin topological recursion. Observe that
			\begin{equation}
				\begin{aligned}
					\frac{1}{y^{gB}(z)}=&\frac{z^2+2J_0}{(1-J_1)z}(1-z^2\frac{J_2}{3(1-J_1)}-z^4\frac{J_3}{15(1-J_1)}+\cdots)^{-1}\\
					=&\frac{z^2+2J_0}{(1-J_1)z}(1+z^2s_1+z^4s_2+\cdots),
				\end{aligned}
			\end{equation}
			where $s_n$ is a polynomial in $J_2,J_3,\cdots,J_{n+1}$ and $(1-J_1)^{-1}$.
			From the topological recursion, one can then deduce that the coefficients of $\frac{\d z}{z^{2m+2}}$ in $W^{gB}_{g,n}$ are polynomials in $J_0,(1-J_1)^{-1},J_2,\cdots$. Moreover, $W_{g,n}^{gB}$ contains only finitely many terms of the form $\frac{\d z}{z^{2m+2}}$, namely those with $m\leq 3g-3+n$. We will exploit this finiteness of $W_{g,1}$ to recover $F^{gB}_{g}$.
			
			From the definition of multidifferentials (\ref{def of Wz}), we have
			\begin{equation}
				\begin{aligned}
					W^{gB}_{g,1}(z)=&\sum_{a=0}^{\infty} \frac{\partial}{\partial t_a} F^{gB}_{g}\frac{(2a+1)!!\d \tilde{z}}{\tilde{z}^{2a+2}}\\
					=&\sum_{a=0}^{\infty}\sum_{b=0}^{\infty}\frac{\partial J_b}{\partial t_a}\frac{\partial }{\partial J_b} F^{gB}_g \frac{(2a+1)!!\d \tilde{z}}{\tilde{z}^{2a+2}}.
				\end{aligned}
			\end{equation}
			Using the Jacobian (\ref{jacobian for J0}) and (\ref{Jac for Jn}), then
			\begin{equation}
				\begin{aligned}
					\sum_{a=0}^{\infty}\frac{\partial J_0}{\partial t_a}\frac{(2a+1)!!\d \tilde{z}}{\tilde{z}^{2a+2}}=&
					\sum_{a=0}^{\infty} (1-J_1)^{-1}\frac{2}{a!}J^{a+1}_0 \frac{(2a+1)!!\d \tilde{z}}{\tilde{z}^{2a+2}}
					\\
					=& (1-J_1)^{-1}2J_0 (\tilde{z}^2-2J_0)^{-\frac{3}{2}}\tilde{z}\d \tilde{z}\\
					=& \frac{2J_0}{1-J_1} \frac{\d z}{z^2},
				\end{aligned}
			\end{equation}
			and for $b\geq 1$,
			\begin{equation}
				\begin{aligned}
					&\sum_{a=0}^{\infty}\frac{\partial J_b}{\partial t_a}\frac{(2a+1)!!\d \tilde{z}}{\tilde{z}^{2a+2}}\\
					&=\sum_{a=0}^{\infty} (1-J_1)^{-1}\frac{2}{a!}J^{a+1}_0 J_{b+1}\frac{(2a+1)!!\d \tilde{z}}{\tilde{z}^{2a+2}}+ \sum_{a=b-1}^{\infty} \frac{2a+1}{(a+1-b)!}J_0^{a+1-b}\frac{(2a+1)!!\d \tilde{z}}{\tilde{z}^{2a+2}}\\
					&=\frac{2J_0J_{b+1}}{1-J_1}\frac{\d z}{z^2}+(2b-1)\frac{(2b-1)!!\d z}{z^{2b}}+2J_0\frac{(2b+1)!!\d z}{z^{2b+2}}.
				\end{aligned}
			\end{equation}
			Consequently, the \textbf{shifted loop operator} (with the differential $\d (\frac{1}{2}z^2)$ for convenience) takes the form
			\begin{equation}\label{loop operator of gB}
				\begin{aligned}
					\nabla^{gB}(z)\frac{\d (z^2)}{2}:=&\sum_{a=0}^{\infty} \frac{(2a+1)!!z\d z}{(z^2+2J_0)^{a+\frac{3}{2}}}\frac{\partial}{\partial t_a}=\sum_{a=0}^{\infty}\frac{(2a+1)!!\d \tilde{z}}{\tilde{z}^{2a+2}}\sum_{b=0}^{\infty} \frac{\partial J_b}{\partial t_a}\frac{\partial }{\partial J_b}\\
					=&\frac{2J_0}{1-J_1}\frac{\d z}{z^2}\sum_{b=0}^{\infty}J_{b+1}\frac{\partial}{\partial J_b}
					+\sum_{b=0}^{\infty} \frac{(2b+1)!!\d z}{z^{2b+2}}2J_0\frac{\partial}{\partial J_b}\\
					&+\sum_{b=1}^{\infty} (2b-1)\frac{(2b-1)!!\d z}{z^{2b}}\frac{\partial}{\partial J_b}.
				\end{aligned}
			\end{equation}
			Thus the differential $W^{gB}_{g,1}(z)$ takes the following form for $g\geq1$:
			\begin{equation}\label{form of Wg1}
				\begin{aligned}
					W_{g,1}^{gB}(z)=&\big( \frac{\partial F^{gB}_g}{\partial J_0}\frac{2J_0}{1-J_1}+\sum_{b=1}^{\infty} \frac{\partial F_g^{gB}}{\partial J_b}\frac{2J_0J_{b+1}}{1-J_1} +\frac{\partial F^{gB}_g}{\partial J_1} \big)\frac{\d z}{z^2} \\
					+& \sum_{b=1}^{\infty}\big(\frac{\partial F_g^{gB}}{\partial J_b}2J_0+\frac{\partial F_g^{gB}}{\partial J_{b+1}}(2b+1)  \big)\frac{(2b+1)!!\d z}{z^{2b+2}}.
				\end{aligned}
			\end{equation}
			On the other hand, we may write
			\begin{equation}\label{Wgb f}
				W^{gB}_{g,1}(z)=\sum_{b=0}^{\infty} f_{g,b}\frac{(2b+1)!!\d z}{z^{2b+2}},
			\end{equation}
			where the coefficients $f_{g,b}$ are polynomials in $J_0,(1-J_1)^{-1},J_2,\cdots$ determined by the topological recursion. For fixed genus $g$, only finitely many of the $f_{g,b}$ are non-zero.
			
			Equating the coefficients in (\ref{form of Wg1}) and (\ref{Wgb f}) yields an infinite-dimensional linear system of PDEs for the derivatives $\frac{\partial F^{gB}_g}{\partial J_b}$:
			\begin{equation}\label{PDE}
				\left(
				\begin{array}{ccccc}
					\frac{2J_0}{1-J_1} & 1+\frac{2J_0J_2}{1-J_1} & \frac{2J_0J_3}{1-J_1} & \frac{2J_0J_4}{1-J_1}& \cdots \\
					0                              & 2J_0                                  &      3                          &0     & \cdots \\
					0                             &  0                                      &     2J_0                     & 5 & \cdots \\
				 \vdots  &	\vdots & \vdots &\vdots &\ddots
				\end{array}
				\right)
				\left(
				\begin{array}{c}
					\frac{\partial F_g}{\partial J_0} \\
					\frac{\partial F_g}{\partial J_1} \\
					\frac{\partial F_g}{\partial J_2} \\
					\vdots
				\end{array}
				\right)
				=
				\left(
				\begin{array}{c}
					f_{g,0}\\
					f_{g,1}\\
					f_{g,2}\\
					\vdots
				\end{array}
				\right)
			\end{equation}			
			As a system of infinite-dimensional linear equations, the solution is not unique. However, by the definition of $W^{gB}_{g,1}(z)$, any solution $\tilde{F}_{g}$ of these PDEs can differ from the true free energy $F_g^{gB}$ only by a function of $S$, where $S$ is regarded as a function of $\mathbf{J}$.
			
			On the other hand, any such solution of these PDEs is integrable. Indeed, writing the system as
			\begin{equation}
				A(\partial F_g)=(f_g).
			\end{equation}
			Let $A^{-1}$ denote a formal inverse matrix of $A$, and let $\partial \tilde{F}_g$ be a formal solution; that is
			\begin{equation}
				(\partial \tilde{F}_g)=A^{-1}(f_g).
			\end{equation}
			Then one has:
			\begin{equation}
				\begin{aligned}
					\frac{\partial}{\partial J_n} (\partial \tilde{F}_g)= &(\frac{\partial}{\partial J_n} A^{-1})(f_g)+A^{-1}\frac{\partial}{\partial J_n} (f_g) \\
					=& (\frac{\partial}{\partial J_n} A^{-1})A(\partial F_g)+A^{-1}(\frac{\partial}{\partial J_n} A)(\partial F_g)+\frac{\partial}{\partial J_n}  (\partial F_g)\\
					=&\frac{\partial}{\partial J_n} (\partial F_g).
				\end{aligned}
			\end{equation}
			Thus, $(\partial \tilde{F}_g)$ is (formally) integrable because $\partial F_g$ is.
			
			For the infinite-dimensional matrix $A$ in the equation (\ref{PDE}), one can take the formal upper-triangular inverse matrix:
			\begin{equation}
				A^{-1}=\left(
				\begin{array}{cccc}
					\frac{1-J_1}{2J_0} & -\frac{1-J_1}{(2J_0)^2}-\frac{J_2}{2J_0} & \frac{3(1-J_1)}{(2J_0)^3}+\frac{3J_2}{(2J_0)^2}-\frac{J_3}{2J_0} & \cdots \\
					0 & \frac{1}{2J_0} & \frac{-3}{2J_0} & \cdots \\
					0 & 0 & \frac{1}{2J_0} &\cdots \\
					\vdots & \vdots & \vdots &\ddots
				\end{array}
				\right).
			\end{equation} 
			Since only finitely many $f_{g,n}$ are non-zero for fixed $g$, the formal solution $(\partial \tilde{F}_g)=A^{-1}(f_g)$ has only finitely many non-zero terms. Thus, for sufficiently large $n$, we have $\partial \tilde{F}_g/\partial J_n=0$. As a result, $\tilde{F}_g$ is a function of only finitely many of the $J_n$. The true free energy $F_g$ differs from $\tilde{F}_g$ by a function of $S$ alone, and this remaining dependence can be fixed using the weighted homogeneity. More precisely, when we restrict to the small phase space, where $t_n=0$ for all $n=0,1,\cdots$, the higher-genus free energies must vanish. In this limit, we have $J_n=0$ for $n\geq 1$ and $J_0=\frac{S^2}{2}$. Weighted homogeneity then yields 
			\begin{equation}\label{F ti F}
				F^{gB}_g(\mathbf{J})=\tilde{F}_{g}(\mathbf{J})-\tilde{F}_{g}(J_0=\frac{S^2}{2},J_1=0,J_2=0,\cdots).
			\end{equation}
			
			In summary, we have proved:
			\begin{thm}\label{finiteness}
				The free energy $F^{gB}_{g}$ of the generalized BGW model can be expressed in terms of $S^2$ and finitely many of the renormalized coordinates $J_0,J_1,J_2,\cdots$ for $g\geq 1$.
			\end{thm}
			
			In Section \ref{section relation}, we will show that $F^{gB}_g$ is a polynomial in $J_0^{\pm 1},(1-J_1)^{\pm 1},J_2,J_3,\cdots,J_{3g-2}$ with a correction term proportional to $S^{-(2g-2)}$ when $g\geq2$.
			
			Let us verify this for $g=1$. From the expression for $W^{gB}_{1,1}(z)$ in (\ref{W11}) and the form (\ref{form of Wg1}), one immediately finds an obvious solution to the linear system:
			\begin{equation}
				\begin{aligned}
					\frac{\partial}{\partial J_0} \tilde{F}_1 = \frac{1}{24}\frac{1}{J_0}, \quad
					\frac{\partial}{\partial J_1} \tilde{F}_1 = \frac{1}{24}\frac{1}{1-J_1},\quad
					\frac{\partial}{\partial J_n} \tilde{F}_1 =0 \quad (n\geq 2).  
				\end{aligned}
			\end{equation}
			Up to an additive constant, the solution is
			\begin{equation}\label{gB g1}
				\tilde{F}_1=\frac{1}{24}\log \frac{J_0}{1-J_1}.
			\end{equation}
			Applying the homogeneity correction (\ref{F ti F}), we obtain the free energy
			\begin{equation}
				F^{gB}_1=\frac{1}{24}\log \frac{2J_0}{S^2(1-J_1)}.
			\end{equation}
			Noting that $J_0=\frac{\partial^2 F^{gB}_0}{\partial t_0^2}$, we can write
			\begin{equation}
				F^{gB}_1=\frac{1}{24}\log \frac{\partial^3 F_0^{gB}}{\partial t_0^3}-\frac{1}{24}\log S^2.
			\end{equation}
			This is analogous to the genus-one free energy of the Witten-Kontsevich model \cite{dijkgraafMeanFieldTheory1990}:
			\begin{equation}
				F^{WK}_1=\frac{1}{24}\log \frac{\partial^3 F^{WK}_{0}}{\partial t_0^3}.
			\end{equation}
			It is the genus-one part of Okuyama-Sakai Conjecture \cite{okuyamaJTSupergravityBrezinGrossWitten2020} and has been proved by Yang and Zhang \cite{yangHodgeBGWCorrespondence2024,yangNewProofOkuyama2023}. See Subsection \ref{subsection Okuyama-Sakai}.
			
			Another point needed to note is $\frac{\partial F^{gB}_1}{\partial J_n}\ne 0$ for $n\geq 2$. 
			Indeed, with the equation (\ref{S der with Jn}), the derivative of $F^{gB}_{1}$ with respect to $J_{n+1}$ for $n\geq1$ is
			\begin{equation}
				\begin{aligned}
					\frac{\partial}{\partial J_{n+1}} F^{gB}_{1}=&-\frac{1}{12S}\frac{\partial S}{\partial J_{n+1}}  \\
					&=\frac{1}{12}\frac{(-2J_0)^{n}}{(2n+1)!!}(1-\sum_{n=0}^{\infty}J_{n+1}\frac{(-2J_0)^n}{(2n+1)!!})^{-1}.
				\end{aligned}
			\end{equation}
			It also illustrates that $F^{gB}_1$ cannot be expressed in terms of finitely many of the $J_0,J_1,J_2,\cdots$ without $S$.

			For higher genera, we will present an alternative method to determine these expressions based on the transition between the Witten-Kontsevich model and the generalized BGW theory. Two initial examples are
				\begin{equation}\label{FgB2}
				\begin{aligned}
					&F^{gB}_2
					=\frac{1}{1152}\frac{(2J_0)^2}{(1-J_1)^3}J_4+\frac{13}{1440}\frac{2J_0}{(1-J_1)^3}J_3\\
					&+\frac{29}{5760}\frac{(2J_0)^2}{(1-J_1)^4}J_2J_3
					+\frac{7}{1440}\frac{(2J_0)^2}{(1-J_1)^5}J_2^3
					+\frac{107}{5760}\frac{2J_0}{(1-J_1)^4}J_2^2\\
					&+\frac{61}{5760}\frac{1}{(1-J_1)^3}J_2-\frac{1}{240}\frac{1}{2J_0(1-J_1)^2} +\frac{1}{240}S^{-2},
				\end{aligned}
			\end{equation}
			and

			\begin{equation}\label{FgB3}
				\begin{aligned}
					&F^{gB}_3=-\frac{\tilde{J}_1^{4}}{1008 (2J_0)^2} + \frac{\tilde{J}_1^{5} J_2}{756 (2J_0)} + \frac{3329 \tilde{J}_1^{6} J_2^2}{580608} + \frac{48253 (2J_0) \tilde{J}_1^{7} J_2^3}{580608} \\
					&+ \frac{215 (2J_0)^2 \tilde{J}_1^{8} J_2^4}{1536} + \frac{505 (2J_0)^3 \tilde{J}_1^{9} J_2^5}{6912} + \frac{245 (2J_0)^{4} \tilde{J}_1^{10} J_2^6}{20736} + \frac{2579 \tilde{J}_1^{5} J_3}{967680} \\
					&+ \frac{40009 (2J_0) \tilde{J}_1^{6} J_2 J_3}{580608} + \frac{176839 (2J_0)^2 \tilde{J}_1^{7} J_2^2 J_3}{967680} + \frac{115}{864} (2J_0)^3 \tilde{J}_1^{8} J_2^3 J_3 \\
					&+ \frac{193 (2J_0)^{4} \tilde{J}_1^{9} J_2^4 J_3}{6912} + \frac{8363 (2J_0)^2 \tilde{J}_1^{6} J_3^2}{362880} + \frac{13379 (2J_0)^3 \tilde{J}_1^{7} J_2 J_3^2}{322560} \\
					&+ \frac{205 (2J_0)^{4} \tilde{J}_1^{8} J_2^2 J_3^2}{13824} + \frac{583 (2J_0)^{4} \tilde{J}_1^{7} J_3^3}{580608} + \frac{4499 (2J_0) \tilde{J}_1^{5} J_4}{483840} \\
					&+ \frac{21449 (2J_0)^2 \tilde{J}_1^{6} J_2 J_4}{580608} + \frac{7877 (2J_0)^3 \tilde{J}_1^{7} J_2^2 J_4}{241920} + \frac{53 (2J_0)^{4} \tilde{J}_1^{8} J_2^3 J_4}{6912} \\
					&+ \frac{5701 (2J_0)^3 \tilde{J}_1^{6} J_3 J_4}{725760} + \frac{1121 (2J_0)^{4} \tilde{J}_1^{7} J_2 J_3 J_4}{241920} + \frac{607 (2J_0)^{4} \tilde{J}_1^{6} J_4^2}{2903040} \\
					&+ \frac{2671 (2J_0)^2 \tilde{J}_1^{5} J_5}{725760} + \frac{983 (2J_0)^3 \tilde{J}_1^{6} J_2 J_5}{193536} + \frac{17 (2J_0)^{4} \tilde{J}_1^{7} J_2^2 J_5}{11520} \\
					&+ \frac{503 (2J_0)^{4} \tilde{J}_1^{6} J_3 J_5}{1451520} + \frac{83 (2J_0)^3 \tilde{J}_1^{5} J_6}{207360} + \frac{77 (2J_0)^{4} \tilde{J}_1^{6} J_2 J_6}{414720} + \frac{(2J_0)^{4} \tilde{J}_1^{5} J_7}{82944}\\
					&+\frac{1}{1008}S^{-4},
				\end{aligned}
			\end{equation}
			where $\tilde{J}_1=(1-J_1)^{-1}$. 
			We will explain how these expressions are obtained in the next section.
			
			\section{Transition between Witten-Kontsevich model and generalized BGW theory}\label{section relation}
			In this section, we establish the connection between the Witten-Kontsevich model and the generalized BGW theory, which together shed considerable light on both theories.
			
			\subsection{Transition between two theories} 
			As noted earlier, the multidifferentials defined in this article generalize the usual Eynard-Orantin multidifferentials. The case $\mathbf{t}\ne 0$ can be interpreted as a deformation, or a transition, of the original theory, and it is this deformation that establishes a direct link between the Witten-Kontsevich model and the generalized BGW theory.
			
			We recall the following theorem of Zhou:
			\begin{thm}[Zhou \cite{zhouEmergentGeometryUnifications2025}]
				For the Witten-Kontsevich partition function 
				$$Z^{WK}(\mathbf{t})=\exp \sum_{g=0}^{\infty} F^{WK}_g(\mathbf{t}),$$
				define the multidifferentials $W^{WK}_{g,n}(z_{[n]})$ by
				\begin{equation}
					\begin{aligned}
						W^{WK}_{g,n}(z_1,z_2,\cdots,z_n;\mathbf{t})
						:=\sum_{i_1,i_2,\cdots,i_n\geq 0}\frac{\partial^n F^{WK}_{g}(\mathbf{t})}{\partial t_{i_1}\partial t_{i_2}\cdots\partial t_{i_n}}\prod_{k=1}^{n}\frac{(2i_k+1)!!z_k\d z_k}{(z_k^2+2I_0)^{i_k+\frac{3}{2}}}.
					\end{aligned}
				\end{equation}
				Then they satisfy the Eynard-Orantin topological recursion for the spectral curve:
				\begin{equation}
					x(z)=\frac{1}{2}z^2+I_0, \qquad  y^{WK}(z)=z\big(1-\sum_{n=0}^{\infty} \frac{z^{2n}}{(2n+1)!!}I_{n+1}\big),
				\end{equation}
				with the Bergman kernel (or equivalent Bergman kernel (\ref{equivalent Bergman kernel})):
				\begin{equation}
					B_0(z_1,z_2)=\frac{\d z_1\d z_2}{(z_1-z_2)^2}.
				\end{equation}								
			\end{thm}
			
			With Theorem \ref{EO TR foe gB}, we observe that, if the two spectral curves are identified through a suitable relation between the coordinates $\mathbf{I}$ and $\mathbf{J}$, then the corresponding multidifferentials will coincide as well. More precisely, for $J_0\ne 0$, 
			\begin{equation}
				\begin{aligned}
					&y^{gB}(z)=\frac{z}{z^2+2J_0}\big(1-\sum_{n=0}^{\infty} \frac{z^{2n}}{(2n+1)!!}J_{n+1}\big)\\
					&=z(\sum_{m=0}^{\infty}\frac{(-1)^m}{(2J_0)^{m+1}}z^{2m})\big(1-\sum_{n=0}^{\infty} \frac{z^{2n}}{(2n+1)!!}J_{n+1}\big)\\
					&=z\Big(1-\sum_{k=0}^{\infty}\big(\sum_{n+m=k}\frac{J_{n+1}}{(2J_0)^{m+1}}\frac{(-1)^m}{(2n+1)!!}-\frac{(-1)^k}{(2J_0)^{k+1}}+\delta_{k,0}\big)z^{2k}\Big).
				\end{aligned}
			\end{equation}
			Hence, if we let 
			\begin{equation}\label{tran from J to I}
				\begin{aligned}
					I_0&=J_0;\\
					I_1&=\frac{J_1-1}{2J_0}+1;\\
				\frac{I_n}{(2n-1)!!}&=(-2J_0)^{-n}\big(1-\sum_{k=1}^{n}\frac{J_k(-2J_0)^{k-1}}{(2k-1)!!} \big),
				\end{aligned}
			\end{equation}
			then the multidifferentials of the two theories are equal:
			\begin{equation}
				W^{WK}_{g,n}(z_{[n]};\mathbf{I}(\mathbf{J}))=W^{gB}_{g,n}(z_{[n]};\mathbf{J}).
			\end{equation}
			The inverse transformation of $\mathbf{I}$ and $\mathbf{J}$ is obtained from
			\begin{equation}
				\begin{aligned}
					-\frac{(-2I_0)^{n+1}I_n}{(2n-1)!!}-2I_0+\delta_{n,1}(2I_0)^2=\sum_{k=1}^{\infty} \frac{J_k}{(2k-1)!!}(-2I_0)^k,	
				\end{aligned}
			\end{equation}
			which yields
			\begin{equation}\label{tran from I to J}
				\begin{aligned}
					&J_0=I_0;\\
					&J_1=2I_0I_1-2I_0+1;\\
					&J_2=2I_0I_2+3I_1-3;\\
					&J_n=2I_0I_n+(2n-1)I_{n-1} \quad (n\geq 3).
				\end{aligned}
			\end{equation}
	
			\begin{remark}
				Notice that the transformations (\ref{tran from J to I}) and (\ref{tran from I to J}) are inhomogeneous. The natural degrees of the $\mathbf{I}$-coordinates in the Witten-Kontsevich model are
				\begin{equation}\label{deg of WK}
					\deg I_n=n-1 \qquad \deg F^{WK}_{g}=3g-3.
				\end{equation}
				This is a characteristic feature of a transition: the degrees of the parameters before and after the transition are inhomogeneous.
			\end{remark}
			
			The transformations of vector fields between two sets of renormalized coordinates are
			\begin{equation}\label{tranform of vector field from I to J}
				\begin{aligned}
					&\frac{\partial}{\partial I_0}=\frac{\partial}{\partial J_0}+2(I_1-1)\frac{\partial }{\partial J_1}+\sum_{n=2}^{\infty} 2I_n\frac{\partial}{\partial J_n},\\
					&\frac{\partial}{\partial I_n}=2I_0\frac{\partial}{\partial J_n}+(2n+1)\frac{\partial}{\partial J_{n+1}}\quad \text{for }n\geq1,
				\end{aligned}
			\end{equation}
			and 
			\begin{equation}
				\begin{aligned}
					&\frac{\partial}{\partial J_n}=-\sum_{m=n}^{\infty} \frac{(2m-1)!!}{(2n-1)!!}(-2I_0)^{n-m-1}\frac{\partial}{\partial I_m} \quad \text{for }n\geq1,\\
					&\frac{\partial }{\partial J_0}=\frac{\partial}{\partial I_0}+2\sum_{m=1}^{\infty}\big(\sum_{n=1}^{m}\frac{(2m-1)!!}{(2n-1)!!}(-2I_0)^{n-m-1}(I_n-\delta_{n,1})\big)\frac{\partial}{\partial I_m}.
				\end{aligned}
			\end{equation}
			
			Remarkably, the shifted loop operators of the Witten-Kontsevich model coincide with those of the generalized BGW theory when expressed in the renormalized coordinates. First, we recall the following result from \cite{zhouPhaseTransitionTwoDimensional2024}: 
			\begin{lemma}[Zhou \cite{zhouPhaseTransitionTwoDimensional2024}]
				The shifted loop operator of the Witten-Kontsevich model has the following expression in the renormalized coordinate:
				\begin{equation}\label{loop operator of WK}
					\begin{aligned}
						\nabla^{WK}(z)\frac{\d (z^2)}{2}:=&\sum_{n=0}^{\infty} \frac{(2n+1)!!z\d z}{(z^2+2I_0)^{n+\frac{3}{2}}}\frac{\partial}{\partial t^{WK}_n}\\
						=&\frac{1}{1-I_1}\frac{\d z}{z^2}\frac{\partial}{\partial I_0}+\sum_{n=1}^{\infty} \frac{I_{n+1}}{1-I_1}\frac{\d z}{z^2}\frac{\partial}{\partial I_n}+\sum_{n=1}^{\infty} \frac{(2n+1)!!\d z}{z^{2n+2}}\frac{\partial }{\partial I_n}.
					\end{aligned}
				\end{equation}
			\end{lemma}
			We then have
			\begin{proposition}
				The shifted loop operator of the Witten-Kontsevich model $\nabla^{WK}(z)$ defined in (\ref{loop operator of WK}) and the shifted loop operator of the generalized BGW theory $\nabla^{gB}(z)$ defined in (\ref{loop operator of gB}) coincide under the transformation of the renormalized coordinates; that is,
				\begin{equation}
					\begin{aligned}
						\nabla^{WK}(z)|_{\mathbf{I}=\mathbf{I}(\mathbf{J})}=\nabla^{gB}(z),
					\end{aligned}
				\end{equation}
			\end{proposition}
			\begin{proof}
				Substituting the transformation of the vector field (\ref{tranform of vector field from I to J}) into the terms of $\nabla^{WK}(z)$, we obtain
				\begin{equation}
					\frac{1}{1-I_1}\frac{\partial}{\partial I_0}=\frac{2J_0}{1-J_1}\frac{\partial}{\partial J_0}-2\frac{\partial}{\partial J_1}+\sum_{n=2}^{\infty}\frac{2I_n}{1-I_1}\frac{\partial}{\partial J_n},
				\end{equation} 
				and
				\begin{equation}
					\sum_{n=1}^{\infty} \frac{I_{n+1}}{1-I_1}\frac{\partial}{\partial I_n}=\sum_{n=1}^{\infty} \frac{2I_0I_{n+1}}{1-I_1}\frac{\partial}{\partial J_n} +\sum_{n=2}^{\infty} \frac{(2n-1)I_{n}}{1-I_1}\frac{\partial}{\partial J_{n}}.
				\end{equation}
				It follows that
				\begin{equation}
					\begin{aligned}
						&\frac{1}{1-I_1}\frac{\partial}{\partial I_0}+\sum_{n=1}^{\infty} \frac{I_{n+1}}{1-I_1}\frac{\partial}{\partial I_n}\\
						=&\frac{2J_0}{1-J_1}\frac{\partial}{\partial J_0}+(\frac{2I_0I_2}{1-I_1}-2)\frac{\partial}{\partial J_1}+\sum_{n=2}^{\infty}\frac{(2n+1)I_n+2I_0I_{n+1}}{1-I_1}\frac{\partial}{\partial J_n}\\
						=&\frac{2J_0}{1-J_1}\frac{\partial}{\partial J_0}+\frac{\partial}{\partial J_1}+\sum_{n=1}^{\infty} \frac{2J_0J_{n+1}}{1-J_1}\frac{\partial}{\partial J_n}.
					\end{aligned}
				\end{equation}
				Therefore
				\begin{equation}
					\begin{aligned}
						\nabla^{WK}(z)\frac{\d (z^2)}{2}=&\frac{\d z}{z^2}\big(\frac{1}{1-I_1}\frac{\partial}{\partial I_0}+\sum_{n=1}^{\infty} \frac{I_{n+1}}{1-I_1}\frac{\partial}{\partial I_n}\big)+\sum_{n=1}^{\infty} \frac{(2n+1)!!\d z}{z^{2n+2}}\frac{\partial}{\partial I_n}\\
						=&\frac{\d z}{z^2}\big( \frac{2J_0}{1-J_1}\frac{\partial}{\partial J_0}+\frac{\partial}{\partial J_1}+\sum_{n=1}^{\infty} \frac{2J_0J_{n+1}}{1-J_1}\frac{\partial}{\partial J_n}\big)\\
						&+\sum_{n=1}^{\infty}\frac{(2n+1)!!\d z}{z^{2n+2}}\big(2J_0\frac{\partial}{\partial J_n}+(2n+1)\frac{\partial}{\partial J_{n+1}}\big)\\
						=&\nabla^{gB}(z)\frac{\d (z^2)}{2}.
					\end{aligned}
				\end{equation}
			\end{proof}

			The equality of the multidifferentials implies a relation between $F^{WK}_{g}$ and $F^{gB}_{g}$ for genus $g\geq1$, namely
			\begin{equation}
				\nabla^{WK}(z)|_{\mathbf{I}=\mathbf{I}(\mathbf{J})}F^{WK}_{g}(\mathbf{I}(\mathbf{J}))=\nabla^{gB}(z)F^{gB}_{g}(\mathbf{J}).
			\end{equation}
			As discussed in the previous section, after substituting $\mathbf{I}=\mathbf{I}(\mathbf{J})$  according to (\ref{tran from J to I}), the function $F^{WK}_{g}(\mathbf{I}(\mathbf{J}))$ differs from $F^{gB}_{g}(\mathbf{J})$ only by a function of $S^2$. Using the weighted homogeneity (\ref{F ti F}), we obtain
			\begin{equation}\label{WK to gB}
				F^{gB}_{g}(\mathbf{J})=F^{WK}_{g}(\mathbf{I}(\mathbf{J}))-F^{WK}_{g}(\mathbf{I}(S^2)),
			\end{equation}
			where 
			\begin{equation}\label{I(S)}
				\mathbf{I}(S^2)=\mathbf{I}(J_0=\frac{S^2}{2}, J_1=0, J_2=0, \cdots),
			\end{equation}
			i.e.
			\begin{equation}
				\begin{aligned}
					I_0=\frac{S^2}{2},\quad
					I_1=1-S^{-2},\quad 
					I_n=(2n-1)!! (-1)^n S^{-2n}
				\end{aligned}
			\end{equation}
			for $n\geq 2$.

			For example, when $g=1$, by the formula (\ref{WK g1}), 
			\begin{equation}
				\begin{aligned}
					&F^{WK}_1(\mathbf{I}(\mathbf{J}))=\frac{1}{24}\log\frac{1}{1-I_1}=\frac{1}{24}\log \frac{2J_0}{1-J_1},\\
					&F^{WK}_1(\mathbf{I}(S^2))=\frac{1}{24}\log S^2. 
				\end{aligned}
			\end{equation}
			This coincides with the equation (\ref{gB g1}) and yields
			\begin{equation*}
				F^{gB}_1(\mathbf{J})=\frac{1}{24}\log\frac{2J_0}{S^2}\frac{1}{1-J_1}.
			\end{equation*}
			
			For $g\geq 2$, we use the variables $\tilde{I}_j$ defined in (\ref{tilde I}) for $j\geq2$:
			\begin{equation}\label{tilde I in J}
				\begin{aligned}
					\tilde{I}_{j}(\mathbf{J})&=(1-I_1)^{-\frac{2j+1}{3}}I_j\\
					&=(\frac{2J_0}{1-J_1})^{\frac{2j+1}{3}}(2j-1)!!(-2J_0)^{-j}\big(1-\sum_{k=1}^{j}\frac{J_{k}(-2J_0)^{k-1}}{(2k-1)!!}\big),
				\end{aligned}
			\end{equation}
				and 
			\begin{equation}\label{tilde I in S}
				\begin{aligned}
					\tilde{I}_j(S^2)=(2j-1)!!(-1)^j S^{\frac{-2j+2}{3}}.
				\end{aligned}
			\end{equation}
			 
			 Recalling the formula (\ref{WK Fg >1}), we arrive at:
			\begin{thm}\label{expression of F}
				For genus $g\geq 2$, the free energy of the generalized BGW theory has the following expression in the renormalized coordinates $\mathbf{J}=\{J_0,J_1,\cdots\}$:
				\begin{equation}\label{formula of FgB in WK}
					\begin{aligned}
						F^{gB}_g(\mathbf{J})=&F^{WK}_g(\mathbf{I}(\mathbf{J}))-F^{WK}_g(\mathbf{I}(S^2))\\
						=&\sum_{\sum_{j=2}^{3g-2} (j-1)l_j=3g-3} \langle \tau_2^{l_2}\tau_{3}^{l_3}\cdots \tau_{3g-2}^{l_{3g-2}}\rangle_g \big(\prod_{j=2}^{3g-2}\frac{1}{l_j!}\tilde{I}_j^{l_j}(\mathbf{J})-\prod_{j=2}^{3g-2}\frac{1}{l_j!}\tilde{I}_j^{l_j}(S^2)\big),
					\end{aligned}
				\end{equation}
				where $\langle\tau_2^{l_2}\tau_{3}^{l_3}\cdots \tau_{3g-2}^{l_{3g-2}}\rangle_g$ are the intersection numbers of $\psi$-classes on the moduli space $\overline{M}_{g,n}$, and $S^2$ is the function $S^2(\mathbf{J})$ defined in equation (\ref{S der with Jn}) as
				 \begin{equation*}
				 	S^2=2J_0\big(1-\sum_{n=0}^{\infty}J_{n+1}\frac{(-2J_0)^n}{(2n+1)!!}\big)^2.
				 \end{equation*}
			\end{thm}

			The correction term has the expression:
			\begin{equation}
				F^{WK}_{g}(\mathbf{I}(S^2))=\sum_{\sum_{j=2}^{3g-2} (j-1)l_j=3g-3}\langle\prod_{j=2}^{3g-2}\frac{1}{l_j!}((-1)^{j}(2j-1)!!\tau_{j})^{l_j}\rangle_{g}S^{-(2g-2)}.
			\end{equation}
			With a conjecture of Kazarian and Norbury \cite{kazarianPolynomialRelationsKappa2024} proved by Yang and Zhang \cite{yangHodgeBGWCorrespondence2024}, this correction term takes the form
			\begin{equation}
				F^{WK}_{g}(\mathbf{I}(S^2))=\frac{(-1)^{g}B_{2g}}{2g(2g-2)}S^{-(2g-2)},
			\end{equation}
			where $B_{2g}$ denotes the $2g$-th Bernoulli number. We discuss further details in Subsection \ref{subsection Okuyama-Sakai}.
			
			For illustration, we list a few of the transformed variables:
			\begin{equation}
				\tilde{I}_2=(1-J_1)^{-\frac{5}{3}}(2J_0)^{\frac{2}{3}} J_2+3(1-J_1)^{-\frac{2}{3}}(2J_0)^{-\frac{1}{3}};
			\end{equation}
			\begin{equation}
				\begin{aligned}
					\tilde{I}_3=(1-J_1)^{-\frac{7}{3}}(2J_0)^{\frac{4}{3}}J_3-5(1-J_1)^{-\frac{7}{3}}(2J_0)^{\frac{1}{3}}J_2-15(1-J_1)^{-\frac{4}{3}}(2J_0)^{-\frac{2}{3}};
				\end{aligned}
			\end{equation}
			\begin{equation}
				\begin{aligned}
					\tilde{I}_4=&(1-J_1)^{-3}(2J_0)^{2}J_4-7(1-J_1)^{-3}(2J_0)J_3\\
					&+35(1-J_1)^{-3}J_2+105(1-J_1)^{-3}(2J_0)^{-1}.
				\end{aligned}
			\end{equation}
			Then 
			\begin{equation}
				\begin{aligned}
					&F^{WK}_2(\mathbf{I}(\mathbf{J}))=\frac{7}{1440}\tilde{I}^3_2+\frac{29}{5760}\tilde{I}_2\tilde{I}_3+\frac{1}{1152}\tilde{I}_4\\
					&=\frac{1}{1152}\frac{(2J_0)^2}{(1-J_1)^3}J_4+\frac{13}{1440}\frac{2J_0}{(1-J_1)^3}J_3+\frac{29}{5760}\frac{(2J_0)^2}{(1-J_1)^4}J_2J_3+\frac{7}{1440}\frac{(2J_0)^2}{(1-J_1)^5}J_2^3\\
					&+\frac{107}{5760}\frac{2J_0}{(1-J_1)^4}J_2^2+\frac{61}{5760}\frac{1}{(1-J_1)^3}J_2-\frac{1}{240}\frac{1}{2J_0(1-J_1)^2}.
				\end{aligned}
			\end{equation}
			One can check that $F^{WK}_{2}(\mathbf{I}(\mathbf{J}))$ matches the differential $W^{gB}_{2,1}(z)$ in (\ref{W2,1}) via formula (\ref{form of Wg1}),
			and 
			\begin{equation}
				\begin{aligned}
					F^{gB}_2(\mathbf{J})=&F^{WK}_2(\mathbf{I}(\mathbf{J}))-F^{WK}_2(\mathbf{I}(S^2))\\
					=&F^{WK}_2(\mathbf{I}(\mathbf{J}))+\frac{1}{240}S^{-2},
				\end{aligned}
			\end{equation}
			which reproduces (\ref{FgB2}). The expression for $F^{gB}_3$ given in (\ref{FgB3}) is obtained similarly with Mathematica.
			
			Recall that the degrees of variables in the generalized BGW theory are
			\begin{equation*}
				\deg J_n=2n-2,\quad \deg S=-1.
			\end{equation*}
			The transformations (\ref{tilde I in J}) and (\ref{tilde I in S}) are weighted homogeneous and proportional to the degrees of these variables in the Witten-Kontsevich model:
			\begin{equation}
				\deg^{gB}\tilde{I}_j=\frac{2j-2}{3}=\frac{2}{3}\deg^{WK}\tilde{I}_j .
			\end{equation}
			This corresponds to 
			\begin{equation}
				\deg^{gB}F^{gB}_g=2g-2=\frac{2}{3}\deg^{WK} F^{WK}_g.
			\end{equation}
			
			
			Formula (\ref{WK to gB}) does not hold in genus zero. Indeed, we only have the equation:
				\begin{equation}
					\nabla^{WK}(z_1)\nabla^{WK}(z_2)\nabla^{WK}(z_3)F^{WK}_0(\mathbf{I}(\mathbf{J}))=\nabla^{gB}(z_1)\nabla^{gB}(z_2)\nabla^{gB}(z_3)F^{gB}_0(\mathbf{J}).
				\end{equation}
				This implies that the genus-zero two-point functions of the two theories differ by a function of $S^2$, which is consistent with the constitutive relation.
				
				 One may also substitute the transformation (\ref{J to t}) directly into formula (\ref{F0}), yielding
				\begin{equation}
					\begin{aligned}
						&F^{gB}_0=\frac{1}{4}S^2\log\frac{2J_0}{S^2}+\frac{3}{4}S^2-\frac{3}{2}J_0+\sum_{n=0}^{\infty}t_n\frac{1}{(n+1)!}\frac{4n^2+6n+3}{2n+1}J_0^{n+1}\\
						&-\sum_{n,m=0}^{\infty} t_nt_m\frac{4m+2n+3}{2(2n+1)(n+m+1)n!m!}J_0^{n+m+1},
					\end{aligned}
				\end{equation}
				with 
				\begin{equation}
					\begin{aligned}
						&\sum_{n=0}^{\infty}t_n\frac{1}{(n+1)!}\frac{4n^2+6n+3}{2n+1}J_0^{n+1}\\
						=&\sum_{n=0}^{\infty} \big( \sum_{m=0}^{\infty} \frac{(-1)^{n-m}}{(m+1)!(n-m)!}\frac{4m^2+6m+3}{(2m+1)^2}\big)J_{n+1}J_0^{n+1}\\
						=&\sum_{n=0}^{\infty}(-1)^n\big(\frac{1}{(n+1)!}+\frac{2^{n+1}}{(2n+1)!!}(\sum_{m=0}^{n}\frac{1}{2m+1})\big)J_{n+1}J_0^{n+1}, 
					\end{aligned}
				\end{equation}
				and
				\begin{equation}
					\begin{aligned}
						&\sum_{n,m=0}^{\infty} t_nt_m\frac{4m+2n+3}{2(2n+1)(n+m+1)m!n!}J_0^{n+m+1}\\
						=&\sum_{n,m,k,l=0}^{\infty}\frac{(-1)^{k+l}(4m+2n+3)J_{n+k+1}J_{m+l+1}J_0^{n+m+k+l+1}}{2(2n+1)^2(2m+1)(n+m+1)!n!m!k!l!}\\
						=&-\sum_{n,m=0}^{\infty}\big(\frac{2^{n+m}}{(2n+1)!!(2m+1)!!}(\sum_{j=0}^{n}\frac{1}{2j+1})\\
						&+\sum_{j=0}^{n}\frac{(-1)^j}{2(2j+1)^2(n-j)!(m+j+1)!}\big)J_{n+1}J_{m+1}(-J_0)^{n+m+1}.
					\end{aligned}
				\end{equation}
				
				\subsection{Transformation of coordinates}
				In the previous subsection, we discussed the transition as a transformation between the renormalized coordinates of the two models. It is natural to examine the corresponding transformation between the original variables $\{t^{WK}_n\}_{n=0}^{\infty}$ and $\{S^2,t^{gB}_{n}\}_{n=0}^{\infty}$; that is, to consider the following diagram:
				
				\begin{center}
						\begin{tikzpicture}
						\node[draw, inner sep=10pt] at (0,0) {$\{t^{WK}\}_{n=0}^{\infty}$};
						\node [draw, inner sep=10pt] at (6,0) {$\{S^2, t^{gB}_n\}_{n=0}^{\infty}$};
						\draw[dashed, <->] (1.5,0) -- (4.5,0) node[midway, above]{$?$};
						\node[draw, inner sep=10pt] at (0,-2.5) {$\{I_n\}_{n=0}^{\infty}$};
						\node [draw, inner sep=10pt] at (6,-2.5) {$\{J_n\}_{n=0}^{\infty}$};
						\draw[<->] (0,-0.7)--(0,-1.7) node[midway, left]{$\alpha$};
						\draw[<->] (6,-0.7)--(6,-1.7) node[midway, right]{$\beta$};
						\draw[<->] (1.5,-2.5)--(4.5,-2.5)  node[midway, above]{$\gamma$};
					\end{tikzpicture}
				
				\end{center}
			
				For convenience, we summarize the transformations $\alpha, \beta, \gamma$ here:
				\begin{equation*}
					\alpha: \qquad t_n^{WK}=\sum_{m=0}^{\infty} I_{n+m}\frac{(-I_0)^m}{m!},\quad I_{n}=\sum_{m=0}^{\infty} t_{n+m}^{WK}\frac{I_0^{m}}{m!}. 
				\end{equation*}
				\begin{equation*}
				\begin{aligned}
						\beta:\qquad &t^{gB}_n= \frac{1}{2n+1}\sum_{m=0}^{\infty}J_{n+m+1}\frac{(-J_0)^{m}}{m!}, \ S^2=2J_0(1-\sum_{k=1}^{\infty} \frac{J_k(-2J_0)^{k-1}}{(2k-1)!!})^2\\
					& J_0=\sum_{m=0}^{\infty}(2m+1)t^{gB}_{m}\frac{J_0^{m+1}}{(m+1)!}+\frac{\partial F^{gB}_0}{\partial t_0}(\mathbf{t}^{gB};S^2),\\
					&J_n=\sum_{m=0}^{\infty} (2n+2m-1)t^{gB}_{n+m-1}\frac{J_0^{m}}{m!} \quad \text{for }n\geq 1.
				\end{aligned}
				\end{equation*}
				\begin{equation*}
					\begin{aligned}
						\gamma: &\quad J_0=I_0,\\
						& \ J_n=2I_0I_n+(2n-1)I_{n-1}+\delta_{n,1}(-3I_0+1)-3\delta_{n,2} \quad \text{for }n\geq 1,\\
						&I_n=(2n-1)!!(-2J_0)^{-n}(1-\sum_{k=1}^{n}\frac{J_k(-2J_0)^{k-1}}{(2k-1)!!})+\delta_{n,1} \quad \text{for }n\geq 1.
					\end{aligned}
				\end{equation*}
				Here $I_0, J_0$ are defined by the expression of order parameters when considering the transformations from $\{t_n^{WK}\}_{n=0}^{\infty}$ and $\{S^2, t^{gB}_n\}_{n=0}^{\infty}$ to them respectively.
				
				Each concrete transformation involves infinitely many terms. They are the transformations of formal power series in infinitely many indeterminates, meaning that one must specify the coefficients of each monomial in these indeterminates. Furthermore, the transformations $\alpha$ and $\beta$ have a finiteness property: 
				if only finitely many of the parameters are non-zero, the transformations reduce to finite-dimensional inversions of power series.
				For instance, if there exists some $N>0$ such that $t_m^{WK}=0$ for all $m>N$, then each variable $I_n(\mathbf{t}^{WK})=I_{n}(t^{WK}_0,t^{WK}_1,\cdots,t^{WK}_N)$ is a well-defined power series for $n\leq N$ and $I_m=0$ for $m>N$. The inverse transformation is also well-defined in this setting.
				
				On the other hand, however, the transformation $\gamma$ does not share this property. Starting from a finite set $\{J_n\}_{n=0}^{N}$, the resulting $\mathbf{I}_n$ are generally non-zero and grow like $O((2n-1)!!)$ for $n>N$, unless extra constraints are enforced. In such situations, the composition of the formal power series transformations may not be well-defined. Even when it is, the composition of the inverse transformations need not be invertible.
		
				In any case, one can formally compute the map from $\{t^{WK}_n\}_{n=0}^{\infty}$ to $\{S^2, t^{gB}_n\}_{n=0}^{\infty}$. A direct calculation yields
				\begin{equation}
					\begin{aligned}
						t_n^{gB}=&\frac{1}{2n+1}\sum_{m=0}^{\infty} J_{n+m+1}\frac{(-J_0)^{m}}{m!}\\
						=&\frac{1}{2n+1}\sum_{m=0}^{\infty}(2I_0I_{n+m+1}+(2n+2m+1)I_{n+m})\frac{(-I_0)^m}{m!}+\delta_{n,0}-\delta_{n,1}\\
						=&\frac{-2}{2n+1}\sum_{m=0}^{\infty}I_{n+m+1}\frac{(-I_0)^{m+1}}{m!}+\frac{2}{2n+1}\sum_{m=1}^{\infty} I_{n+m}\frac{(-I_0)^m}{(m-1)!}\\
						&+\sum_{m=0}^{\infty}I_{n+m}\frac{(-I_0)^m}{m!}+\delta_{n,0}-\delta_{n,1}\\
						=&t^{WK}_n+\delta_{n,0}-\delta_{n,1},
					\end{aligned}
				\end{equation}
				together with 
				\begin{equation}
					\begin{aligned}
						\frac{S}{\sqrt{2J_0}}=&1-\sum_{k=1}^{\infty}\frac{J_k(-2J_0)^{k-1}}{(2k-1)!!}\\
						=&1-\sum_{k=1}^{\infty} (2I_0I_k+(2k-1)I_{k-1})\frac{(-2I_0)^{k-1}}{(2k-1)!!}-(-3I_0+1)+(-2I_0)\\
						=&I_0+\sum_{k=1}^{\infty}I_k\frac{(-2I_0)^k}{(2k-1)!!}-\sum_{k=1}^{\infty}I_{k-1}\frac{(-2I_0)^{k-1}}{(2k-3)!!}\\
						=&0.
					\end{aligned}
				\end{equation}
				
				\begin{remark}
					Because the transformation from $\{t^{WK}_n\}_{n=0}^{\infty}$ to $\{S^2, t^{gB}_n\}_{n=0}^{\infty}$ implies $S^2=0$, 
					one cannot express the free energy $F^{gB}_g$ directly in terms of the Witten-Kontsevich coupling constants. More precisely, the expression of $F^{gB}_g$ in the renormalized coordinates $\{J_n\}$ contains a term proportional to $S^{-(2g-2)}$, where $S^2=S^2(\mathbf{J})$ is viewed as a function of the $J_n$. Substituting $I_n$ for $J_n$ yields $S^2(\mathbf{J}(\mathbf{I}))=0$, rendering the expression singular. Nevertheless, the formal relation remains useful if one restricts to terms that do not involve negative powers of $S^2$.
				\end{remark}
				
				Conversely, the inverse transformation is ill-defined.
				Notice that
				\begin{equation}
					\begin{aligned}
						\frac{(-2J_0)^{n}(I_n-\delta_{n,1})}{(2n-1)!!}&=1-\sum_{k=1}^{n}\frac{J_k(-2J_0)^{k-1}}{(2k-1)!!}\\
						&=1-\sum_{k=1}^{n}\sum_{m=0}^{\infty} \frac{(2k+2m-1)(-2)^{k-1}t^{gB}_{k+m-1}}{(2k-1)!!m!}J_0^{k+m+1}\\
						&=1-\sum_{l=0}^{\infty}\sum_{k=1}^{\min\{l+1,n\}}\frac{(2l+1)(-2)^{k-1}}{(2k-1)!!(l+1-k)!}t^{gB}_lJ_0^l\\
						&=1-\sum_{l=0}^{\infty}(\frac{1}{l!}-\frac{(-2)^{n}}{(2n-1)!!(l-n)!})t^{gB}_lJ_0^l\\
						&=1-\sum_{l=0}^{\infty}t^{gB}_l\frac{J_0^l}{l!}+\sum_{l=n}^{\infty} \frac{(-2)^n}{(2n-1)!!(l-n)!}t^{gB}_lJ_0^l\\
						&=\frac{S}{\sqrt{2J_0}}+\sum_{l=n}^{\infty} \frac{(-2)^n}{(2n-1)!!(l-n)!}t^{gB}_lJ_0^l,
					\end{aligned}
				\end{equation}
				where we have used the combinatorial identity
				\begin{equation}
					\sum_{k=1}^{n}\frac{(2l+1)(-2)^{k-1}}{(2k-1)!!(l+1-k)!}=\frac{1}{l!}-\frac{(-2)^n}{(2n-1)!!(l-n)!}
				\end{equation}
				with $(-a-1)!=\Gamma(-a)=\infty$ for $a\in \mathbb{Z}_{\geq0}$, and with this identity for $n=l$ one has
				\begin{equation}
					\frac{S}{\sqrt{2J_0}}=1-\sum_{k=1}^{\infty}\frac{J_k(-2J_0)^{k-1}}{(2k-1)!!}=1-\sum_{l=0}^{\infty}t^{gB}_l\frac{J_0^l}{l!}.
				\end{equation}
				It follows that for $n\geq 1$
				\begin{equation}
					\begin{aligned}
						t^{WK}_n=&\sum_{m=0}^{\infty}\frac{I_{n+m}}{m!}(-I_0)^m\\
						=& \delta_{n,1} +\sum_{m=0}^{\infty} \frac{(2m+2n-1)!!}{2^{m+n}m!}(-J_0)^{-n}\Big(\frac{S}{\sqrt{2J_0}}\Big.\\
						&\Big.+\sum_{l=n+m}^{\infty} \frac{(-2)^{n+m}}{(2(n+m)-1)!!(l-n-m)!}t^{gB}_lJ_0^l\Big)\\
						=&\delta_{n,1}+\sum_{m=0}^{\infty}  \frac{(2m+2n-1)!!}{2^{m+n}m!}(-J_0)^{-n}\frac{S}{\sqrt{2J_0}}+t^{gB}_n.
					\end{aligned}
				\end{equation}
				It is easy to verify the series
				\begin{equation}
					\sum_{m=0}^{\infty}\frac{(2m+2n-1)!!}{(2m)!!} =\infty.
				\end{equation}
				It has the same problem when computing $t^{WK}_0$. Hence the inverse transformation makes no sense.
				
				These coordinate transformations illustrate an important point: before renormalization and transition, one is simply manipulating formal power series. After renormalization, however, the coordinates begin to resemble genuine charts on an infinite-dimensional manifold—though one that exhibits properties quite distinct from the finite-dimensional manifold. For example, the two theories (the Witten-Kontsevich model and the generalized BGW theory) cannot be transformed into one another using their original coordinates $\{t^{WK}_n\}_{n=0}^{\infty}$ and $\{S^2,t^{gB}_n\}_{n=0}^{\infty}$ (at least in the present setting). Yet after renormalization, the transition becomes possible. Even then, the Witten-Kontsevich model cannot be mapped to the ordinary BGW theory (which corresponds to $J_0=0$). This suggests that the hypothetical infinite-dimensional manifold is far from trivial. We leave this geometric perspective for future research.

			\section{Application}\label{section application}
			In this section, we provide some applications of the emergent geometry of the generalized BGW theory.
			
			\subsection{Alexandrov conjecture for genus-zero expression}\label{subsection alexandrov conjecture}
			In \cite{alexandrovCutandjoinDescriptionGeneralized2019}, Alexandrov conjectured that the genus-zero free energy of the generalized BGW theory has the expression
			\begin{equation}\label{Alexandrov conjecture}
				F^{gB}_0=\sum_{k=0}^{\infty}\frac{S^{2k+2}}{2^k}\sum_{p=1}^{\infty}\sum_{j_a} \frac{(p+2k-1)!}{(2k+2)!}\prod_{a=0}^{\infty}\frac{1}{j_a!}(\frac{t_a}{a!})^{j_a},
			\end{equation}
			where the sum $\sum_{j_a}$ runs over all sequences ${\{(j_a)_{a=0}^{\infty}:\sum_a aj_a=k,\sum_a j_a=p\}}$. It has been proved by Okuyama and Sakai in \cite{okuyamaJTSupergravityBrezinGrossWitten2020}. In this subsection, we use the constitutive relation to prove this conjecture.
			
			Starting from the Landau-Ginzburg equation (\ref{Q equ}) for the order parameter $Q$, we consider the generalized Lagrange inversion formula for $Q^n$.
			Let
			\begin{equation}
				z=\frac{Q^n}{(\sum_{k=0}^{\infty} \frac{t_k}{(2k)!!}Q^{2k+1}+S )^{n}}.
			\end{equation}
			Then $Q^n$ has a Puiseux-series expansion
			\begin{equation}
				Q^n=\sum_{m\geq 1,m\in \frac{1}{n}\mathbb{Z}} a_m z^m.
			\end{equation}
			$a_m$ can be formally computed by the residue
			\begin{equation}
				\begin{aligned}
					a_m&=\text{res}_{z=0} \frac{Q^n}{z^{m+1}}\d z=-\frac{1}{m}\text{res}_{z=0}Q^n\d \frac{1}{z^m}=\frac{1}{m}\text{res}_{Q^n=0}\frac{1}{z^m}\d (Q^n)\\
					&=\frac{1}{m}\text{res}_{w=0}w^{-m}\big(\sum_{k=0}^{\infty}\frac{t_k}{(2k)!!}w^{2k+1}+S \big)^{nm}\d w\\
					&=\frac{1}{m}\sum_{l=0}^{nm} \binom{nm}{l} S^l\sum_{2a_1+2a_2+\cdots+2a_{nm-l}=l-n}\prod_{i=1}^{nm-l}\frac{t_{a_i}}{(2a_i)!!}.
				\end{aligned}
			\end{equation}
			Taking $z=1$ yields the expression
			\begin{equation}
				\begin{aligned}
					Q^n&=\sum_{m\geq1,m\in \frac{1}{n}\mathbb{Z}}\frac{1}{m}\sum_{l=0}^{nm} \binom{nm}{l} S^l\sum_{2a_1+2a_2+\cdots+2a_{nm-l}=l-n}\prod_{i=1}^{nm-l}\frac{t_{a_i}}{(2a_i)!!}\\
					&=\sum_{k=0}^{\infty} \frac{S^{2k+n}}{2^k} \sum_{p=0}^{\infty}\frac{n}{p+2k+n}\binom{p+n+2k}{n+2k}\sum_{a_1+\dots+a_p=k}\prod_{i=1}^{p}\frac{t_{a_i}}{a_i!}\\
					&=\sum_{k=0}^{\infty}\frac{S^{2k+n}}{2^k}\sum_{p=0}^{\infty} \sum_{j_a}\frac{n(p+n+2k-1)!}{(n+2k)!}\prod_{a=0}^{\infty}\frac{1}{j_a!}(\frac{t_a}{a!})^{j_a},
				\end{aligned}
			\end{equation}
				where the sum $\sum_{j_a}$ runs over all sequences ${\{(j_a)_{a=0}^{\infty}:\sum_a aj_a=k,\sum_a j_a=p\}}$. 
				
				In particular, by Theorem \ref{Zhang's main thm}, 
				\begin{equation}
					\begin{aligned}
						\frac{\partial^2 F^{gB}_0}{\partial S\partial t_n} =&\frac{Q^{2n+1}}{2^n(2n+1)n!}\\
						=&\sum_{k=0}^{\infty}\frac{S^{2k+2n+1}}{2^{k+n}n!}\sum_{p=0}^{\infty} \sum_{j_a}\frac{(p+2n+2k)!}{(2n+2k+1)!}\prod_{a=0}^{\infty}\frac{1}{j_a!}(\frac{t_a}{a!})^{j_a}.
					\end{aligned}
				\end{equation}
				Integrating it with respect to $S$, one obtains
				\begin{equation}
					\begin{aligned}
							\frac{\partial F_0^{gB}}{\partial t_n}=&\int_{0}^{S}\frac{\partial^2 F^{gB}_0}{\partial S\partial t_n} \d S\\
							=&\sum_{k=0}^{\infty}\frac{S^{2k+2n+2}}{2^{k+n}n!}\sum_{p=0}^{\infty} \sum_{j_a}\frac{(p+2n+2k)!}{(2n+2k+2)!}\prod_{a=0}^{\infty}\frac{1}{j_a!}(\frac{t_a}{a!})^{j_a}.
					\end{aligned}
				\end{equation}
				It is easy to verify that the expression in (\ref{Alexandrov conjecture}) satisfies those derivatives with the initial value zero. This completes the proof of the Alexandrov conjecture.

				\subsection{Okuyama-Sakai conjecture}\label{subsection Okuyama-Sakai}
				In the theory of integrable hierarchies, there exists another coordinate system for the Gromov-Witten theory called the \textbf{jet coordinates} (refer to \cite{dubrovinNormalFormsHierarchies2001}). For the Witten-Kontsevich model, the jet coordinates are the variables
				\begin{equation}
					u:=I_0=\frac{\partial^2 F_0^{WK}}{(\partial t_0^{WK})^2},\quad u^{(1)}:=\frac{\partial I_0}{\partial t^{WK}_0},  \cdots, u^{(n)}:=\frac{\partial^n I_0}{(\partial t^{WK}_0)^n},\cdots.
				\end{equation}
				
				Eguchi, Yamada and Yang proved in \cite{eguchiGenusExpansionTopological1995} that, when $g\geq 2$, the genus $g$ free energy of the Witten-Kontsevich model is a function of $3g-2$ jet variables. More precisely, the free energy takes the form
				\begin{equation}\label{WK jet expression}
					F_g^{WK}=\sum_{\sum_{k}(k-1)l_k=3g-3}a_{l_2\cdots l_{3g-2}} \frac{[u^{(2)}]^{l_2}\cdots[u^{(3g-2)}]^{l_{3g-2}}}{[u^{(1)}]^{2-2g+\sum_k l_k}},
				\end{equation}
				with the constant coefficients $a_{l_2\cdots l_{3g-2}}$. 
				
				In \cite{okuyamaJTSupergravityBrezinGrossWitten2020}, Okuyama-Sakai conjectured that the genus $g$ free energy of the generalized BGW theory is the same function of the corresponding jet variables with a correction term in $S^2$. More precisely, define the jet variables of the generalized BGW theory by
				\begin{equation}
					v:=J_0=\frac{\partial^2 F_0^{gB}}{(\partial t^{gB}_0)^2},\quad v^{(1)}:=\frac{\partial J_0}{\partial t_0^{gB}}, \cdots,v^{(n)}:=\frac{\partial^n J_0}{(\partial t^{gB}_0)^n},\cdots.
				\end{equation}
				Then, one has
				\begin{thm}[Okuyama-Sakai Conjecture\cite{okuyamaJTSupergravityBrezinGrossWitten2020}, Yang-Zhang Theorem\cite{yangHodgeBGWCorrespondence2024,yangNewProofOkuyama2023}]\label{Okuyama-Sakai conjecture}
					When $g\geq 2$, the genus $g$ free energy of the generalized BGW theory has the expression:
					\begin{equation}
						\begin{aligned}
									F_g^{gB}=&\sum_{\sum_{k}(k-1)l_k=3g-3}a_{l_2\cdots l_{3g-2}} \frac{[v^{(2)}]^{l_2}\cdots[v^{(3g-2)}]^{l_{3g-2}}}{[v^{(1)}]^{2-2g+\sum_k l_k}}-\frac{(-1)^{g}B_{2g}}{2g(2g-2)}S^{-(2g-2)}\\
									=&F^{WK}_g|_{u^{(n)}\to v^{(n)}}-\frac{(-1)^{g}B_{2g}}{2g(2g-2)}S^{-(2g-2)}.
						\end{aligned}
					\end{equation}
					Here $B_{2g}$ denotes the $2g$-th Bernoulli number.
				\end{thm}
				It has been proved by Yang and Zhang in \cite{yangHodgeBGWCorrespondence2024,yangNewProofOkuyama2023}. Here we provide another proof using Theorem \ref{expression of F}. 
				First, we have the lemma:
				\begin{lemma}
					With the transition (\ref{tran from I to J}) and (\ref{tran from J to I}), the derivatives of a function with respect to $t^{WK}_0$ and $t^{gB}_0$ are the same. More precisely, 
					\begin{equation}
						\frac{\partial f(\mathbf{J})}{\partial t^{gB}_0}=\sum_{n}\frac{\partial J_n}{\partial t^{gB}_0}\frac{\partial f(\mathbf{J})}{\partial J_n}=\sum_{n}\frac{\partial I_n}{\partial t^{WK}_0}\frac{\partial f(\mathbf{J}(\mathbf{I}))}{\partial I_n}=\frac{\partial f(\mathbf{J}(\mathbf{I}))}{\partial t_0^{WK}},
					\end{equation}
					where $\frac{\partial J_n}{\partial t^{gB}_0}$ and $\frac{\partial I_n}{\partial t^{WK}_0}$ are understood as the functions of $\mathbf{J}$ and $\mathbf{I}$ respectively.
				\end{lemma}
				\begin{proof}
					We only need to show, for $n\geq 0$, 
					\begin{equation}
						\frac{\partial J_n}{\partial t^{gB}_0}=\sum_{m=0}^{\infty} \frac{\partial I_m}{\partial t^{WK}_0} \frac{\partial J_n}{\partial I_m}.
					\end{equation}
					Recall (\ref{jacobian for J0}) and (\ref{Jac for Jn}),
					\begin{equation}
						\frac{\partial J_0}{\partial t_0^{gB}}=\frac{2J_0}{1-J_1},\quad \frac{\partial J_n}{\partial t_0^{gB}}=\frac{2J_0J_{n+1}}{1-J_1}+\delta_{n,1} \ \forall n\geq 1.
					\end{equation}
					It has been computed in \cite{zhouTopological1DGravity2014}, 
					\begin{equation}
						\frac{\partial I_0}{\partial t^{WK}_0}=\frac{1}{1-I_1}, \quad \frac{\partial I_n}{\partial t^{WK}_0}=\frac{I_{n+1}}{1-I_1}\ \forall n\geq1.
					\end{equation}
					Consequently, with the transformation (\ref{tran from I to J}), for all $n\geq 2$, 
					\begin{equation}
						\begin{aligned}
							\sum_{m=0}^{\infty} \frac{\partial I_m}{\partial t^{WK}_0}\frac{\partial J_n}{\partial I_m}= \frac{2I_{n}}{1-I_1}+\frac{2I_{n+1}}{1-I_1}+\frac{(2n-1)I_{n}}{1-I_1}=\frac{\partial J_n}{\partial t^{gB}_0}.
						\end{aligned}
					\end{equation}
					It also holds true for $n=0,1$. This completes the proof.
				\end{proof}
				
					It implies that, under the transformation, the jet variables of the two models are equal, i.e.,
					\begin{equation}
						u^{(n)}|_{\mathbf{I}=\mathbf{I}(\mathbf{J})}=v^{(n)}.
					\end{equation} 
					
					It follows that, by Theorem \ref{expression of F},
					\begin{equation}
						F^{gB}_g+F^{WK}_g(\mathbf{I}(S^2))=F^{WK}_g(\mathbf{I}(\mathbf{J}))=F^{WK}_g(u^{(n)})|_{\mathbf{I}=\mathbf{I}(\mathbf{J})}=F^{WK}_g|_{u^{(n)}\to v^{(n)}}.
					\end{equation}
					
				This completes the main part of the Okuyama-Sakai conjecture. The additional term corresponds to the identity for the intersection numbers
				\begin{equation}\label{identity of psi and B}
					\sum_{\sum_{j}(j-1)l_j=3g-3}\langle\prod_{j=2}^{3g-2}\frac{1}{l_j!}((-1)^{j}(2j-1)!!\tau_{j})^{l_j}\rangle_{g}=\frac{(-1)^{g}B_{2g}}{2g(2g-2)}.
				\end{equation}
				To the best of our knowledge, we do not have a simple proof. We refer to the proof of Yang and Zhang \cite{yangHodgeBGWCorrespondence2024}. We note that it is equivalent to a conjecture of Kazarian and Norbury \cite{kazarianPolynomialRelationsKappa2024}.
				
				For a sequence $(p_k)_{k=2}^{\infty}$, define a sequence $(s_i)_{i=1}^{\infty}$ related to it by the series
				\begin{equation}
					1-\sum_{k=2}^{\infty} p_k w^k=\exp\big(-\sum_{i=1}^{\infty} s_iw^i\big).
				\end{equation}
				Then, by a well-known result (see \cite{liuRecursionFormulaeHigher2008}), there is an identity for intersection numbers of $\psi$-classes and $\kappa$-classes
				\begin{equation}
					\langle \exp(\sum_{k=2}^{\infty} p_k\tau_k)\rangle_{g}=\langle \exp(\sum_{i=1}^{\infty} s_i\kappa_i)\rangle_g.
				\end{equation}
				In \cite{kazarianPolynomialRelationsKappa2024}, Kazarian and Norbury take $p_k=(-1)^k(2k-1)!!$ and define the cohomological class
				\begin{equation}
					\mathbb{K}:=\exp(\sum_{i=1}^{\infty}s_i\kappa_i)=K_1+K_2+\cdots\in H^*(\overline{M}_g,\mathbb{Q}). 
				\end{equation}
				 Then, the identity (\ref{identity of psi and B}) is precisely the conjecture of Kazarian and Norbury, proved by Yang and Zhang (also proved by Chidambaram, Garcia-Failde and Giacchetto in \cite{chidambaramRelations$overlinemathcalM_gn$Negative2025}):
				 \begin{equation}
				 	\langle \exp(\sum_{k=2}^{\infty}p_k\tau_k)\rangle_g=\int_{\overline{M}_g}\mathbb{K}=\int_{\overline{M}_g}K_{3g-3}=\frac{(-1)^{g}B_{2g}}{2g(2g-2)}=|\chi(M_g)|,
				 \end{equation}
				 where $\chi(M_g)$ is the orbifold Euler characteristic of the moduli space of curves.

			\section{Conclusion}\label{section last}
			In this article, we have studied the emergent geometry of the generalized BGW theory. Starting from the suitable order parameter $J_0=\frac{\partial^2 F^{gB}_0}{\partial t_0^2}$, 
			we derived the constitutive relation
				\begin{equation*}
					\begin{aligned}
							&\frac{\partial^2 F^{gB}_0}{\partial S^2}=\frac{1}{2}\log\frac{2J_0}{S^2},\quad \frac{\partial^2 F_0^{gB}}{\partial S\partial t_m}=\frac{\sqrt{2J_0}J_0^m}{(2m+1)m!},\\
							&\frac{\partial^2 F^{gB}_0}{\partial t_n\partial t_m}=\frac{J_0^{n+m+1}}{(n+m+1)n!m!}.
					\end{aligned}
			\end{equation*}
			Using its Landau-Ginzburg potential, we defined the renormalized coordinates $J_n$ of the generalized BGW theory.
			The spectral curve defined by the one-point special deformation was shown to take the form
			\begin{equation*}
				\begin{aligned}
					y^{gB}(w)=& w^{-\frac{1}{2}}-\sum_{n=0}^{\infty} \frac{t_n}{(2n-1)!!}w^{n-\frac{1}{2}}-\sum_{n=0}^{\infty}\frac{(2n+1)!!}{w^{n+\frac{3}{2}}} \frac{\partial F^{gB}_0}{\partial t_n}\\
					=&\frac{\sqrt{w-2J_0}}{w}(1-\sum_{n=0}^{\infty}\frac{(w-2J_0)^{n}}{(2n+1)!!}J_{n+1}).
				\end{aligned}
			\end{equation*}
			Under the coordinate change $w=z^2+2J_0$, the multidifferentials defined by the shifted loop operator
			\begin{equation*}
				\begin{aligned}
						&W^{gB}_{g,n}(z_1,z_2,\cdots,z_n):=\frac{1}{2^n}\nabla^{gB}(z_1)\nabla^{gB}(z_2)\cdots\nabla^{gB}(z_n)F^{gB}_g\prod_{i=1}^{n}\d (z_i^2)\\
						&=\sum_{i_1,i_2,\cdots,i_n\geq 0}\frac{\partial}{\partial t_{i_1}}\frac{\partial}{\partial t_{i_2}}\cdots\frac{\partial}{\partial t_{i_n}}F^{gB}_{g}
					\cdot \prod_{k=1}^{n}\frac{(2i_k+1)!!z_k\d z_k}{(z_k^2+2J_0)^{i_k+\frac{3}{2}}}.
				\end{aligned}
			\end{equation*}
			satisfy the Eynard-Orantin topological recursion with the spectral curve $y^{gB}(z)$ and $x^{gB}(z)=\frac{1}{2}z^2+J_0$, using the standard Bergman kernel $B_0(z_1,z_2)$ on the Riemann sphere.
			
			 Through the Eynard-Orantin topological recursion, we established a transition between the Witten-Kontsevich model and the generalized BGW theory:
			\begin{equation*}
					\begin{aligned}
					I_0&=J_0;\\
						\frac{I_n}{(2n-1)!!}&=(-2J_0)^{-n}\big(1-\sum_{k=1}^{n}\frac{J_k(-2J_0)^{k-1}}{(2k-1)!!} \big)+\delta_{n,1}.
				\end{aligned}
			\end{equation*}
			Consequently, the multidifferentials of the two theories are equal for all stable $(g,n)$:
			\begin{equation*}
				W^{WK}_{g,n}(z_{[n]};\mathbf{I}(\mathbf{J}))=W^{gB}_{g,n}(z_{[n]};\mathbf{J}).
			\end{equation*}
			In particular, for $g\geq1$, we obtained the relation
			\begin{equation*}
				\nabla^{WK}(z)F^{WK}_{g}(\mathbf{I})\big|_{\mathbf{I}=\mathbf{I}(\mathbf{J})}=\nabla^{gB}(z)F^{gB}_{g}(\mathbf{J}),
			\end{equation*}
			which yields the explicit formula
			\begin{equation*}
					F^{gB}_{g}(\mathbf{J})=F^{WK}_{g}(\mathbf{I}(\mathbf{J}))-F^{WK}_{g}(\mathbf{I}(S^2)),
			\end{equation*}
			
			We close with two future directions:\\
			First, the methods developed here illustrate a general emergent geometry framework that may be applicable to other Gromov-Witten type theories. We expect this perspective to stimulate further research in the field.\\
			Second, there may be a new sequence of variables as the descendants of the parameter $S$. It could extend the generalized BGW theory to a more general theory corresponding to a two-dimensional Frobenius manifold.

			\appendix

				\section{Identification in genus zero via Landau-Ginzburg Theory}\label{appendix identification}
			
			In Remark \ref{tran from LG}, 
			we noted that the Witten-Kontsevich model and the generalized BGW theory can be identified at genus zero via their Landau-Ginzburg potentials. We now elaborate on this point.
			
			It was pointed out in \cite{zhouTopological1DGravity2014} that the spectral curve (\ref{spec of WK}) satisfies the relation:
			\begin{equation}
				(y^{WK}(w)^2)_{-}=0.
			\end{equation}
			That is, the negative part of the square of the spectral curve (with respect to the degree of $w$) vanishes, which is precisely equivalent to the Virasoro constraints for the genus-zero free energy: 
			\begin{equation}
				\begin{aligned}
					(2m+3)!!\frac{\partial F^{WK}_0}{\partial t_{m+1}}=&\sum_{k=0}^{\infty}\frac{(2k+2m+1)!!}{(2k-1)!!}t_k\frac{\partial F^{WK}_0}{\partial t_{k+m}}\\
					&+\frac{1}{2}\sum_{a+b=m-1} (2a+1)!!(2b+1)!!\frac{\partial^2 F^{WK}_0}{\partial t_a\partial t_b}+\delta_{m,-1}\frac{t_0^2}{2}. 
				\end{aligned}
			\end{equation}
			The relationship between $\frac{\partial F^{WK}_0}{\partial t_n}$ and $t_n$ is unique. More precisely:
			\begin{proposition}\label{y2-=0}
				For a Puiseux series 
				\begin{equation}
					y_+(w)=w^{\frac{1}{2}}-\sum_{n=0}^{\infty} a_n w^{n-\frac{1}{2}}
				\end{equation}
				in infinitely many indeterminates $a_0,a_1,\cdots$, there exists a unique sequence of formal power series $b_0,b_1,\cdots\in \mathbb{C}[[a_0,a_1,\cdots]]$ such that the series
				\begin{equation}\label{P series}
					y(w)=w^{\frac{1}{2}}-\sum_{n=0}^{\infty} a_n w^{n-\frac{1}{2}}-\sum_{m=0}^{\infty} b_m w^{-m-\frac{3}{2}}
				\end{equation}
				satisfies $(y(w)^2)_-=0$.
			\end{proposition}
			\begin{proof}
				A direct computation yields
				\begin{equation}
					\begin{aligned}
						y(w)^2_-=&a_0^2 w^{-1}-2\sum_{m=0}^{\infty} b_m w^{-m-1}+2\sum_{n-m-2<0}a_n b_m w^{n-m-2}+\sum_{n,m=0}^{\infty}b_n b_m w^{-(n+m+3)}\\
						=&(a_0^2-2b_0+2\sum_{m=0}^{\infty}a_{m+1}b_m )w^{-1}+(-2b_1+2\sum_{m=0}^{\infty}a_mb_m)w^{-2}\\
						&+\sum_{n=1}^{\infty}\big(-2b_{n+1}+2\sum_{m=0}^{\infty}a_mb_{m+n}+\sum_{i+j=n-1}b_ib_j\big)w^{-n-2}.
					\end{aligned}
				\end{equation}
				So $(y(w)^2)_-=0$ is equivalent to the infinite system
				\begin{equation}
					\begin{aligned}
						b_0=& \frac{1}{2}a_0^2+& a_1b_0 +& a_2b_1+&a_3b_2+ \cdots; &\\
						b_1= &            &a_0b_0 + & a_1b_1+& a_2b_2 + \cdots;&\\
						b_2=&              &                  & a_0b_2+&a_1b_3+\cdots+&\frac{1}{2} b_0^2;\\
						b_3=&               &                  &    &a_0b_3+\cdots+&b_0b_1;\\
						\cdots\cdots
					\end{aligned}
				\end{equation}
				One can solve for the coefficients recursively by expanding in ordinary degree. More precisely, write
				\begin{equation}
					b_n=b_n^{(0)}+b_n^{(1)}+\cdots+b_n^{(j)}+\cdots,
				\end{equation}
				where $b_n^{(j)}$ is homogeneous of degree $j$ in the variables $a_0,a_1,\cdots$ with $\text{deg }a_n=1$. Using such decompositions, one can deduce by induction from the
				above system of equations:
				\begin{equation}
					b_n^{(j)}=0, \qquad j=0,1,\cdots,n+1.
				\end{equation}
				Furthermore, for any fixed degree,  the expression of $b_n$ is uniquely determined.
			\end{proof}
			Viewing $\{a_n\}$ as coordinates, there exists a renormalized coordinate system determined by a Landau-Ginzburg potential that brings the series (\ref{P series}) into a form involving only positive powers. Concretely:
			\begin{proposition}\label{universal LG prop}
				For the series 
				\begin{equation}\label{universal spec}
					y(w)=w^{\frac{1}{2}}-\sum_{n=0}^{\infty} \frac{\hat{t}_n}{(2n-1)!!}w^{n-\frac{1}{2}}-\sum_{m=0}^{\infty} f_m w^{-m-\frac{3}{2}}
				\end{equation}
				satisfying $(y(w)^2)_-=0$, there exists a sequence $\hat{\mathbf{I}}=\{\hat{I}_0,\hat{I}_1,\cdots\}$ defined by the Landau-Ginzburg potential:
				\begin{equation}\label{univerasl LG}
					W(x)=-\frac{1}{2}x^2+\sum_{n=0}^{\infty} \frac{\hat{t}_n}{n!} x^n,
				\end{equation}  
				with
				\begin{equation}
					W'(\hat{I}_0)=0;
				\end{equation}
				and
				\begin{equation}
					W(x)=W(\hat{I}_0)-\frac{1}{2}(x-\hat{I}_0)^2+\sum_{n=0}^{\infty}\hat{I}_{n+1}\frac{(x-\hat{I}_0)^{n+1}}{(n+1)!},
				\end{equation}
				such that 
				\begin{equation}
					y(w)=(w-2\hat{I}_0)^{\frac{1}{2}}\big(1-\sum_{n=0}^{\infty}\hat{I}_{n+1}\frac{(w-2\hat{I}_0)^n}{(2n+1)!!}\big).
				\end{equation}
			\end{proposition}
			\begin{proof}
				By Proposition \ref{y2-=0}, the sequence $\{f_n\}$ is unique. 
				The remaining statements follow from the properties of the spectral curve and the Landau-Ginzburg potential of the Witten-Kontsevich model.
			\end{proof}
			In the Landau-Ginzburg theory (\ref{univerasl LG}) and the series (\ref{universal spec}), taking
			\begin{equation}\label{t in WK}
				\hat{t}_n=t^{WK}_n; \qquad f_n=(2n+1)!!\frac{\partial F^{WK}_0}{\partial t_n}
			\end{equation}
			yields the Witten-Kontsevich model.
			
			Taking instead
			\begin{equation}\label{t in gB}
				\hat{t}_0=\frac{\partial F^{gB}_0}{\partial t_0}; \ \hat{t}_n=(2n-1)t_{n-1}^{gB}, \quad (n\geq 1),
			\end{equation}
			(note that $\frac{\partial F^{gB}_0}{\partial t_0}$ here should be viewed as an explicit function of $\mathbf{t}^{gB}$ and $S^2$) and
			\begin{equation}\label{f in gB}
				f_n=(2n+3)!!\frac{\partial F^{gB}_0}{\partial t_{n+1}}, \quad n=0,1,\cdots,
			\end{equation}
			the condition $(y(w)^2)_-=0$ translates into the genus-zero Virasoro constraints for the generalized BGW theory with $m\geq 1$, namely (\ref{genus zero part of gB}):
			\begin{equation*}
				\begin{aligned}
					&(2m+1)!!\frac{\partial F^{gB}_0}{\partial t_m} =\sum_{k=0}^{\infty} \frac{(2k+2m+1)!!}{(2k-1)!!} t_{k}\frac{\partial F^{gB}_0}{\partial t_{m+k}}\\
					&+\frac{1}{2}\sum_{a+b=m-1}(2a+1)!!(2b+1)!!\frac{\partial F^{gB}_0}{\partial t_a}\frac{\partial F^{gB}_0}{\partial t_b}.
				\end{aligned}
			\end{equation*}
			In this case, the critical point $\hat{I}_0=J_0$ satisfies the Landau-Ginzburg equation (\ref{LG equ for J0}). The spectral curve of the generalized BGW theory (\ref{def of spectral curve}) corresponds to
			\begin{equation}
				y^{gB}(w)=y(w)w^{-1},
			\end{equation}
			with $J_n=\hat{I}_n$. This yields another proof of Proposition \ref{spec curve in J}.
			
			Now, we give a proof of Proposition \ref{cons rel of F in J}, the constitutive relation of the generalized BGW theory. First, by the weighted homogeneity and the constitutive relation of the Witten-Kontsevich model, see ((\ref{deg of WK}) and (\ref{cons of WK})), 
			\begin{equation}
				\begin{aligned}
					\frac{\partial F^{WK}_0}{\partial t_m}=&\frac{1}{-2-m}\sum_{n=0}^{\infty} (n-1)t^{WK}_n\frac{I_0^{m+n+1}}{(n+m+1)n!m!}\\
					=&-\frac{1}{m+2}\sum_{n=0}^{\infty} t_n^{WK}\frac{I_0^{n+m+1}}{n!m!}+\sum_{n=0}^{\infty}t_{n}^{WK} \frac{I_0^{n+m+1}}{(n+m+1)n!m!}\\
					=&-\frac{I_0^{m+2}}{(m+2)m!}+\sum_{n=0}^{\infty}t_n^{WK}\frac{I^{n+m+1}_0}{(n+m+1)n!m!}.
				\end{aligned}
			\end{equation}
			This relation is universal, i.e.
			\begin{lemma}
				With the notation of Proposition \ref{universal LG prop}, we have
				\begin{equation}
					\begin{aligned}
						f_m=&-\frac{(2m+1)!!}{m+2}\sum_{n=0}^{\infty}(n-1)\hat{t}_n\frac{\hat{I}_0^{n+m+1}}{(n+m+1)n!m!}\\
						=&-\frac{(2m+1)!!}{(m+2)m!}\hat{I}_0^{m+2}+(2m+1)!!\sum_{n=0}^{\infty}\hat{t}_n\frac{\hat{I}_0^{n+m+1}}{(n+m+1)n!m!}.
					\end{aligned}
				\end{equation}
			\end{lemma}
			With this lemma, we proceed to the proof of the constitutive relation of the generalized BGW theory.
			
			\begin{proof}[Proof of the constitutive relation of the generalized BGW theory, Proposition \ref{cons rel of F in J}]
				Using the substitution (\ref{t in gB}) and (\ref{f in gB}), one has
				\begin{equation}
					\begin{aligned}
						\frac{\partial F^{gB}_0}{\partial t_{m+1}}=&-\frac{J_0^{m+2}}{(2m+3)(m+2)m!}+\frac{\partial F_0^{gB}}{\partial t_0}\frac{J_0^{m+1}}{(2m+3)(m+1)!}\\
						&+\frac{1}{(2m+3)}\sum_{n=1}^{\infty}(2n-1)t^{gB}_{n-1}\frac{J_0^{n+m+1}}{(n+m+1)n!m!}.
					\end{aligned}
				\end{equation}
				Together with the Landau-Ginzburg equation (\ref{LG equ for J0}) (which can be taken as the definition of $J_0$ in this proof):
				\begin{equation*}
					\frac{\partial F^{gB}_0}{\partial t_0}=J_0-\sum_{n=0}^{\infty} (2n+1)t_n^{gB}\frac{J_0^{n+1}}{(n+1)!},
				\end{equation*}
				this yields
				\begin{equation}\label{der F with tm}
					\begin{aligned}
						\frac{\partial F^{gB}_0}{\partial t_{m+1}}=&-\frac{J_0^{m+2}}{(2m+3)(m+2)m!}+\frac{1}{(2m+3)}\sum_{n=1}^{\infty}(2n-1)t^{gB}_{n-1}\frac{J_0^{n+m+1}}{(n+m+1)n!m!}\\
						&+\frac{J_0^{m+2}}{(2m+3)(m+1)!}-\sum_{n=0}^{\infty}(2n+1)t_{n}^{gB}\frac{J_0^{n+m+2}}{(2m+3)(n+1)!(m+1)!}\\
						=&\frac{J_0^{m+2}}{(2m+3)(m+2)!}-\sum_{n=0}^{\infty}\frac{2n+1}{2m+3}t_{n}^{gB}\frac{J_0^{n+m+2}}{(n+m+2)n!(m+1)!}.
					\end{aligned}
				\end{equation}
				Differentiating both sides with respect to $t_n$ yields
				\begin{equation}
					\begin{aligned}
						\frac{\partial^2 F^{gB}_0}{\partial t_n \partial t_{m+1}}=&\frac{J_0^{m+1}}{(m+1)!(2m+3)}\frac{\partial J_0}{\partial t_n}-\frac{2n+1}{2m+3}\frac{J_0^{m+n+2}}{(n+m+2)(m+1)!n!}\\
						&-\sum_{k=0}^{\infty} \frac{2k+1}{2m+3}t_{k}^{gB}\frac{J_0^{m+k+1}}{(m+1)!k!}\frac{\partial J_0}{\partial t_n}\\
						=& \frac{J_0^{m+1}}{(m+1)!(2m+3)}(1-J_1)\frac{\partial J_0}{\partial t_n}-\frac{2n+1}{2m+3}\frac{J_0^{m+n+2}}{(n+m+2)(m+1)!n!}.
					\end{aligned}
				\end{equation}
				In particular, for $n=0$
				\begin{equation}
					\frac{\partial^2 F^{gB}_0}{\partial t_0\partial t_{m+1}}=\frac{J_0^{m+1}}{(m+1)!(2m+3)}(1-J_1)\frac{\partial J_0}{\partial t_0}-\frac{J_0^{m+2}}{(2m+3)(m+2)!}.
				\end{equation}
				Differentiating the Landau-Ginzburg equation with respect to $t_0$ yields
				\begin{equation}
					\frac{\partial^2 F^{gB}_0}{\partial t_0^2}=(1-J_1)\frac{\partial J_0}{\partial t_0}-J_0.
				\end{equation}
				Therefore,
				\begin{equation}\label{sec der of F0 with 0 and m}
					\begin{aligned}
						\frac{\partial^2 F^{gB}_0}{\partial t_0\partial t_m}=&\frac{J_0^{m}}{m!(2m+1)}(\frac{\partial^2 F^{gB}_0}{\partial t_0^2}+J_0)-\frac{J_0^{m+1}}{(2m+1)(m+1)!}\\
						=&\frac{J_0^{m}}{m!(2m+1)}\frac{\partial^2 F^{gB}_0}{\partial t_0^2}+m\frac{J_0^{m+1}}{(m+1)!(2m+1)},
					\end{aligned}
				\end{equation}
				for $m\geq 1$, and one checks that the formula also holds for $m=0$.
				
				On the other hand, the Virasoro constraint
				\begin{equation*}
					\frac{\partial F^{gB}_0}{\partial t_0}=\sum_{n=0}^{\infty} (2n+1)t_n^{gB}\frac{\partial F^{gB}_0}{\partial t_n}+\frac{S^2}{2},
				\end{equation*}
				implies
				\begin{equation}
					\frac{\partial^2 F_0^{gB}}{\partial t_0^2}=\frac{\partial F^{gB}_0}{\partial t_0}+\sum_{n=0}^{\infty}(2n+1)t_n^{gB}\frac{\partial^2 F^{gB}_0}{\partial t_0\partial t_n}.
				\end{equation}
				Using the Landau-Ginzburg equation and equation (\ref{sec der of F0 with 0 and m}), we obtain 
				\begin{equation}
					\begin{aligned}
						&\frac{\partial^2 F^{gB}_0}{\partial t_0^2}=J_0-\sum_{n=0}^{\infty}(2n+1)t_n^{gB}\frac{J_0^{n+1}}{(n+1)!}\\
						&+\sum_{n=0}^{\infty} (2n+1)t_n^{gB}\frac{J_0^n}{n!(2n+1)}\frac{\partial^2 F^{gB}_0}{\partial t_0^2}+\sum_{n=0}^{\infty}(2n+1)t_{n}^{gB}n\frac{J_0^{n+1}}{(n+1)!(2n+1)}\\
						&=J_0\big(1-\sum_{n=0}^{\infty}t_n^{gB}\frac{J_0^n}{n!}\big)+\sum_{n=0}^{\infty}t_{n}^{gB}\frac{J_0^{n}}{n!}\frac{\partial^2 F^{gB}_0}{\partial t_0^2}.
					\end{aligned}
				\end{equation}
				It follows that
				\begin{equation*}
					\frac{\partial^2 F_0^{gB}}{\partial t_0^2}=J_0,
				\end{equation*}
				and therefore one gets the constitutive relation (\ref{cons for J}):
				\begin{equation*}
					\frac{\partial^2 F_0^{gB}}{\partial t_n\partial t_m}=\frac{J_0^{n+m+1}}{n!m!(n+m+1)}.
				\end{equation*}
				To verify the formula for $\frac{\partial^2 F^{gB}_0}{\partial S\partial t_n}$, we only need to show $2J_0=Q^2:=(\frac{\partial^2 F^{gB}_0}{\partial S\partial t_0})^2$.
				
				Substituting the equation (\ref{der F with tm}) into the Virasoro constraint yields
				\begin{equation}\label{lins1}
					\begin{aligned}
						\frac{S^2}{2}=&\frac{\partial F^{gB}_0}{\partial t_0}-\sum_{n=0}^{\infty}(2n+1)t_n^{gB}\frac{\partial F^{gB}_0}{\partial t_n}\\
						=&J_0-\sum_{n=0}^{\infty}(2n+1)t_n^{gB}\frac{J_0^{n+1}}{(n+1)!}-\sum_{n=0}^{\infty}(2n+1)t_n^{gB}\frac{J_0^{n+1}}{(2n+1)(n+1)!}\\
						&+\sum_{n,k=0}^{\infty}(2k+1)t_n^{gB}t_k^{gB}\frac{J_0^{n+k+1}}{(n+k+1)n!k!}\\
						=&J_0-2\sum_{n=0}^{\infty} t_n^{gB}\frac{J_0^{n+1}}{n!}+\sum_{n,k=0}^{\infty}t_n^{gB}t_k^{gB}\frac{J_0^{n+k+1}}{n!k!}\\
						=&J_0\big(1-\sum_{n=0}^{\infty}t_n^{gB}\frac{J_0^n}{n!}\big)^2,
					\end{aligned}
				\end{equation}
				where in the third line, we used 
				\begin{equation}
					\begin{aligned}
						&\sum_{n,k=0}^{\infty} (2k+1)t_n^{gB}t_k^{gB}\frac{J_0^{n+k+1}}{(n+k+1)n!k!}\\
						=&\frac{1}{2}	\big(\sum_{n,k=0}^{\infty} (2k+1)t_n^{gB}t_k^{gB}\frac{J_0^{n+k+1}}{(n+k+1)n!k!}
						+	(k\leftrightarrow n)\big)\\
						=&\sum_{n,k=0}^{\infty}t_k^{gB}t_n^{gB}\frac{J_0^{n+k+1}}{k!n!}.
					\end{aligned}
				\end{equation}
				On the other hand, differentiating the equation (\ref{der F with tm}) with respect to $S$ yields, for $m\geq1$,
				\begin{equation}
					\begin{aligned}
						\frac{\partial^2 F^{gB}_0}{\partial S\partial t_m}=&\frac{J_0^{m}}{(2m+1)m!}\frac{\partial J_0}{\partial S}-\sum_{n=0}^{\infty}\frac{2n+1}{2m+1}t_n^{gB}\frac{J_0^{m+n}}{n!m!}\frac{\partial J_0}{\partial S}\\
						=&\frac{J_0^{m}}{(2m+1)m!}(1-J_1)\frac{\partial J_0}{\partial S}.
					\end{aligned}
				\end{equation}
				Differentiating the Landau-Ginzburg equation with respect to $S$ yields
				\begin{equation}
					Q:=\frac{\partial^2 F^{gB}_0}{\partial S \partial t_0}=(1-J_1)\frac{\partial J_0}{\partial S}.
				\end{equation}
				Then differentiating the Virasoro constraint with respect to $S$ produces
				\begin{equation}
					\begin{aligned}
						S=&\frac{\partial^2 F^{gB}_0}{\partial S\partial t_0}-\sum_{n=0}^{\infty}(2n+1)t_n^{gB}\frac{\partial^2 F^{gB}_0}{\partial S\partial t_n}\\
						=&Q(1-\sum_{n=0}^{\infty}t_n^{gB}\frac{J_0^n}{n!}).
					\end{aligned}
				\end{equation}
				Comparing with the equation (\ref{lins1}), one has:
				\begin{equation*}
					J_0=\frac{Q^2}{2}.
				\end{equation*}
				This completes the proof of Proposition \ref{cons rel of F in J}.
			\end{proof}
			
			This identification via the Landau-Ginzburg potential provides a second link between the Witten-Kontsevich model and the generalized BGW theory at genus zero. At higher genera, the differing degrees of the spectral curves lead to different quantizations (see Subsection \ref{quantization of the spectral curve}). This interplay merits further investigation.					
			
			\begin{remark}\label{mod eGUE}
				There is another model, the modified even GUE model (see \cite{zhouEmergentGeometryMatrix2019} or \cite{zhangApplication2020}), which also fits into this framework. Taking
				\begin{equation}
					\hat{t}_0=2t; \ \hat{t}_n=(2n-1)!!2ns_{2n}, \ n\geq1 ;\ f_n=4\frac{\partial F^{eG}_0}{\partial s_{2n+2}},
				\end{equation}
				one finds that $y^{eG}(w)=\frac{1}{4}y(w)w^{-\frac{1}{2}}$ reproduces the spectral curve in \cite{zhouEmergentGeometryMatrix2019}. 
				This suggests that the modified even GUE model likewise possesses an order parameter, constitutive relations, and a topological recursion of the same type. We leave this topic for future investigation.
			\end{remark}

\end{document}